\documentclass[12pt,journal,onecolumn]{IEEEtran}

\usepackage[latin1]{inputenc}
\usepackage{amsmath}
\usepackage{algorithmic}
\usepackage{graphicx}
\usepackage{amsfonts}
\usepackage[dvipsnames]{color}
\definecolor{rot}{named}{RedOrange}
\definecolor{blau}{named}{NavyBlue}
\definecolor{orange}{named}{Dandelion}
\definecolor{weiss}{named}{White}
\definecolor{gelb}{rgb}{1,1,0.5}
\definecolor{gruen}{rgb}{.5,1,.5}
\definecolor{grau}{rgb}{.95,.95,.95}
\definecolor{background}{rgb}{.43,.91,.97} 
\definecolor{schwarz}{named}{Black}

\newtheorem{Cor}{Corollary}
\newtheorem{Def}{Definition}

\newtheorem{Lm}{Lemma}
\newtheorem{Ex}{Example}

\newtheorem{Prop}{Proposition}

\newtheorem{Alg}{Algorithm}

\newcommand{\diag}{{\mathit{diag}}}
\newcommand{\lmax}{{\lambda_{\textit{max}}}}

\newcommand{\N}{{\mathbb{N}}}

\newcommand{\D}{{\mathbb{D}}}
\renewcommand{\P}{{\mathbb{P}}}

\newcommand{\R}{{\mathbb{R}}}
\newcommand{\C}{{\mathbb{C}}}

\newcommand{\Ks}{{\mathcal{K}}}
\newcommand{\Fs}{{\mathcal{F}}}
\newcommand{\Ns}{{\mathcal{N}}}

\newcommand{\As}{{\mathcal{A}}}
\newcommand{\Bs}{{\mathcal{B}}}
\newcommand{\Cs}{{\mathcal{C}}}
\newcommand{\Ds}{{\mathcal{D}}}
\newcommand{\Es}{{\mathcal{E}}}

\newcommand{\Ls}{{\mathcal{L}}}
\newcommand{\Ms}{{\mathcal{M}}}

\newcommand{\Ps}{{\mathcal{P}}}

\newcommand{\Xs}{{\mathcal{X}}}

\newcommand{\sr}{^{\frac{1}{2}}}
\newcommand{\ma}[1]{{\boldsymbol{#1}}}
\newcommand{\bma}[1]{{\bar{\boldsymbol{#1}}}}

\newcommand{\ve}[1]{{\boldsymbol{#1}}}
\newcommand{\bve}[1]{{\bar{\boldsymbol{#1}}}}

\newcommand{\bs}[1]{{\boldsymbol{#1}}}

\newcommand{\sir}{{\mathrm{SIR}}}

\newcommand{\la}{\langle}
\newcommand{\ra}{\rangle}

\usepackage{url}
\begin{document}

\title{A new graph perspective on max-min fairness in Gaussian parallel channels}

\author{Marcin Wiczanowski$^{\dag}$, Holger Boche$^{\ast\dag}$\\ 
  {\centering
    $^{\ast}$Heinrich-Hertz Group for Mobile Communications, EECS,\\
    Berlin University of Technology,
    Einsteinufer 25, 10587 Berlin, Germany\\
    $^{\dag}$Fraunhofer German-Sino Lab
    for Mobile Communications (MCI)\\
    Einsteinufer 37, 10587 Berlin, Germany\\ Email:    {\tt\{marcin.wiczanowski,boche\}@hhi.fraunhofer.de}\\
    Phone: +4930-314-28462, Fax: +4930-314-28320}
}

\maketitle

\centerline{\footnotesize \textit{Submitted to IEEE Transactions on Information Theory, August the 6th, 2008}}
\vspace{0.5 cm}

\begin{abstract}
In this work we are concerned with the problem of achieving max-min fairness in Gaussian parallel channels with respect to a general performance function, including channel capacity or decoding reliability as special cases. As our central results, we characterize the laws which determine the value of the achievable max-min fair performance as a function of channel sharing policy and power allocation (to channels and users).
In particular, we show that the max-min fair performance behaves as a specialized version of the Lovasz function, or Delsarte bound, of a certain graph induced by channel sharing combinatorics. We also prove that, in addition to such graph, merely a certain $2$-norm distance dependent on the allowable power allocations and used performance functions, is sufficient for the characterization of max-min fair performance up to some candidate interval. Our results show also a specific role played by odd cycles in the graph induced by the channel sharing policy and we present an interesting relation between  max-min fairness in parallel channels and optimal throughput in an associated interference channel. 
\end{abstract}


\section{Introduction}

Fairness represents an important goal 
in the design of power, bandwidth and time allocation policies for multi-user channels. 
It is usually desired to achieve fairness with respect to communications and information theory metrics, such as spectral efficiency, decoder reliability, etc. \cite{TaS02}, \cite{RaL00}.
The mostly used notion of fairness is the max-min fairness, which is an instance of equity in terms of economy markets and consists in the maximal possible improvement of the worst performance metric \cite{Fol67} \cite{Var74}, \cite{Eur07}.

The single-user communication over parallel channels is a well-studied topic both from the viewpoint of information theoretic optimality as well as suboptimal practical power allocation approaches, see e.g. \cite{LTV06}, \cite{PaF05}, \cite{FoU98}, \cite{BCC95}, \cite{LTV061} and references therein. 
However, the max-min fair allocation of power, bandwidth and time to multiple users sharing the parallel channels access still poses practical problems and needs a deeper understanding \cite{TaS02}, \cite{ToK04}.
The issue of max-min fairness
in (multi-user) parallel channels has been addressed in \cite{ToK04}, \cite{Isi06}, \cite{Itw06}, \cite{ZhB05} and references therein.
Concurrently, a cellular downlink or uplink using Orthogonal Frequency Division Multiplex (OFDM) appears to be the most relevant example of parallel channels shared among multiple users. 
In \cite{ToK04}, the max-min fair carrier and antenna assignment is studied for a multiple antenna OFDM downlink.
More generally, in \cite{Isi06}, \cite{Itw06} the performance of max-min fair power allocation and max-min fair assignment of parallel channels is characterized within the framework of so-called  blocking and antiblocking polyhedra \cite{Ful70}, \cite{Ful72}. The characterization of user performance achieved under max-min fairness is provided in \cite{Isi06}, \cite{Itw06} in the form of bounds and duality-like optimization problems.

Several important aspect of the fairness problem in parallel channels, such as e.g. the optimum interrelations between the combinatorics of channel sharing and the real-valued power allocation, are still open in the general case. Also the essential straight questions such as
"what is the user performance under max-min fairness in parallel channels equal to?" or "what is the power/time/bandwidth function describing it?" remain unanswered so far.
In this work we make a step towards satisfying answers to the above questions in Gaussian parallel channels, when the interest is in max-min fairness with respect to user performance measured by a \textit{general} performance/QoS function; this includes the most celebrated cases of channel capacity, spectral efficiency, decoder reliability (unity minus decoder error rate), etc. We state insightful optimistic and pessimistic bounds on the user performance (Sections \ref{lbounds}, \ref{ubounds}). The essence of our results is that, under constraints on transmit power, the max-min fair performance behaves as a specialized version of the Lovasz function \cite{Lov79}, \cite{GLS86} of
a specific graph, which is induced by the channel sharing topology, or combinatorics. We prove further that, in addition to such graph description, a certain $2$-norm distance dependent on the allowable power allocations and users' performance functions is sufficient for enclosing the max-min fair performance by some lower and upper bounds. These bounds prove to be especially insightful as they offer a separation of influences of the channel sharing policy and the real-valued problem of power allocation. 
We aid the interpretations of the introduced channel sharing topologies and the proposed bounds by several parallel channel examples and visualizations.

Our results exhibit a specific role of odd cycles in the graph induced by the channel sharing policy. We present also an interesting relation between max-min fairness in parallel channels and optimal throughput in an associated interference channel. Furthermore, the presented bounds show a link between the user performance under max-min fairness and (zero-error) graph capacity \cite{KoO98}, \cite{Lov79}. The proofs of our results are constructive and allow for the design of several novel power and time allocation algorithms for parallel channels with predetermined channel sharing topology; this can be motivated by regulations on processing of traffic classes or standardization/hardware constraints (Section \ref{policies}). The proposed algorithms offer a better performance-complexity trade off than conventional solution methods and achieve user performance within some specified distance from the max-min fairness.

\section{Model and Preliminaries}
\label{secmo}

We consider the set of Gaussian (in the sense of Additive-White-Gaussian-Noise) parallel channels, treated as one multi-user channel\footnote{Notation: The nonnegative and positive orthants in $\R^{K\times N}$, where we set $\R^{K\times 1}=\R^{K}$ and $\mathbb{C}^{K\times 1}=\mathbb{C}^K$, are denoted as $\R_+^{K\times N}$ and $\R_{++}^{K\times N}$, respectively. By $\mathbb{S}^K$ we denote the set of symmetric matrices in $\R^{K\times K}$ and the cones of doubly nonnegative matrices and completely positive matrices in $\R^{K\times K}$ are denoted as $\D^K$ and $\P^K$, respectively (see Appendix \ref{app} for the definitions). By $S_{\epsilon}(\ma X)$ we denote a ball with radius $\epsilon$ centered at $\ma X\in\mathbb{C}^{K\times N}$. 
By $\succeq$ we denote the usual partial order on the set of symmetric matrices and $\ma X\circ\ma Y$, denotes the Kronecker product of $\ma X,\ma Y\in\mathbb{C}^{K\times N}$. For any vector $\ve x=(x_1,\ldots,x_{K})\in\C^K$ we define ${(\ve x)}_k=x_k$. Given a matrix $\ma X\in\C^{K\times N}$ with elements $x_{kl}$, $1\leq k\leq K$, $1\leq l\leq N$, we write simply $\ma X=(x_{kl})$ and define similarly ${(\ma X)}_{kl}=x_{kl}$. By $\ma X'$ we denote the conjugate transpose of $\ma X\in\C^{K\times N}$. Given $\ma X\in\C^{K\times K}$, $\mathit{diag}(\ma X)\in\C^{K\times K}$ is such that ${(\mathit{diag}(\ma X))}_{kk}={(\ma X)}_{kk}$ and ${(\mathit{diag}(\ma X))}_{kl}=0$, $k\neq l$, $1\leq k,l\leq K$.
Further, given $\ve x=(x_1,\ldots,x_{K})\in\R_+^K$, a vector $\ve x\sr$ is defined as ${(\ve x\sr)}_k=\sqrt{x_k}$. The identity matrix is denoted by $\ma I$, $\ve e_k$ is the unit vector such that ${(\ve e_k)}_k=1$ and ${(\ve e_k)}_l=0$, $k\neq l$,
and we also define vector $\ve 1$ as ${(\ve 1)}_k=1$, where in all three cases the matrix/vector dimension follows from the context.
By $\langle\ve x,\ve y\rangle$ we denote the inner product of $\ve x,\ve y\in\C^K$. 
Without introducing ambiguity, we do not differ in the notation between random values and deterministic values.
The mean of a random matrix (variable) $\ma X\in\C^{K\times N}$ is denoted as $E(\ma X)$.}. The transmitter-receiver pairs communicating with each other over this channel are referred to abstractly as users and are grouped in the set $\Ks=\{1,\ldots,K\}$. 
The parallel channels are assumed to be deterministic and frequency-flat. 

\subsection{The parallel channels}

The set of parallel channels is denoted as $\Ns=\{1,\ldots,N\}$. 
Let $\ve x_k=(x_{k1},\ldots,x_{kN})\in\C^{N}$ 
be a random vector grouping the independent (zero-mean) symbols of user $k\in\Ks$ transmitted over the channels $n\in\Ns$ equidistantly, at distance $T_s$. Then, the sampled signal of user
$k\in\Ks$ received over the parallel channels can be written 
as $\ve y_k=(y_{k1},\ldots,y_{kN})$, with
\begin{equation*}
\label{ }
y_{kn}=h_{kn}x_{kn} + n_{kn},\quad n\in\Ns,
\end{equation*} 
where $\ve h_k=(h_{k1},\ldots,h_{kN})\in\C^N$ 
collects the path
coefficients between the transmitter and receiver of user $k\in\Ks$ on channels $n\in\Ns$ and
$\ve n_k=(n_{k1},\ldots,n_{kN})\in\C^N$ is a random vector which contains (zero-mean, independent of $\ve x_k$) Gaussian noise variables perceived at the receiver of user $k\in\Ks$ on channels $n\in\Ns$, where we assume $\sigma_{kn}^2=E(\vert n_{kn}\vert^2)>0$.
The transmit power allocation to users and channels (in short, power allocation) can be written as $\ma P=(\ve p_1,\ldots,\ve p_K)'\in\R_+^{K\times N}$, where vector $\ve p_k=(p_{k1},\ldots,p_{kN})$ is such that
$p_{kn}=E(\vert x_{kn} \vert^2)$ is the transmit power allocated to user $k\in\Ks$ on channel
$n\in\Ns$. 

Let $\ma A=(\ve a_1,\ldots,\ve a_K)'\in\R_+^{K\times N}$
denote the sharing matrix of the channels among users such that $\ve a_k=(a_{k1},\ldots,a_{kN})$ collects the relative fractions of time which are assigned to user $k\in\Ks$ for the exclusive access to channels $n\in\Ns$.
Thus, as in practice the operation time is partitioned into frames of some fixed duration $T\gg T_s$, the collection
of times $T\ve a_k$ is reserved for user $k\in\Ks$ for the exclusive access to the respective channels $n\in\N$ within each frame.
The set of allowed sharing matrices of the parallel channels takes the form
\begin{equation}
\label{A1}
\As(\ve r)=\{\ma A\in\R_+^{K\times N}:
{\Vert\ve a_k\Vert}_1\leq r_k,k\in\Ks,\quad
{(\sum_{k\in\Ks}a_{kn})}\leq 1,n\in\Ns\},\quad \ve r\in\R_{++}^K.
\end{equation}
According to the first constraint in (\ref{A1}),
a predefined vector $\ve r=(r_1,\ldots,r_K)\in\R_{++}^K$, with ${\Vert\ve r\Vert}_1\leq N$, is such that $r_k/N$ represents the fraction of the set of parallel channels which is assigned to user $k\in\Ks$ over time (over each frame).
It proves useful in the remainder to introduce also $\ma R\in\R_+^{K\times K}$ such that ${(\ma R)}_{kl}=0$, $k\neq l$, and ${(\ma R)}_{kk}=r_k$, $k,l\in\Ks$. 
For instance, under $\ve r=\frac{N}{K}\ve 1$ any user is assigned an equal $1/K$-fraction of the ensemble of parallel channels over time (over each frame).
The second inequality in the definition (\ref{A1}) models then the obvious constraint that the aggregate time of exclusive uses of a single channel $n\in\Ns$ by the users $k\in\Ks$ does not exceed the total operation time (the total duration of each frame). 
Currently, the most celebrated instance of the considered  parallel channels is the multi-tone/multi-carrier channel accessed by multiple users, as considered e.g. in \cite{QiB05}, \cite{YCB99}.
In this case, $\ve a_k$ groups user's $k\in\Ks$ relative times of exclusive uses of carriers $n\in\Ns$ and $r_k/N$ represents the fraction of the multi-carrier spectrum which he is assigned over time \cite{Boe06}.

Given a sharing matrix $\ma A \in\As(G,\ve r)$ under use, we assume an arbitrary set $\Ps(\ma A)$ of allowed power allocations, requiring merely that
\begin{equation}
\label{cse}
\Ps(\ma A)\supseteq (S_{\epsilon}(\ve 0)\cap\R_+^{K\times N}),\quad\text{for some }\epsilon>0,\quad\epsilon=\epsilon(\ma A),
\end{equation}
Such condition means, broadly, that all power allocations which are sufficiently small for the used sharing matrix are allowable.
In particular, assuming frames of duration $T$, we can take either of the sets
\begin{subequations}
\label{epc}
\begin{align}
&\Ps(\ma A)=\{\ma P\in\R_+^{K\times N}:
\sum_{k\in\Ks}T\la\ve a_k,\ve p_k\ra\leq E\},\quad \ma A\in\As(\ve r),\\
&\Ps(\ma A)=\{\ma P\in\R_+^{K\times N}:
T\la\ve a_k,\ve p_k\ra\leq E_k,\quad k\in\Ks\},\quad \ma A\in\As(\ve r),
\end{align}
\end{subequations}
for some $E,E_k>0$, $k\in\Ks$, which mirror the limitations of energy per frame as a crucial constraint in current and future wireless communication systems \cite{Eut07}, \cite{Eut08}. 
The latter set corresponds to conventional limitations of energy per frame in a multi-user Gaussian channel with user energy per frame budgets constrained by $E_k$, $k\in\Ks$.
The first set models the possibility of energy coordination among all users under the joint energy per frame budget constrained by $E$. This is the case, for instance, when the considered parallel channels are a means of representation of the orthogonalized broadcast channel which applies, suboptimally, single-user precoding instead of multi-user precoding \cite{Cov98}, \cite{JiG03} (for the combination of parallel channels and the broadcast channel see also \cite{Tse97}). 

Complementarily to frame energy constraints it is sometimes desired to account for transmit power constraints at \textit{any} time in a frame. In analogy to (\ref{epc}), under limitation of transmit power of any user $k\in\Ks$ by $P_k>0$ and under the joint transmit power budget of all users constrained by $P>0$ we take, respectively,
\begin{subequations}
\begin{align}
\label{ppc}
&\Ps(\ma A)=\Ps=\{\ma P\in\R_+^{K\times N}:{\vert\vert\vert\ma P\vert\vert\vert}_1\leq P\},\\
&\Ps(\ma A)=\Ps=\{\ma P\in\R_+^{K\times N}:{\Vert\ve p_k\Vert}_1\leq P_k,\quad k\in\Ks\}.
\end{align}
\end{subequations}
It is interesting to note that transmit power constraints at any time within a frame make the set of allowable power allocations independent of sharing matrix $\ma A\in\As(\ve r)$ under use, which will be of key importance at several points in the remainder.

\subsection{The user performance}
\label{uspsec}

For any user $k\in\Ks$ accessing the parallel channels, we assume a general vector-valued \textit{performance/ QoS function}
\begin{equation*}
\ve p\mapsto f_k(\ve p)\in\R^N,\quad \ve p\in\R_+^N,
\end{equation*}
where we have 
$f_k(\ve p)=(f_{k1}(p_{1}),\ldots,f_{kN}(p_{N}))$, with
$p\mapsto f_{kn}(p)$, $p\geq 0$, $n\in\Ns$.
Function $f_{kn}$ expresses the performance of user $k\in\Ks$ on channel $n\in\Ns$, as a function of power allocated to channel $n\in\Ns$, when the user accesses this channel exclusively throughout the operation time. We restrict us to nonnegative QoS functions
\begin{equation}
\label{coi}
f_k(\ve p)\in\R_+^N,\quad\ve p\in\R_+^N,\quad k\in\Ks,
\end{equation}
and to avoid later technical queerness we assume that
$\frac{\partial}{\partial p_k}f_k(\ve p)>0$, $k\in\Ks$, for $\ve p\in S_{\epsilon}(\ve 0)\cap\R_+^K$ and some $\epsilon>0$ (that is, performance functions are componentwise Frechet-differentiable and increasing at least for sufficiently small power allocations).

Due to (\ref{coi}) and the assumed independent symbols of a user on each one of the parallel channels, it is reasonable to consider
\begin{equation*}
\label{ }
(\ve a,\ve p)\mapsto\langle\ve a, f_k(\ve p)\rangle,
\quad(\ve a,\ve p)\in\R_+^N\times\R_+^{N},\quad {\Vert\ve a\Vert}_1\leq r_k,
\end{equation*}
as the \textit{performance/QoS metric} of user $k\in\Ks$. Such metric represents the aggregate performance achieved by user $k\in\Ks$ on the entire channel ensemble, throughout the operation time (throughout each frame), as a function of powers allocated to channels $n\in\Ns$ and relative time fractions of exclusive channel uses. We refer to a value of the performance metric of a user, for some $\ma A\in\As(\ve r)$ and $\ma P\in\Ps(\ma A)$, as \textit{user performance} under policy $(\ma A,\ma P)$.

Let any predefined requirement/expectation of user $k\in\Ks$ with respect to the user performance be denoted as $\gamma_k>0$, $k\in\Ks$.
Then, $\min_{k\in\Ks}\frac{\langle\ve a_k, f_k(\ve p_k)\rangle}{\gamma_k}$ 
can be seen as the worst relative performance among the users accessing the parallel channels under a policy $(\ma A,\ma P)\in\As(\ve r)\times\Ps(\ma A)$. Hereby, we implicitly assume that a smaller user performance implies a worse perceived service quality at the corresponding receiver. Such assumption complies with the nature of the very most QoS functions used in communications and information theory, but does not necessarily require strict componentwise increasingness of $f_k$, $k\in\Ks$. We give a few celebrated examples of such performance functions.

\begin{Ex}[Symbol decoding reliability] 
\label{ex1}
Let user $k\in\Ks$ access channel $n\in\Ns$ and use uncoded constant-envelope modulation. Then, the achieved 
probability of error-free symbol decoding is
\begin{equation}
\label{ber}
f_{kn}(p)=1-Q(\sqrt{\frac{c}{\log_2 M}\frac{p \vert h_{kn}\vert^2}{\sigma_{kn}^2}}),
\end{equation}
with $Q$ denoting the Marcum $Q$-function, $M$ denoting the constellation size, and $c>0$ as some constant (e.g., $c=2$ for binary Phase Shift Keying or $c=1$ for binary Frequency Shift Keying) \cite{TsV05}. By (\ref{ber}) and the uniform symbol distance $T_s$, the map $(\ve a_k,\ve p_k)\mapsto\frac{T}{T_s}\la\ve a_k,f_k(\ve p_k)\ra$, $(\ma A,\ma P)\in\As(\ve r)\times\Ps(\ma A)$, expresses the aggregate (over channels $n\in\Ns$) average number of error-free decoded symbols of user $k\in\Ks$ in a frame as a function of policy.
\end{Ex}

\begin{Ex}[Mean square detection reliability]
\label{ex2}
If the receiver of user $k\in\Ks$ utilizes the Minimum Mean Square Error (MSE) receiver and the user accesses channel $n\in\Ns$, then the achieved MSE can be expressed as 
$\frac{1}{1+\frac{p_{kn}\vert h_{kn}\vert^2}{\sigma_{kn}^2}}$ \cite{Ver98}. As a consequence, 
\begin{equation}
\label{mse}
f_{kn}(p)=1-\frac{1}{1+\frac{p \vert h_{kn}\vert^2}{\sigma_{kn}^2}},
\end{equation} 
can be regarded as a kind of symbol detection reliability in the mean square sense. Thus, given (\ref{mse}), $(\ve a_k,\ve p_k)\mapsto\frac{T}{T_s}\la\ve a_k,f_k(\ve p_k)\ra$, $(\ma A,\ma P)\in\As(\ve r)\times\Ps(\ma A)$,
describes the mean square detection reliability of user $k\in\Ks$, aggregated over all symbols received in a frame, as a function of policy.
\end{Ex}

\begin{Ex}[Spectral efficiency]
\label{ex3}
Let the modulation constellation size of user $k\in\Ks$ which accesses channel $n\in\Ns$ be constrained by $M$. The spectral efficiency, in the sense of maximum number of reliably decodable bits/nats per symbols under given modulation constellation\footnotemark[3], is not expressible analytically but can be approximated by a function
\begin{equation*}
f_{kn}(p)=g(\frac{p \vert h_{kn}\vert^2}{\sigma_{kn}^2})
\end{equation*}
such that the map $x\mapsto g(x)$, $x\geq 0$, is nondecreasing, and $g(0)=0$ and $g(x)=\log_2 M$, $x\geq x_0$, for some $x_0=x_0(M)$ \cite{FoU98}.
Then, it is easily seen that map $(\ve a_k,\ve p_k)\mapsto\frac{T}{T_s}\la\ve a_k,f_k(\ve p_k)\ra$, $(\ma A,\ma P)\in\As(\ve r)\times\Ps(\ma A)$, describes the achievable number of reliably decoded bits/nats in a frame, as a function of policy.
\end{Ex}

\begin{Ex}[Capacity]
\label{ex3_}
If user $k\in\Ks$ utilizes the Maximum Likelihood (ML) receiver, then  
\begin{equation}
\label{cap}
f_{kn}(p)=\log(1+\frac{p \vert h_{kn}\vert^2}{\sigma_{kn}^2})
\end{equation}
represents the (information) capacity\footnotemark[3], achievable by user $k\in\Ks$ when accessing the channel $n\in\Ns$, that is, the overall maximum number of reliably decodable bits/nats per symbol. Thus, given (\ref{cap}), the function $(\ve a_k,\ve p_k)\mapsto\frac{T}{T_s}\la\ve a_k,f_k(\ve p_k)\ra$, $(\ma A,\ma P)\in\As(\ve r)\times\Ps(\ma A)$, corresponds to the achievable (under Gaussian codebook) number of reliably decoded bits/nats per frame.
\end{Ex}

\footnotetext[3]{Obviously, the notions of spectral efficiency and capacity are meaningful only when the duration of the considered channel access is sufficiently long, in the sense $Ta_{kn}\gg T_s$.}
With the given assumptions on user performance, the maximum attainable performance of the worst-case user accessing the parallel channels among policies from $\As(\ve r)\times\Ps$ can be expressed as
\begin{equation}
\label{mmf}
\max_{(\ma A,\ma P)\in\As(\ve r)\times\Ps(\ma A)}\min_{k\in\Ks}\frac{\langle\ve a_k, f_k(\ve p_k)\rangle}{\gamma_k}.
\end{equation}
According to the common understanding of fairness in various multi-user channels, see e.g. \cite{TaS02}, \cite{RaL00}, \cite{ToK04}, \cite{ZhB05}, we refer to (\ref{mmf}) as the max-min fair performance (in/of the considered parallel channels), and we say that a pair 
$(\ma A,\ma P)=\arg\max_{(\ma A,\ma P)\in\As(\ve r)\times\Ps(\ma A)}\min_{k\in\Ks}\frac{\langle\ve a_k, f_k(\ve p_k)\rangle}{\gamma_k}$,
is a max-min fair policy, which is, in general, not unique. 

\section{Graph of parallel channels sharing}
\label{sec_4}

For any sharing matrix $\ma A\in\As(\ve r)$ we define an undirected graph of parallel channels sharing, in short a \textit{sharing graph}, which is induced by $\ma A$.
For the definition, recall that any graph is a pair, say $G=(\Ks,\Es)$, where $\Ks$ is the set of graph vertices, and $\Es$ is the set of edges; any edge is represented by a pair $(k,l)\in\Es$ such that $k,l\in\Ks$ are the vertices which are joined/connected by this edge (are adjacent) \cite{Bol98}.

\begin{Def}
\label{or}
\textit{For $N\geq K$ and any $\ma A\in\As(\ve r)$, a corresponding sharing graph $G=G(\ma A)$ is such that $G=(\Ks,\Es)$ where $(k,l)\in\Es$, $k\neq l$, if
$\langle\ve a_k,\ve a_l\rangle>0$.}
\end{Def}
The proposed induction of a sharing graph $G$ by $\ma A$ is a version of orthogonal graph labeling from \cite{Knu94}, which further differs slightly from the original concept of \textit{orthonormal representation} of a graph
in \cite{Lov79}.
Precisely, a (not necessarily nonnegative) matrix $\ma A\in\R^{K\times N}$ is referred to as an orthonormal representation of graph $G=G(\ma A)=(\Ks,\Es)$, which we write as $\ma A\in\As^0(G)$, if $\la\ve a_k,\ve a_k\ra=1$, $k\in\Ks$, and $(k,l)\in\Es$, $k\neq l$, whenever $\la\ve a_k,\ve a_l\ra\neq 0$. 
By Definition \ref{or}, any two vertices $k,l\in\Ks$, $k\neq l$, of the sharing graph are adjacent if some of the parallel channels are shared by users $k,l$, where a channel is said to be shared by some two users if both users access this channel exclusively some fraction of time. The converse is also clear: If two nodes $k,l\in\Ks$, $k\neq l$, of the sharing graph are nonadjacent, then no one of the channels $n\in\Ns$ is shared by users $k,l$.

It is readily seen that, for any fixed $\ma A\in\As(\ve r)$, an induced sharing graph $G=G(\ma A)$ is in general not unique.
Besides this, the graph characterization of parallel channels sharing provides merely the information on the topology, or combinatorics, of 
sharing relationships.
Thus, given any graph $G=(\Ks,\Es)$, different sharing matrices induce $G$ as a sharing graph and we can group them in the set
\begin{equation*}
\As(G,\ve r)=\{\ma A\in\As(\ve r):\quad G=G(\ma A)\},\quad \ve r\in\R_{++}^K.
\end{equation*}
The illustration is provided in the following example.

\begin{figure}
\centering
\scalebox{0.45}{\input{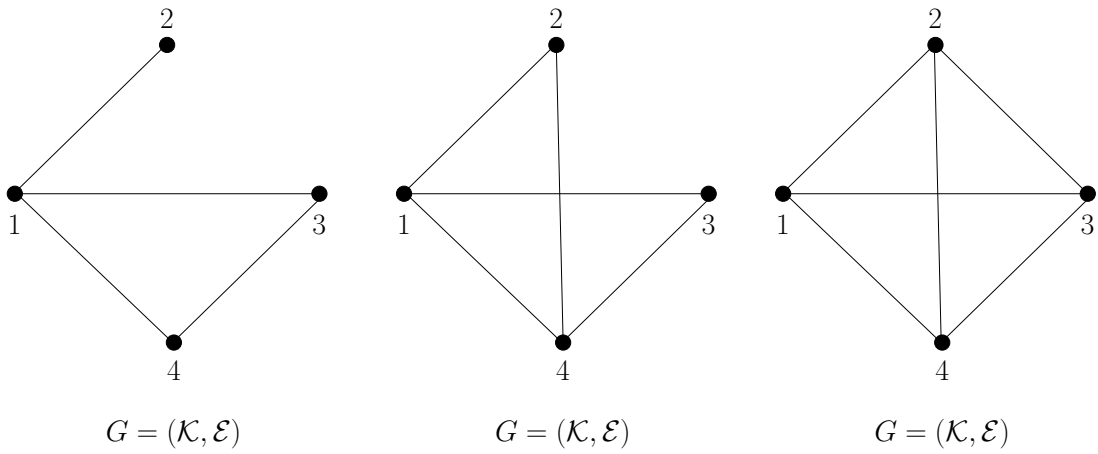}}
\caption{Three sharing graphs of the multi-carrier channel with $K=4$ users considered in Example 5; such graphs are induced, in the sense $G=G(\ma A)$, by the matrices $\ma A$ from Example 5.}
\label{fig:figone}
\end{figure}

\begin{Ex}
\label{ex6}
Consider parallel channels as a multi-tone/multi-carrier channel with $N=7$ tones accessed by $K=4$ users in the proportions $\ve r=(2,2,1,2)$.
Let the sharing of the tones be described by the sharing matrix
\begin{equation*}
\label{ }
\ma A=
\left(\begin{array}{ccccccc}
0 &0.2 &0 &0.5 &0.5 &0 &0.8\\
1 &0 &0 &0.5 &0.5 &0 &0\\
0 &0.3 &0.3 &0 &0 &0.3 &0.1\\
0 &0.5 &0.7 &0 &0 &0.7 &0.1\\
\end{array}\right)\in\As(2,2,1,2),
\end{equation*}
Then, the three possible sharing graphs $G=G(\ma A)$ are depicted in Fig. \ref{fig:figone}, with the graph on the right hand side as the sharing graph with the minimal number of edges. On the other hand, for $G$ as the minimum sharing graph from Fig. \ref{fig:figone}, the set $\As(G,(2,2,1,2))$ of sharing matrices inducing it includes, in particular, column permutations of all matrices of the form
\begin{equation*}
\ma A=
\left(\begin{array}{ccccccc}
0 &a_{12} &0 &a_{14} &a_{15} &0 &a_{17}\\
a_{21} &0 &0 &a_{24} &a_{25} &0 &0\\
0 &a_{32} &a_{33} &0 &0 &a_{36} &a_{37}\\
0 &a_{42} &a_{43} &0 &0 &a_{46} &a_{47}\\
\end{array}\right),
\end{equation*}
with $a_{ij}>0$, $1\leq i\leq 4$, $1\leq j\leq 7$.
\end{Ex}

A special role in our considerations of sharing graphs is played by the subgraphs called \textit{cycles}. 

\begin{figure}
\centering
\scalebox{0.45}{\input{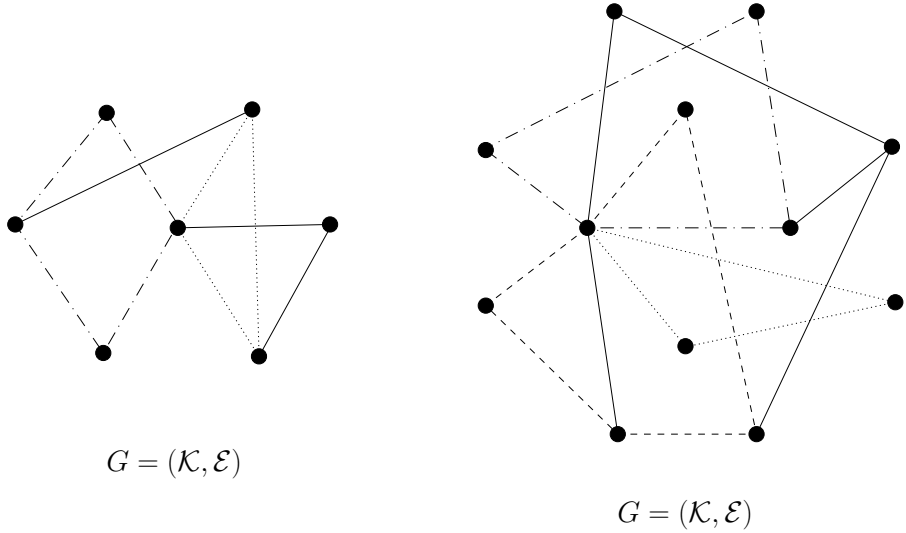}}
\caption{Two exemplary sharing graphs of parallel channels with $K=7$ and $K=12$ users with marked exemplary cycles of length $3$ (dotted edges), $4$ (dash-dotted edges) and $5$ (dashed edges).}
\label{fig:figtwo}
\end{figure}

\begin{Def}[\cite{Bol98}, \cite{CDS80}]
\textit{A cycle of length $M$ in a graph $G=(\Ks,\Es)$ is a sequence of distinct graph vertices $k_i\in\Ks$, $1\leq i\leq M$, which satisfy $(k_i,k_{i+1})\in\Es$, $1\leq i\leq M-1$ and $(k_M,k_1)\in\Es$.}
\end{Def}
In simple words, a cycle represents a simple closed path in a graph\footnote{In the context of undirected graphs, some works prefer the notion of a circuit to the notion of a cycle used here. In such a convention, the cycle is understood as the analog to the circuit in directed graphs.}. Note that the length of a cycle is the number of {edges}, or equivalently vertices, constituting the cycle.
As an illustration, in Fig. \ref{fig:figtwo} particular cycles are emphasized in two exemplary sharing graphs. 
A cycle of a sharing graph has an easy interpretation in terms of sharing policies: A cycle of length, say, $M$ corresponds to a chain/sequence of $M$ users accessing the parallel channels such that any pair of subsequent users shares some channel and the last user shares a channel with the first user.

\begin{figure}
\centering
\scalebox{0.45}{\input{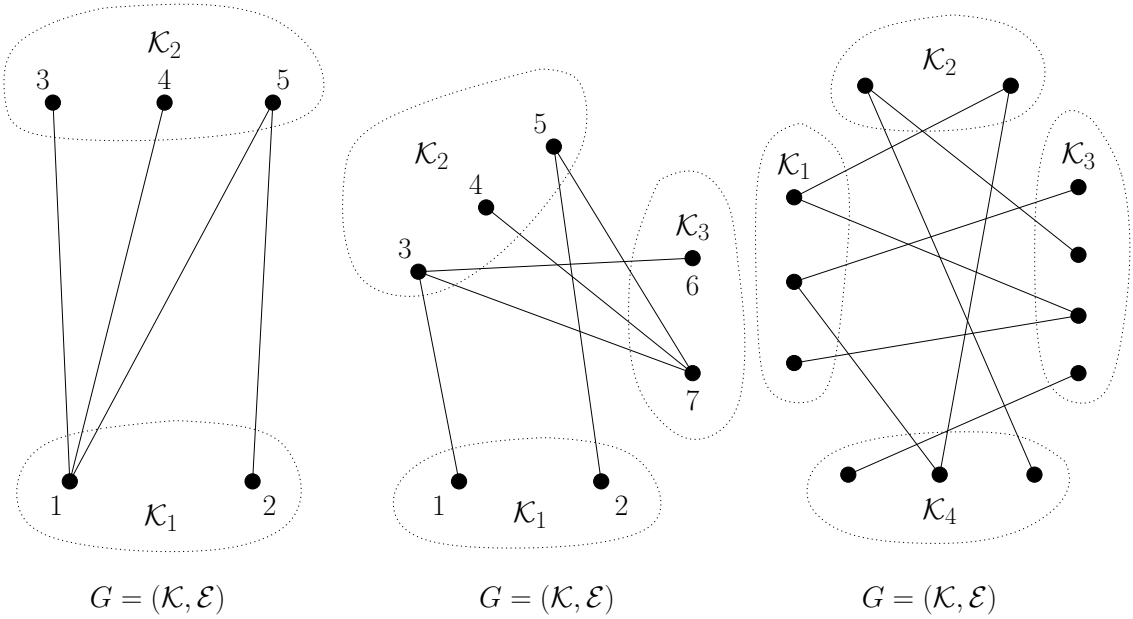}}
\caption{Three exemplary $M$-partite sharing graphs of parallel channels with $K=5,7,12$ users and for $M=2,3,4$, respectively. The graph on the left-hand side is a sharing graph of the multi-carrier channel from Example 6
and is induced (in the sense $G=G(\ma A)$) in particular by the matrix $\ma A$ from Example 6. The graph in the middle can be a sharing graph of the multi-carrier channel considered in Example 7.}
\label{fig:figthree}
\end{figure}

In Fig. \ref{fig:figthree} we show examples of $M$-partite sharing graphs for $M=2,3,4$. As can be seen from the figure, such graphs contain only edges between some disjoint vertex subsets: The vertex set of an $M$-partite graph $G=(\Ks,\Es)$ is divided into partitions $\Ks_i$, $1\leq i\leq M$, such that $(k,l)\notin\Es$ whenever $k,l\in\Ks_i$, $1\leq i\leq M$.
It is easily deduced that an $M$-partite graph can not contain any cycle longer than $M$.
An $M$-partite sharing graph is induced by sharing policies of parallel channels which distinguish $M$ classes of users with the property that users within one class are not allowed, or not able, to share any channels over time. Such constraint is likely to be imposed by 
traffic processing regulations and/or the implementation effort, as is illustrated by the following examples.

\begin{Ex}
\label{ex7}
Consider parallel channels as a multi-carrier channel with $N=9$ carriers and $K=5$ users accessing the carrier set in the proportions $\ve r=(2,2,0.5,1.5,3)$ and having a common transmitter. 
Let the traffic of users $1,2\in\Ks_1$ be the real-time  traffic, like voice or multimedia, while users $3,4,5\in\Ks_2$ transmit and receive so-called background traffic, such as file transfer, signaling or system information. From the viewpoint of percepted QoS and traffic processing complexity, it may be required to assign a carrier to real-time traffic for a large relative time fraction, say, no less than half of the total time, so that no carrier can be shared by two real-time users. 
Due to the processing effort, it may be also
undesired to share a carrier between multiple users carrying the minor background traffic.
These conditions enforce that the single carriers are either user-specific or carry mostly the real-time traffic of users $\Ks_1$ with 
some add-on background traffic of users $\Ks_2$ from time to time.
Thus, an exemplary sharing matrix can take the form
\begin{equation*}
\ma A=
\left(\begin{array}{ccccccccc}
0 	&0.75 	&0.8 	&0.7 	&0 	&0 	&0	&0.75 	&0\\
0.6 	&0 	&0 	&0 	&0.6 	&0 	&0	&0 	&0.8\\
0 	&0 	&0 	&0 	&0 	&0 	&0	&0.25 	&0\\
0 	&0.25 	&0 	&0 	&0 	&1 	&0	&0 	&0\\
0.4 	&0 	&0.2 	&0.3 	&0.4 	&0 	&1	&0 	&0.2\\
\end{array}\right)
\in\As(3,2,0.25,1.25,2.5),
\end{equation*}
which induces the bipartite sharing graph $G=G(\ma A)$ on the left hand side of Fig. \ref{fig:figthree}. 
\end{Ex}

\begin{Ex}
\label{ex8}
Let a multi-carrier channel with $N=13$ carriers accessed by $K=7$ users in the proportions $\ve r=(3,2,2,1,1,3,1)$ be the considered parallel channels with a common receiver.
Let the users be grouped in classes $1,2\in\Ks_1$, $3,4,5\in\Ks_2$ and $6,7\in\Ks_3$ such that for any two users within one class the difference between their propagation
times is larger than some critical propagation time difference (note that such classification is not always possible). Then, the sharing of a carrier between two users from one class can be undesired due to the required effort of time- and frequency synchronization to come up with the propagation time difference. This implies that the carriers are either user-specific or shared only across the classes $\Ks_1$, $\Ks_2$, $\Ks_3$, and that a particular sharing matrix $\ma A\in\As(3,2,2,1,1,3,1)$ can induce the $3$-partite sharing graph $G=G(\ma A)$ in the middle of Fig. \ref{fig:figthree}.
\end{Ex}
  
As shown in the remainder, the description of the channel sharing topology by a sharing graph plays a key role in the problem of ensuring max-min fairness (\ref{mmf}). 

\subsection{Selected algebraic graph characterizations}
\label{sec_4_2}

We make use of the description of a graph by its so-called \textit{feasible matrix}, which is a symmetric matrix indicating an edge by a nonzero entry \cite{DJL94}, \cite{KoB93}, \cite{BeH87}, \cite{ChX02}: The set of feasible matrices of a graph $G=(\Ks,\Es)$ is
\begin{equation*}
\Cs(G)=\{\ma C=(c_{kl})\in\mathbb{S}^{K}: c_{kl}\neq 0, k\neq l,\quad\textit{iff}\quad (k,l)\in\Es\}.
\end{equation*}
Given $G$, another set of interest here is parameterized by a vector $\ve v\in\R_+^K$ of its vertex weights and can be written as
\begin{equation}
\label{B0}
\Bs^0(G,\ve v)=\{\ma B=(b_{kl})\in\mathbb{S}^{K} :
b_{kl}=(v_kv_l)^{\frac{1}{2}},\quad (k,l)\notin\Es\textit{ or }k=l\}.
\end{equation}
For $\ve v=\ve 1$, this concept has its origin in the seminal work \cite{Lov79} where it was used in an approach to the problem of graph capacity. The generalization to the case $\ve v\in\R_+^K$ was provided later in the study of relaxations of the vertex packing problem \cite{GLS86}). 
The graph capacity problem, as the key problem of zero-error information theory, remains still unsolved in the general case \cite{KoO98}. The set (\ref{B0}) is, however, a central element of the concept of so-called weighted Lovasz function, which in unweighted form (i.e. for $\ve v=\ve 1$) represents a general upper bound on graph capacity and is equal to the capacity for a certain graph class, including e.g. 
self-complementary graphs with vertex-transitive automorphism groups \cite{Lov79}, \cite{Bol98}. 
Precisely, the weighted Lovasz function (later, simply Lovasz function) of a graph $G=(\Ks,\Es)$ is defined as the map 
\begin{equation}
\label{lfd}
(G,\ve v)\mapsto\theta^0(G,\ve v)=\min_{\substack{\ma A\in\As^0(G),\\\ve c\in\R^L:{\Vert\ve c\Vert}_2=1}}\max_{k\in\Ks}\frac{v_k}{\langle\ve a_k,\ve c\rangle^2},\quad\ve v\in\R_+^K,
\end{equation}  
with $\As^0(G)$ as the set of orthonormal representations of $G$, and it has the following property.

\begin{Prop}[\cite{GLS86}, \cite{Knu94}]
\label{lt}
\textit{For any graph $G=(\Ks,\Es)$, we have
\begin{equation*}
\theta^0(G,\ve v)=\min_{\ma B\in\Bs^0(G,\ve v)}\lambda_{\text{max}}(\ma B),\quad\ve v\in\R_+^K.
\end{equation*}}
\end{Prop}
In \cite{GLS86}, \cite{Knu94}, \cite{LuS05} one can find further interesting characterizations of the Lovasz function.

A similar set which we make use of is
\begin{equation}
\label{B1}
\Bs^1(G,\ve v)=\{\ma B=(b_{kl})\in\mathbb{S}^{K} :
b_{kl}\geq(v_kv_l)^{\frac{1}{2}},\quad (k,l)\notin\Es\textit{ or }k=l\},
\end{equation}
for any graph $G=(\Ks,\Es)$ and $\ve v\in\R_+^K$ is a vector of its vertex weights. For the case $\ve v=\ve 1$, the concept of the set (\ref{B1}) is known from the considerations on the Delsarte bound, or Delsarte number, in \cite{Sch79}, and the generalization to the case $\ve v\in\R_+^K$ is straightforward. The unweighted (i.e. for $\ve v=\ve 1$) Delsarte bound was proposed within the algebraic framework of coding theory in \cite{Del73}, as an upper bound on the cardinality of an $\Ms$-clique, $\Ms\subset\{1,\ldots,M\}$, in an association scheme with $M$ associate classes denoted here as $(\Ks,\{\Es_i\}_{i=1}^M)$.
As the notion of association scheme is only loosely related with our topic, we refer here to Appendix \ref{app_} for an outline of the theory. 
One can relate the Delsarte bound for an association scheme $(\Ks,\{\Es_i\}_{i=1}^M)$ to the graph $G=(\Ks,\Es)$, $\cup_{i=1}^M\Es_i$, i.e. the graph whose edge set corresponds to the union of associate classes: The unweighted Delsarte number upper bounds the independence number of such graph \cite{Sch79} and the weighted Delsarte number (later, simply Delsarte number/bound), denoted as map $\theta^1$, has then the following characterization.

\begin{Prop}[\cite{Sch79}]
\label{lt2}
\textit{For any graph $G=(\Ks,\Es)$, we have
\begin{equation*}
\theta^1(G,\ve v)=\min_{\ma B\in\Bs^1(G,\ve v)}\lambda_{\text{max}}(\ma B),\quad\ve v\in\R_+^K.
\end{equation*}}
\end{Prop}
Further formulations of the Delsarte number which are direct analogs of the original characterizations of the Lovasz function from \cite{Lov79} can be found, partly without proof, in \cite{ChR00} and \cite{Gal00}.
In particular, the authors apply the framework of graph Laplacians and identify the Delsarte number with the so-called $\sigma$-function of a graph and the Lovasz function with a related version of it. 
Similar characterizations of the Lovasz function and Delsarte bound and their properties in terms of edge orbits are studied in \cite{MRR78}.

For our purposes, we define two further 
sets of the type (\ref{B0}), (\ref{B1}) and two related graph functions in the spirit of Propositions \ref{lt} and \ref{lt2}. First, we associate with a graph $G=(\Ks,\Es)$ and a weight vector $\ve v\in\R_+^K$, the set
\begin{equation}
\label{B2}
\begin{split}
\Bs^2(G,\ve v)=\{\ma B=(b_{kl})\in\mathbb{S}^{K} :\text{ }
&b_{kl}=(v_kv_l)^{\frac{1}{2}},\quad (k,l)\notin\Es\textit{ or }k=l,\\
&b_{kl}\leq (v_kv_l)^{\frac{1}{2}},\quad k,l\in\Ks\}.
\end{split}
\end{equation}
In analogy to Proposition \ref{lt}, we define for any graph $G$ the map
\begin{equation}
\label{lfd2}
(G,\ve v)\mapsto\theta^2(G,\ve v)=\min_{\ma B\in\Bs^2(G,\ve v)}\lambda_{\textit{max}}(\ma B),\quad\ve v\in\R_+^K.
\end{equation}
Second, also the set
\begin{equation}
\label{B3}
\begin{split}
\Bs^3(G,\ve v)=\{\ma B=(b_{kl})\in\mathbb{S}^{K} :\text{ }
&b_{kl}=(v_kv_l)^{\frac{1}{2}},\quad (k,l)\notin\Es\textit{ or }k=l\\
&\ve v\sr{\ve v\sr}'-\ma B+\lmax(\ma B)\ma I\in\P^K\}
\end{split}
\end{equation}
associated with any $G=(\Ks,\Es)$ and $\ve v\in\R_+^K$  proves to be of key use in the remainder. By the definition of the class $\P^K$ of completely positive matrices in $\R^{K\times K}$ (Appendix \ref{app}), the latter condition in (\ref{B3}) can be written equivalently as
\begin{equation}
\label{cpa}
\ve v\sr{\ve v\sr}'-\ma B+\lmax(\ma B)\ma I=\ma V'\ma V,\quad\text{for some}\quad\ma V\in\R_+^{M\times K},M\in\N.
\end{equation}
Furthermore, it is worth noting here that the condition 
$\ve v\sr{\ve v\sr}'-\ma B+\lmax(\ma B)\ma I\in\P^K$ is implied by a slightly stronger requirement that $\ve v\sr{\ve v\sr}'-\ma B$ is included in the closure of $\P^K$: This is an immediate consequence of $\lmax(\ma B)\ma I=\sum_{k\in\Ks}\sqrt{\lmax(\ma B)}\ve e_k\sqrt{\lmax(\ma B)}\ve e'_k$ and the characterization in Appendix \ref{app}.
By analogy to Proposition \ref{lt}, for any $G=(\Ks,\Es)$ we define a further map
\begin{equation}
\label{lfd3}
(G,\ve v)\mapsto\theta^3(G,\ve v)=\min_{\ma B\in\Bs^3(G,\ve v)}\lambda_{\textit{max}}(\ma B),\quad\ve v\in\R_+^K.
\end{equation}

The relations between (\ref{B0}), (\ref{B1}) and the proposed sets (\ref{B2}), (\ref{B3}) are readily seen.
It is immediate that
$\Bs^0(G,\ve v)\subset\Bs^1(G,\ve v)$
and that the second condition in (\ref{B2}) can be written as $\ve v\sr{\ve v\sr}'-\ma B\in\R_+^{K\times K}$.
Thus, by the form (\ref{cpa}) of the second condition in (\ref{B3}) and by inspection of (\ref{B0}) and (\ref{B2}), it can be seen that 
\begin{equation*}
\Bs^3(G,\ve v)\subset\Bs^2(G,\ve v)\subset\Bs^0(G,\ve v)\subset\Bs^1(G,\ve v),
\end{equation*}
and thus
\begin{equation}
\label{Bes}
\theta^3(G,\ve v)\geq\theta^2(G,\ve v)\geq\theta^0(G,\ve v)\geq
\theta^1(G,\ve v)
\end{equation}
for any $G=(\Ks,\Es)$ and $\ve v\in\R_+^K$ on hand.

It is worth noting here that, given $\ve v=\ve 1$, the classes of matrices (\ref{B0}), (\ref{B1}) generalize the set of so-called $(1,\delta)$-adjacency matrices of graph $G$ introduced in \cite{JoN80}. Any $(1,\delta)$-adjacency matrix is further affinely transformable to a Seidel adjacency matrix \cite{CDS80}. 

\subsection{Some relations of the characterizations}

The algebraic graph descriptions introduced in Section \ref{sec_4_2} have some simple properties which turn out to be central to our results.
The first lemma below can be partially deduced from the proof of Theorem 3.5 in \cite{BeH87}. 
We give the proof for completeness and refer to Appendix \ref{app} for the notions related to the set of completely positive matrices $\P^K$, such as the cp-rank.

\begin{Lm}
\label{lm1}
\textit{Given any graph $G=(\Ks,\Es)$, we have
\begin{equation*}
\Cs(G)\cap\P^K\neq\o.
\end{equation*}
and the cp-rank satisfies
\begin{equation*}
\phi(\ma C)\leq \frac{K(K+1)}{2},\quad\ma C\in\Cs(G)\cap\P^K.	
\end{equation*}}
\end{Lm}

\begin{proof}
Associate any edge $(k,l)\in\Es$ with $e=e(k,l)$, $1\leq e\leq \vert\Es\vert$, and
let $\ma B=(b_{kl})\in\R_+^{K\times\vert\Es\vert}$ be defined as\footnote{In the particular case $b_{ek}=1,b_{el}=1$ iff $e=e(k,l)$, matrix $\ma B$ represents the so-called incidence matrix of graph $G$ \cite{Bol98}, \cite{CDS80}.}
\begin{equation*}
b_{ek}>0,\quad b_{el}>0\quad\text{iff}\quad e=e(k,l).
\end{equation*}
Then it is readily seen that ${(\ma B\ma B')}_{kl}>0$, $k\neq l$ iff $(k,l)\in\Es$, so that $\ma C=\ma B\ma B'$ satisfies
$\ma C\in\Cs(G)$ and, by Definition \ref{cpmd}, we also have $\ma C\in\P^K$. This proves $\Cs(G)\cap\P^K\neq \o$.

According to the known bound on cp-rank, see e.g. Section 1 in  \cite{DJL94}, if additionally $\ma C\in\P^K$, then we can find $\ma B\in\R_+^{K\times N}$ such that $\ma C=\ma B\ma B'$ for some $N\leq K(K+1)/2$, which completes the proof.
\end{proof}

The lemma says essentially that the set of feasible matrices includes a completely positive matrix for any graph on hand and any such matrix remains completely positive if all diagonal elements are replaced by the largest eigenvalue. 
Furthermore, for any graph with $K$ vertices, e.g. a sharing graph of parallel channels accessed by $K$ users, any of its completely positive feasible matrices has a cp-rank no larger than $K(K+1)/2$. The latter bound on the cp-rank is the best known, but likely not the best possible bound \cite{DJL94}. 

\begin{Lm}
\label{lm2}
\textit{Given any graph $G=(\Ks,\Es)$ and $\ve v\in\R_+^K$, 
consider the set 
\begin{equation*}
\Ds_{i}(G,\ve v)=\{\ve v\sr{\ve v\sr}'-\ma B+\lmax(\ma B):\ma B\in\Bs^i(G,\ve v)\},
\quad i=0,2,3.
\end{equation*}
Then, we have
\begin{equation*}
\Ds_{0}(G,\ve v)\subset\cup_{G'\subset G}\Cs(G'),\quad\!
\Ds_{2}(G,\ve v)\subset\cup_{G'\subset G}\Cs(G')\cap\R_+^{K\times K},\quad\!
\Ds_{3}(G,\ve v)\subset\cup_{G'\subset G}\Cs(G')\cap\P^K,
\end{equation*}
where $G'\subset G$ denotes that $G'=(\Ks',\Es')$ is a subgraph of $G$ in the sense that $\Ks'\subseteq\Ks$ and $\Es'\subseteq\Es$. Moreover, we have
\begin{equation*}
\Ds_2(G,\ve v)=\Ds_3(G,\ve v),\quad\text{equivalent to}\quad
\Bs^2(G,\ve v)=\Bs^3(G,\ve v), 	
\end{equation*}
if either $K\leq 4$ or $G$ has no odd cycles longer than $4$.}
\end{Lm}

\begin{proof}
Let any $\ma C=(c_{kl})\in\R^{K\times K}$ such that
\begin{equation}
\label{cmat}
\ma C=\ve v\sr{\ve v\sr}'-\ma B+\lmax(\ma B),
\end{equation}
for an arbitrary $\ma B\in\Bs^i(G,\ve v)$, $i=0,2,3$, be given. 
Then, by the definitions (\ref{B0}), (\ref{B2}), (\ref{B3}) we have
\begin{equation*}
c_{kl}=0,\quad k\neq l,\quad\text{if}\quad(k,l)\notin\Es,
\end{equation*} 
but also $c_{kl}=0$, $k\neq l$, if $(k,l)\in\Fs$, for some $\Fs\subseteq\Es$, where $\Fs=\o$ is allowed. This implies 
\begin{equation*}
c_{kl}\neq 0,\quad k\neq l,\quad\text{iff}\quad(k,l)\in\Es',
\end{equation*}
with $\Es'=\Es\setminus\Fs$. 
Thus, for any matrix $\ma C$ we have $\ma C\in\Cs(G')$ for some subgraph $G'\subset G$, with $G'=(\Ks',\Es')$.

If now $i=2$, then it is evident by the definition (\ref{B2}) and by\footnote{This is readily seen by the feature that for any $\ma C=\ma B\ma B'$ we have 
$\lmax(\ma C)=
\max_{\ve x\in\R^K: {\Vert\ve x\Vert}_2=1}\ve x'\ma B(\ve x'\ma B)'\geq\ve e'_k\ma B(\ve e'_k\ma B)'={(\ma B\ma B')}_{kk}\geq 0$, $k\in\Ks$.} $\lmax(\ma B)\geq 0$ that
$\ma C\in\D^K$ (see Definition \ref{dnmd}). By the result in \cite{GrW80} this implies $\ma C\in\P^K$ whenever $K\leq 4$ (see Appendix \ref{app}). Since $\ma C\in\Cs(G')$ for some $G'\subset G$ is proven for any (\ref{cmat}), we have further by Theorem 3.1 in \cite{KoB93}, or by \cite{Ber93}, that $\ma C\in\P^K$ holds also  
if $G$ has no odd cycles longer than $4$. By the definition (\ref{B3}), this completes the proof.
\end{proof}

An implication of the lemma is that any matrix $\ma C\in\Ds_i(G,\ve v)$, $\ve v\in\R_+^K$, is a feasible matrix of some subgraph of $G$. Further, \textit{any} matrix $\ma C\in\Ds_2(G,\ve v)$ is completely positive whenever either a graph $G$ with no more than $4$ vertices is considered or when the maximum odd cycle length in the graph is no longer than $4$ edges (the existence of \textit{some} completely positive matrix $\ma C\in\Ds_2(G,\ve v)$ is ensured already by Lemma \ref{lm1}). 
In particular, any such matrix is completely positive for $G$ as a sharing graph if the parallel channels are accessed by no more than $4$ users, or if there are   
$M\leq 4$ classes of parallel channels users, where channel sharing within a class is not allowed/possible due to restrictions on implementation or QoS. Recall that such parallel channels are illustrated by Examples \ref{ex7}, \ref{ex8} and their graphs are given in Fig. \ref{fig:figthree}.

\section{Upper bounds on max-min fair performance}
\label{lbounds}

In this section we derive several upper bounds on the  worst-case user performance in the considered parallel channels. According to our performance model, an upper bound represents an optimistic case, i.e. a better value of user performance than the upper bounded one.
The bounds in this section are not proven to be tight and thus, are not very interesting when considered alone. They become, however, interesting and lead to the central conclusions of this work when considered together with the lower bounds from Section \ref{ubounds}. 

\subsection{Upper bounds}

In the following Proposition, a policy-specific bound on the worst performance within the user population is proposed.

\begin{Prop}
\label{lbprop}
\textit{Given $N\geq K$, any $G=(\Ks,\Es)$ and any $(\ma A,\ma P)\in\As(G,\ve r)\times\Ps(\ma A)$, $\ve r\in\R_{+}^K$, we have
\begin{equation}
\label{lb}
\min_{k\in\Ks}\frac{\langle\ve a_k,f_k(\ve p_k)\rangle^2}{\gamma_k^2}\leq
\frac{\min_{\ve f\in\check{\Fs}(\ma A,\ma P)}\langle\ve f,\ve f\rangle}{\theta^i(G(\ma A),\ve w)},\quad i=0,1,2,
\end{equation}
with $\ve w\in\R_+^K$ such that
\begin{equation}
\label{wgr}
w_k=\frac{\gamma_k^2}{r_k^2},\quad k\in\Ks,
\end{equation}
and where we defined
\begin{equation*}
\check{\Fs}(\ma A,\ma P)=\{\ve f\in\R_+^N:\langle\bve a_k,\ve f\rangle\geq\langle\ve a_k,f_k(\ve p_k)\rangle,k\in\Ks,\quad\text{for some}\quad\bma A\in\As(G,\ve r)\}.
\end{equation*}}
\end{Prop}

\begin{proof}
Given any $\ma A\in\As(G,\ve r)$, $\ve f\in\R_+^N$ and any
$\ma P\in\Ps(\ma A)$ such that $\la\ve a_k,f_k(\ve p_k)\ra\neq 0$, $k\in\Ks$ (which by our assumptions in Section \ref{secmo} exists) , let us define $\ma Z=(\ve z_1,\ldots,\ve z_K)'\in\R^{K\times N}$ such that
\begin{equation*}
\ve z_k=\ve z_k(\bve a_k)=\sqrt{\frac{w_k}{\langle\ve f,\ve f\rangle}}\ve f-
\frac{\sqrt{w_k\langle\ve f,\ve f\rangle}}{\langle\ve a_k,f_k(\ve p_k)\rangle}\bve a_k,\quad k\in\Ks,
\end{equation*}
with an arbitrary $\bma A=(\bve a_1,\ldots,\bve a_K)'\in\R^{K\times N}$. 
Then, we have
\begin{equation}
\label{zkzl}
\langle\ve z_k,\ve z_l\rangle=\sqrt{w_kw_l}-\frac{\sqrt{w_kw_l}\langle\bve a_k,\ve f\rangle}{\langle\ve a_k, f_k(\ve p_k)\rangle}-
\frac{\sqrt{w_kw_l}\langle\bve a_l,\ve f\rangle}{\langle\ve a_l, f_l(\ve p_l)\rangle}+
\frac{\sqrt{w_kw_l}\langle\ve f,\ve f\rangle\langle\bve a_k,\bve a_l\rangle }{\langle\ve a_k,f_k(\ve p_k)\rangle\langle\ve a_l, f_l(\ve p_l)\rangle},\quad k,l\in\Ks,
\end{equation}
for any $\bma A\in\R^{K\times N}$.
Let now $\ve f\in\check{\Fs}(\ma A,\ma P)$, and note that then we can find a particular $\bma A\in\As(G,\ve r)$ which satisfies
\begin{equation}
\label{teq}
\la\bve a_k,\ve f\ra=\la\ve a_k,f_k(\ve p_k)\ra,\quad k\in\Ks
\end{equation}
(in fact, the system (\ref{teq}) has always a solution $\bma A\in\R^{K\times N}$ and since by $\ve f\in\check{\Fs}(\ma A,\ma P)$ there exists some $\tilde{\ma A}\in\As(G,\ve r)$ such that $\la\tilde{\ve a}_k,\ve f\ra\geq\la\ve a_k,f_k(\ve p_k)\ra$, $k\in\Ks$, it is implied that $\bma A\leq\tilde{\ma A}$, and thus $\bma A\in\As(G,\ve r)$).
When $\bma A\in\As(G,\ve r)$ satisfying (\ref{teq}) is taken in (\ref{zkzl}), we yield
\begin{equation}
\label{zklin}
\la\ve z_k,\ve z_l\ra\geq -\sqrt{w_kw_l},\quad k,l\in\Ks,
\end{equation}
where in particular $\la\ve z_k,\ve z_k\ra=-w_k+\frac{w_k\la\bve a_k,\bve a_k\ra\la\ve f,\ve f\ra}{\la\ve a_k,f_k(\ve p_k)\ra^2}$, and, since by Definition \ref{or} $(k,l)\notin\Es$ implies $\la\ve a_k,\ve a_l\ra=0$, also
\begin{equation}
\label{zkleq}
\la\ve z_k,\ve z_l\ra=-\sqrt{w_kw_l},\quad (k,l)\notin\Es,\quad k\neq l.
\end{equation}
Thus, by the definition (\ref{B2}), we can
write 
\begin{equation}
\label{bz}
-\ma B=\ma Z\ma Z'-\la\ve f,\ve f\ra\ma G(\bma A),\quad\text{for some}\quad\ma B\in\Bs^2(G,\ve w),
\end{equation}
where the map $\tilde{\ma A}\mapsto\ma G(\tilde{\ma A})$, $\tilde{\ma A}\in\R_+^{K\times N}$, follows 
by the definition of $\ve w$ as 
${(\ma G(\tilde{\ma A}))}_{kk}=
\frac{\gamma_k^2\la\tilde{\ve a}_k,\tilde{\ve a}_k\ra}{r_k^2\la\ve a_k,f_k(\ve p_k)\ra^2}$,
$k\in\Ks$,
and
${(\ma G(\tilde{\ma A}))}_{kl}=0$ for
$k,l\in\Ks$,
$k\neq l$.
Feature (\ref{bz}) implies then
\begin{equation}
\label{giz}
\max_{k\in\Ks}\frac{\gamma_k^2\la\ve f,\ve f\ra}{\la\ve a_k,f_k(\ve p_k)\ra^2}\ma I-\ma B\succeq\ma Z\ma Z',
\end{equation}
since by such definition of $\ma G$ and by the property $\frac{\la\bve a_k,\bve a_k\ra}{r_k^2}=\frac{\la\bve a_k,\bve a_k\ra}{\la\bve a_k,\ve 1\ra^2}\leq 1$ we have ${(\ma G)}_{kk}\leq\max_{k\in\Ks}\frac{\gamma_k^2}{\la\ve a_k,f_k(\ve p_k)\ra^2}$, $k\in\Ks$.
For the particular $\ma B\in\Bs^2(G,\ve w)$ in (\ref{bz}) we have then 
\begin{equation*}
\label{ }
\max_{k\in\Ks}\frac{\gamma_k^2}{\la\ve a_k,f_k(\ve p_k)\ra^2}\geq\frac{\lmax(\ma B)}{\la\ve f,\ve f\ra},\quad\ve f\in\check{\Fs}(\ma A,\ma P),
\end{equation*}
so that for $i=2$ the result follows by the definition (\ref{B2}). For the cases $i=0,1$ the proposition follows then from the definitions (\ref{B0}), (\ref{B1}) and from the property (\ref{Bes}), which completes the proof.
\end{proof}

By the proposition, the worst squared user performance achieved under any policy $(\ma A,\ma P)\in\As(\ve r)\times\Ps(\ma A)$ in parallel channels is no better 
than the ratio of the least $2$-norm achieved among the vectors within the set $\check{\Fs}(\ma A,\ma P)$ and the function $\theta^2$ evaluated for a sharing graph induced by $\ma A$ and for the vector $\ve w$ such that (\ref{wgr}).
According to (\ref{Bes}), when such value of $\theta^2$ is 
replaced by the Lovasz function value or Delsarte bound value assumed by the sharing graph and the vector $\ve w$, the bound from Proposition \ref{lbprop} is loosened.
Since $\gamma_k$ is a predefined performance requirement and $r_k$ the fraction of the
channel set $\Ns$ assigned to user $k\in\Ks$ over time, $\ve w$ can be interpreted as the vector of squared user performance requirements normalized by assigned channel fractions.

It is readily seen that $\check{\Fs}(\ma A,\ma P)$ is the set of values of performance functions\footnote{The value of the performance function $f_k$, $k\in\Ks$, is a vector in $\R_+^N$ and shall not be confused with the user performance, see our performance model in Section \ref{uspsec}.} which
\begin{itemize}
\item are equal for any user accessing the parallel channels and,
\item for some sharing matrix which induces the same sharing graph as $\ma A$ (i.e. under fixed sharing graph),
attain user performance no worse than under policy $(\ma A,\ma P)$. 
\end{itemize}
Thus, in some sense, $\check{\Fs}(\ma A,\ma P)$ can be seen as a set of dominating values of QoS functions for the policy $(\ma A,\ma P)$.
Note that a QoS function value $\ve f\in\check{\Fs}(\ma A,\ma P)$ may be not achievable by an allowable power allocation from $\Ps(\ma A)$, as such value leads to a superior multi-user performance under the penalty of being equal for all users. The set $\check{\Fs}(\ma A,\ma P)$ is not a polyhedron for a general $(\ma A,\ma P)\in\As(\ve r)\times\Ps(\ma A)$. Nevertheless, for any given policy $(\ma A,\ma P)$, $\check{\Fs}(\ma A,\ma P)$ contains the polyhedron
\begin{equation}
\label{polsub}
\{\ve f\in\R_+^N:\la\ve a_k,\ve f\ra\geq\la\ve a_k,f_k(\ve p_k) \ra,\quad k\in\Ks\}
\end{equation}
and its further polyhedral subset
$\{\ve f\in\R_+^N:\ve f\geq f_k(\ve p_k),\quad k\in\Ks\}$
which depends merely on $\ma P$. 
Both polyhedra give rise to obvious simplifications of (\ref{lb}): In particular, for any $(\ma A,\ma P)\in\As(\ve r)\times\Ps(\ma A)$ and for $f_{\text{max}}(\ma P)=(\max_{k\in\Ks}f_{k1}(\ve p_k),\ldots,\max_{k\in\Ks}f_{kN}(\ve p_k))$, we have 
\begin{equation*}
\min_{k\in\Ks}\frac{\langle\ve a_k,f_k(\ve p_k)\rangle^2}{\gamma_k^2}
\leq
\frac{\langle f_{\text{max}}(\ma P),f_{\text{max}}(\ma P)\rangle}{\theta^i(G(\ma A),\ve w)}
\leq
\frac{\langle\sum_{k\in\Ks}f_k(\ve p_k),\sum_{k\in\Ks}f_k(\ve p_k)\rangle}{\theta^i(G(\ma A),\ve w)},\quad i=0,1,2,
\end{equation*}
with $\ve w$ such that (\ref{wgr}).
Thus, given any policy in parallel channels, the worst squared user performance can be no better than the squared $2$-norm of the channel-wise maximum, respectively sum, of performance functions of users divided by the function $\theta^2$ (or the Lovasz function or the Delsarte bound) evaluated for the induced sharing graph and the vector of squared user performance requirements per assigned channel fraction. 

The technicality of the bound (\ref{lb}) lies in the structure of the optimization domain $\check{\Fs}(\ma A,\ma P)$, while the weight vector $\ve w$ is easily interpretable. As Corollary \ref{lbcor} in Appendix \ref{ares}, we prove an alternative version of Proposition \ref{lbprop} which simplifies the optimization domain in the bound at the expense of a more complex weight vector structure.
The bounds from Proposition \ref{lbprop} and Corollary \ref{lbcor} yield the following implication on the max-min fair performance under given sharing topology of parallel channels.

\begin{Cor}
\label{lbcorf}
\textit{Given $N\geq K$, any $G=(\Ks,\Es)$ and $\ve r\in\R_{++}^K$, we have
\begin{equation}
\label{lbf}
\max_{(\ma A,\ma P)\in\As(G,\ve r)\times\Ps(\ma A)}\min_{k\in\Ks}\frac{\langle\ve a_k,f_k(\ve p_k)\rangle^2}{\gamma_k^2}\leq
\frac{\min_{\ve f\in\check{\Fs}(G,\ve r)}\langle\ve f,\ve f\rangle}{\theta^i(G,\ve w)},\quad i=0,1,2,
\end{equation}
where $\ve w$ is such that (\ref{wgr}) and where
\begin{equation*}
\check{\Fs}(G,\ve r)=\{\ve f\in\R_+^N:\la\bve a_k,\ve f\ra\geq\la\hat{\ve a}_k,f_k(\hat{\ve p}_k)\ra,\quad k\in\Ks,\quad\textit{for some}\quad
\bma A\in\As(G,\ve r)\},
\end{equation*}
with
\begin{equation*}
(\hat{\ma A},\hat{\ma P})=\arg\max_{(\ma A,\ma P)\in\As(G,\ve r)\times\Ps(\ma A)}\min_{k\in\Ks}\frac{\la\ve a_k,f_k(\ve p_k)\ra^2}{\gamma_k^2}.
\end{equation*}}
\end{Cor}
By Proposition \ref{lbprop} it is evident that $\check{\Fs}(G,\ve r)$ is equivalent to 
the set of dominating QoS function values 
$\check{\Fs}(\hat{\ma A},\hat{\ma P})$, where
$(\hat{\ma A},\hat{\ma P})\in\As(G,\ve r)\times\Ps(\hat{\ma A})$ is a max-min fair policy under a fixed sharing graph $G$.
As $\check{\Fs}(\hat{\ma A},\hat{\ma P})$ contains the polyhedron (\ref{polsub}) for $\ma P=\hat{\ma P}$, we get the following loosened version of (\ref{lbf}).

\begin{Cor}
\label{ftilcor}
\textit{Given $N\geq K$, any $G=(\Ks,\Es)$ and $\ve r\in\R_{++}^K$, we have
\begin{equation}
\label{lbff}
\max_{(\ma A,\ma P)\in\As(G,\ve r)\times\Ps(\ma A)}\min_{k\in\Ks}\frac{\langle\ve a_k,f_k(\ve p_k)\rangle^2}{\gamma_k^2}\leq
\frac{\min_{\ve f\in\tilde{\Fs}(G,\ve r)}\langle\ve f,\ve f\rangle}{\theta^i(G,\ve w)},\quad i=0,1,2,
\end{equation}
where $\ve w$ is such that (\ref{wgr}) and where, with $\hat{\ma A}$ defined as in Corollary \ref{lbcorf},
\begin{equation*}
\tilde{\Fs}(G,\ve r)=\{\ve f\in\R_+^N:{\Vert\ve f\Vert}_1\geq {\Vert f_k(\ve p_k)\Vert}_1,\quad k\in\Ks,\quad\ma P\in\Ps(\hat{\ma A})\}.
\end{equation*}}
\end{Cor}

\begin{proof}
First notice that for the policy $(\hat{\ma A},\hat{\ma P})$ defined in Corollary \ref{lbcorf} we necessarily have ${\Vert\hat{\ve a}_k\Vert}_1=r_k$, $k\in\Ks$. Further, as for any $\ma A\in\As(G,\ve r)$ such that ${\Vert{\ve a}_k\Vert}_1=r_k$, $k\in\Ks$, it follows that $\sum_{k\in\Ks}\ve a_k=\ve 1$, we can write the condition $\la\ve f,\ve 1\ra\geq\la f_k(\ve p_k),\ve 1\ra$, $k\in\Ks$, $\ma P\in\Ps(\hat{\ma A})$, specifically as
\begin{equation*}
\sum_{k\in\Ks}\la\ve f,\ve a_k\ra\geq\max_{l\in\Ks}\la f_l(\ve p_l),\sum_{k\in\Ks}\hat{\ve a}_k\ra=\sum_{k\in\Ks}
\la\max_{l\in\Ks} f_l(\ve p_l),\hat{\ve a}_k\ra,\quad \ma A\in\As(G,\ve r),\quad \ma P\in\Ps(\hat{\ma A}).
\end{equation*}
This further implies for a particular $\ma P=\hat{\ma P}$ that
\begin{equation}
\label{ftilp1}
\sum_{k\in\Ks}\la\ve a_k,\ve f\ra\geq
\sum_{k\in\Ks}\la\hat{\ve a}_k,f_k(\hat{\ve p}_k)\ra,\quad
\ma A\in\As(G,\ve r).
\end{equation}
Let now $\ma A$ be defined as $\ve a_k=\alpha f_k(\hat{\ve p}_k)\circ\hat{\ve a}_k\circ\ve f^{-1}$, $k\in\Ks$, where $\ve f^{-1}=(1/f_1,\ldots,1/f_N)$ and $\alpha>0$ is chosen to ensure ${\Vert\ve a_k\Vert}_1\leq r_k$, $k\in\Ks$, and thus $\ma A\in\As(G,\ve r)$ (it is evident that any sufficiently small $\alpha$ satisfies such condition). For this particular $\ma A$ we have $\la\ve a_k,\ve f\ra=\alpha\la\hat{\ve a}_k,f_k(\hat{\ve p}_k)\ra$, $k\in\Ks$, so that together with (\ref{ftilp1}) it is implied that $\alpha\geq 1$ and on the other hand
\begin{equation*}
\la\ve a_k,\ve f\ra\geq\la\hat{\ve a}_k,f_k(\hat{\ve p}_k)\ra,\quad k\in\Ks.	
\end{equation*}
Consequently, $\tilde{\Fs}(G,\ve r)\subseteq\check{\Fs}(G,\ve r)$ which, by Corollary \ref{lbcorf}, completes the proof.
\end{proof}

The set $\tilde{\Fs}(G,\ve r)$ includes all QoS function values, equal for all users, which are in the sum over all channels superior to any QoS function value achieved by an allowable (for some $\hat{\ma A}\in\As(G,\ve r)$) power allocation. 
Thus, $\tilde{\Fs}(G,\ve r)$ can be seen as a hull of any user dimension of the \textit{feasible QoS/performance set} of parallel channels, which we define 
in analogy to the theory for channels with interference as \cite{Lnc06}
\begin{equation}
\label{qosr}
\{(f_1(\ve p_1)),\ldots,f_K(\ve p_K)):\ma P\in\Ps(\hat{\ma A})\},
\end{equation}
(equivalently, $\times_{k\in\Ks}\tilde{\Fs}(G,\ve r)$ is a hull of the feasible QoS set).

Corollary \ref{lbcorf} implies that a squared max-min fair performance under the condition of a fixed sharing graph $G$ in parallel channels can never exceed the ratio of the minimum squared $2$-norm within 
the hull $\tilde{\Fs}(G,\ve r)$ of any user dimension of (\ref{qosr}) and the value of the function $\theta^2$ (or the Lovasz function, or the Delsarte number) assumed by $G$ and the vector $\ve w$
satisfying (\ref{wgr}).

Consider now constraints on transmit power at any time (in a frame), as expressed e.g. by (\ref{ppc}), in which case we have $\Ps(\ma A)=\Ps$, $\ma A\in\As(\ve r)$ (allowable power allocations are independent of sharing matrices and sharing graphs). In such case it is readily seen that also set $\tilde{\Fs}(G,\ve r)$ is independent of the sharing graph on hand, i.e. $\tilde{\Fs}(G,\ve r)=\tilde{\Fs}(\ve r)$ regardless of $G$, and thus the bound (\ref{lbff}) assumes a specific separated structure.
Precisely, the max-min fair performance under a fixed sharing graph is upper-bounded by a ratio of a value dependent solely on this graph and a vector norm determined completely by the the attainable power allocations.
Thus, (\ref{lbff}) provides a separation between
the influence of the combinatorial topology induced by the channel sharing policy via Definition \ref{or} and
the impact of (the structure of) the set of allowable power allocations. 
The optimistic bound (\ref{lbff}), although looser than the one from Corollary \ref{lbcorf}, proves in the next section to be particularly insightful, since a complementary pessimistic bound of the same type can be given.
Again, recall that according to (\ref{Bes}), Corollary \ref{lbcorf} and (\ref{lbff}) provide the tightest bounds when the extension $\theta^2$ of the Lovasz function and the Delsarte bound is incorporated.

Obviously, we can reformulate Corollary \ref{lbcorf} and (\ref{lbff}) for the max-min fair performance nonrestricted in term of the sharing graph. Precisely,
\begin{equation*}
\max_{(\ma A,\ma P)\in\As(\ve r)\times\Ps(\ma A)}\min_{k\in\Ks}\frac{\langle\ve a_k,f_k(\ve p_k)\rangle^2}{\gamma_k^2}
\leq
\frac{\min_{\ve f\in\check{\Fs}(\hat{G},\ve r)}\langle\ve f,\ve f\rangle}{\theta^i(\hat{G},\ve w)}
\leq
\frac{\min_{\ve f\in\tilde{\Fs}(G,\ve r)}\langle\ve f,\ve f\rangle}{\theta^i(\hat{G},\ve w)},\quad i=0,1,2,
\end{equation*}
with $\hat{G}$ as the \textit{max-min fair sharing graph} in the sense that 
$\hat{\ma A}\in\As(\hat{G},\ve r)$ (equivalently, $\hat{G}=G(\hat{\ma A})$), where now
\begin{equation}
\label{polglob}
(\hat{\ma A},\hat{\ma P})=\arg\max_{(\ma A,\ma P)\in\As(\ve r)\times\Ps(\ma A)}\min_{k\in\Ks}\frac{\langle\ve a_k,f_k(\ve p_k)\rangle^2}{\gamma_k^2}
\end{equation}
is the (graph-nonrestricted) max-min fair policy of the parallel channels.

\subsection{Relations to coding and zero-error capacity}

Relations of max-min fair performance in parallel channels to coding and zero-error information theory results are obtained in the setting 
\begin{equation}
\label{gr1}
\frac{\gamma_k}{r_k}=1,\quad k\in\Ks.
\end{equation}
This can be assumed for a homogeneous user population, that is, if an equal fraction of the parallel channels is to be assigned (over time) to any user and all users have equal performance requirements. By the celebrated result in \cite{Lov79}, the Lovasz function of $G,\ve w$ satisfies in such case
\begin{equation*}
\theta^0(G,\ve w)\geq\Theta(G),
\end{equation*}
where $\Theta(G)=\lim_{n\to\infty}\sqrt[n]{\alpha(G^n)}$ represents the (zero-error) capacity of $G$; $\alpha$ expresses hereby the independence number of a graph and $G^n$ denotes an $n$-fold concatenation, or power, of graph $G$ \cite{Bol98}. The capacity interpretation of $\Theta(G)$ originates from the fact that $\alpha(G^n)$ represents the maximum number of $n$-letter messages which will not be confounded when $k\in\Ks$ correspond to alphabet letters and any edge $(k,l)\in\Es$ models the (danger of) confusion of letters $k,l$ \cite{Lov79}.
As a consequence of Corollary \ref{lbcorf}, (\ref{lbff}) and the result of Lovasz we yield for any sharing graph $G$ that
\begin{equation*}
\max_{(\ma A,\ma P)\in\As(G,\ve 1)\times\Ps(\ma A)}\min_{k\in\Ks}\langle\ve a_k,f_k(\ve p_k)\rangle^2
\leq
\frac{\min_{\ve f\in\check{\Fs}(G,\ve 1)}\langle\ve f,\ve f\rangle}{\Theta(G)}
\leq
\frac{\min_{\ve f\in\tilde{\Fs}(G,\ve r)}\langle\ve f,\ve f\rangle}{\Theta(G)}.
\end{equation*}
In words, under a homogeneous user population accessing the parallel channels and under sharing graph fixed to $G$, the max-min fair performance never exceeds the minimum $2$-norm within the set $\check{\Fs}(G,\ve 1)$ (respectively, within the hull $\tilde{\Fs}(G,\ve r)$) divided by the square root of the sharing graph capacity.
This means also that the max-min fair performance scales at most with the capacity of the corresponding sharing graph $G=(\Ks,\Es)$, i.e. with the \textit{effective} size of the alphabet needed for error-free communication of the letters $\Ks$ where the letter pairs $\Es$ are confusable \cite{Lov79}. 

Given (\ref{gr1}), we have also the central relation of the Delsarte bound of $G$ and the graph's independence number according to \cite{Sch79}
\begin{equation*}
\theta^1(G,\ve w)\geq\alpha(G),
\end{equation*}
(recall that by (\ref{Bes}) we have additionally 
$\theta^0(G,\ve w)\geq\theta^1(G,\ve w)$).
Thus, 
\begin{equation*}
\max_{(\ma A,\ma P)\in\As(G,\ve 1)\times\Ps(\ma A)}\min_{k\in\Ks}\langle\ve a_k,f_k(\ve p_k)\rangle^2
\leq
\frac{\min_{\ve f\in\check{\Fs}(G,\ve 1)}\langle\ve f,\ve f\rangle}{\alpha(G)}
\leq
\frac{\min_{\ve f\in\tilde{\Fs}(G,\ve r)}\langle\ve f,\ve f\rangle}{\alpha(G)},
\end{equation*}
which, with definition of the independence number,
means that the ratio of
$\min_{\ve f\in\check{\Fs}(G,\ve 1)}\la\ve f,\ve f\ra$ (respectively, $\min_{\ve f\in\tilde{\Fs}(G,\ve r)}\la\ve f,\ve f\ra$) and the maximum cardinality of a vertex subset of a sharing graph $G$ such that no two vertices in it are adjacent upper bounds the max-min fair performance under fixed sharing graph.
This implies that the max-min fair performance in parallel channels scales at most with the independence number of the sharing graph.

We close the discussion of the upper bounds by pointing out two crucial issues. First, the given upper bounds on the max-min fair performance apply to the case $N\geq K$, i.e. to the parallel channel instances with the channel ensemble no smaller than the user population accessing them. Thus, the bounds apply to, in some sense, non-overloaded parallel channels, which allow the possibility of permanent (i.e. in each frame) access to a channel for any user. 
Second, the generality of the upper bounds has to be underlined. The bounds apply to any performance function for which the formulation of the max-min fair performance 
according to (\ref{mmf}) is meaningful, that is, when a larger user performance implies a better perceived service quality level at the user receiver (Examples \ref{ex1}-\ref{ex3_}).

\section{Lower bounds on max-min fair performance}
\label{ubounds}

The lower bounds on max-min fair performance presented in this section correspond to pessimistic values, in the sense that the max-min fair performance is guaranteed to be no worse.
These bounds are analogs, or complements, of the optimistic bounds from Section \ref{lbounds}, and together embrace the max-min fair performance in parallel channels.

\subsection{Some notes on matrix scalings}
\label{scals}

The proposed bounds make use of some novel elements of the theory of matrix similarity and matrix scaling which are outlined in the following.
Let us define a scaling of a nonnegative matrix by straightforwardly extending the idea of scaling of a square positive matrix from \cite{MaO68}. 

\begin{Def}
\label{scald}
\textit{A matrix $\ma A\in\R_+^{K\times N}$ is said to be $(\ve r,\ve c)$-scalable, where $\ve r\in\R_{++}^K$ and $\ve c\in\R_{++}^N$,
if ${\Vert\ve r\Vert}_1={\Vert\ve c\Vert}_1$ and if there exist $\ma X=\diag(\ma X)\in\R_{+}^{K\times K}$ and $\ma Y=\diag(\ma Y)\in\R_{+}^{N\times N}$ such that
\begin{equation}
\label{scal}
\ma X\ma A\ma Y\ve 1=\ve r,\quad\quad\quad
\ve 1'\ma X\ma A\ma Y=\ve c'.
\end{equation}
The pair $(\ma X,\ma Y)$ is then referred to as an $(\ve r,\ve c)$-scaling of $\ma A$.}
\end{Def}
Thus, an $(\ve r,\ve c)$-scaling of a nonnegative matrix collects scaling factors of rows and columns, in the form of two diagonal matrices, such that row sums grouped in $\ve r$ and column sums grouped in $\ve c$ are obtained under row-wise and column-wise scaling.
A related notion which proves useful in later considerations is the set
\begin{equation*}
\Xs(\ma A,\ve r,\ve c)=\{\ve x=\ma X\ve 1,\ve y=\ma Y\ve 1:(\ma X,\ma Y)\textit{ is }(\bve r,\bve c)\textit{-scaling of }\ma A\in\R_+^{K\times N},\quad\!\!(\bve r,\bve c)\!\leq\!(\ve r,\ve c)\}.
\end{equation*}
In words, $\Xs(\ma A,\ve r,\ve c)$ consists of vector pairs 
which collect diagonal entries of those $(\bve r,\bve c)$-scalings of $\ma A\in\R_+^{K\times N}$ which are no larger than\footnote{Here and hereafter we refer to a scaling as larger/smaller than an other scaling if the obtained column and row sums are componentwise larger/smaller.} $(\ve r,\ve c)$.

Given predefined $\ve r\in\R_{++}^K$ and $\ve c\in\R_{++}^N$, it is obvious that matrices which are not $(\ve r,\ve c)$-scalable exist in $\R_{+}^{K\times N}$.  
Nevertheless, for any nonnegative matrix we can always find a scaling which leads to row and column sums no larger than the predefined ones. 

\begin{Lm}
\label{lm4}
\textit{Given any $\ma A\in\R_+^{K\times N}$ and any $\bve r\in\R_{++}^K$, $\bve c\in\R_{++}^N$, there exist $\ve r\leq\bve r$ and $\ve c\leq\bve c$ such that $\ma A$ is $(\ve r,\ve c)$-scalable.}
\end{Lm}

\begin{proof}
Let $\ma A=(\ve a_1,\ldots,\ve a_K)'$, with $\ve a_k\in\R_+^N$, $k\in\Ks$, and define $\bma A=\ma X\ma A$, where $\ma X=\diag(\ma X)$ is such that $\ma X\ve 1=\ve x$ and 
\begin{equation*}
x_k=\frac{\bar{r}_k}{\langle\ve a_k,\ve 1\rangle},\quad k\in\Ks.	
\end{equation*}
Then, letting $\bma A=(\bve a_1,\ldots,\bve a_K)'$, we have 
$\bma A\ve 1=\bve r$,
so that if $\ve 1'\bma A\leq\bve c$, the proof is completed.  
Otherwise, 
let
$\hat{\ma A}=\bma A\ma Y$, where $\ma Y=\diag(\ma Y)$ is such that $\ma Y\ve 1=\ve y$, with
\begin{equation*}
y_n=\min_{n\in\Ns}\frac{\bar{c}_n}{{(\ve 1'\bma A)}_n},\quad n\in\Ns.
\end{equation*}
Then, it is evident that $\ve 1'\hat{\ma A}\leq\bve c'$.
Further, as ${(\ve 1'\bma A)}_n>\bar{c}_n$ for some $n\in\Ns$ (by assumption), we have $\ve 0< \ve y<\ve 1$, which implies also
$\hat{\ma A}\ve 1<\bma A\ve 1=\bve r$	
and completes the proof.
\end{proof}

The original characterization of a scaling (of a square positive matrix) was given in \cite{MaO68} in terms of a nonlinear program. The currently known descriptions of scalings of nonnegative matrices are mostly in terms of optimization problems, see e.g. \cite{RoS89} and references therein. 
In the following we provide a novel (to the best of our knowledge) characterization which extends the concept from \cite{Lon71}.

\begin{Lm}
\label{lm3}
\textit{Let $\ma A\in\R_+^{K\times N}$ be $(\ve r,\ve c)$-scalable for some $\ve r\in\R_{++}^K$,
$\ve c\in\R_{++}^N$. 
Then, if we define $\bve r=(\ve r'\text{ }\ve 0)'\in\R_{+}^N$ and if $\ve y\in\R_{++}^N$ satisfies
\begin{equation}
\label{phiin}
\nabla\varphi(\ve y)\leq 0
\end{equation}
for the function
\begin{equation*}
\ve z\mapsto\varphi(\ve z)=-\sum_{n\in\Ns}\log\frac{z_n^{c_n}}{{(\ma A\ve z)}_n^{\bar{r}_n}},\quad \ve z\in\R_{++}^N,
\end{equation*}
and if $\ve x\in\R_{++}^K$ is such that
\begin{equation}
\label{xy}
x_k=x_k(\ve y)=\frac{r_k}{{(\ma A\ve y)}_k},\quad k\in\Ks,
\end{equation}
then $(\ma X,\ma Y)$ 
such that 
$\ma X\ve 1=\ve x$ and $\ma Y\ve 1=\ve y$ is an $(\ve r,\ve c)$-scaling of $\ma A$.
Moreover, (\ref{phiin}) is satisfied if and only if $\ve y$ is a global minimizer 
\begin{equation}
\label{locfmin}
\ve y=\arg\min_{\ve z\in\R_{++}^N}-\sum_{n\in\Ns}\log\frac{z_n^{c_n}}{{(\ma A\ve z)}_n^{\bar{r}_n}}.
\end{equation}}
\end{Lm}

\begin{proof}
By the definition, we can write
${(\nabla\varphi(\ve z))}_n=\sum_{k\in\Ks}a_{kn}\frac{r_k}{{(\ma A\ve z)}_k}-\frac{c_n}{z_n}$, $n\in\Ns$, for any $\ve z\in\R_{++}^N$,
so that with (\ref{xy}) we have in particular for $\ve z=\ve y$ that
\begin{equation*}
{(\nabla\varphi(\ve y))}_n={(\ma A'\ve x)}_n-\frac{c_n}{y_n}=
{(\ma A'\ma X\ve 1)}_n-\frac{c_n}{y_n},\quad n\in\Ns.
\end{equation*}
This implies together with (\ref{phiin}) that
\begin{equation}
\label{ls}
\ve 1'\ma X\ma A\ma Y\leq\ve c.
\end{equation}
Further, we have
\begin{equation*}
\label{ls_}
{(\ma X\ma A\ma Y\ve 1)}_k=
x_k{(\ma A\ve y)}_k=r_k,\quad k\in\Ks,
\end{equation*}
by the definition (\ref{xy}), and thus $\ve 1'\ma X\ma A\ma Y\ve 1=\ve 1'\ve c$, since $\ve 1'\ve c=\ve 1'\ve r$ holds by assumption (Definition \ref{scald}). Consequently, (\ref{ls}) is satisfied only if $\ma X\ma A\ma Y=\ve c$, and thus $\nabla\varphi(\ve y)\leq 0$ only if $\nabla\varphi(\ve y)=0$. To prove that the latter condition 
is equivalent to (\ref{locfmin}), apply the transform $\ve v=\log\ve z$, $\ve z\in\R_{++}^N$, and then rewrite $\varphi$ with the properties of the logarithm as
\begin{equation*}
\varphi(e^{\ve v})=-\sum_{n\in\Ns}\bar{r}_n\log\frac{e^{v_n}}{{(\ma Ae^{\ve v})}_n}-
\sum_{n\in\Ns}{(c_n-\bar{r}_n)v_n},\quad\ve v\in\R^N.
\end{equation*} 
As $\bve r\in\R_+^N$ and the map $\ve v\mapsto\frac{e^{v_n}}{{(\ma Ae^{\ve v})}_n}$, $\ve v\in\R^N$, is known to be log-concave (see, e.g., \cite{Lnc06}, Chapter 6), it is immediate that $\ve v\mapsto\varphi(e^{\ve v})$ is convex for $\ve v\in\R^N$. 
Thus, $\nabla\varphi(e^{\ve w})=0$ is equivalent to $\ve w=\arg\min_{\ve v\in\R^N}\varphi(e^{\ve v})$, which by the one-to-one setting $\ve w=\log\ve y$ gives (\ref{locfmin}) and completes the proof. 
\end{proof}

It is worth mentioning that function $\varphi$ from the lemma 
is multiplicatively homogeneous in the sense that $\varphi(\ve z)=\varphi(\alpha\ve z)$ for any $\ve z\in\R_{++}^K$ and $\alpha>0$ (so that any minimizer (\ref{locfmin}) scaled by some $\alpha>0$ is a minimizer of $\varphi$ as well).
This is readily seen from the exponential transformation 
\begin{equation*}
e^{\varphi(\ve z)}=\frac{\prod_{k=1}^K{(\ma A\ve z )}_k^{r_k}}{\prod_{n=1}^N z_n^{c_n}},
\end{equation*}
as used originally in \cite{Lon71}, and from the condition ${\Vert\ve r\Vert}_1={\Vert\ve c\Vert}_1$. 
Furthermore, there is a surprising relation of function $\varphi$ to the throughput optimization under interference. Let us interpret $\ve z\in\R_+^N$ as a transmit power vector of the user population $\Ns$ accessing the interference channel which has $(\ma A'\text{ }\ma 0)'\in\R_+^{N\times N}$ as its interference matrix, defined in the usual way as e.g. in \cite{Lnc06}, \cite{FTC05} (this implies that the channel gains of $N-K$ users are zero). Then, by defining the Signal-to-Interference functions of users in the interference channel as $\ve z\mapsto\sir_n(\ve z)=\frac{z_n}{(\ma A\ve z)_n}$, $n\in\Ns$ \cite{FTC05}, we can write 
\begin{equation}
\label{ifu}
\varphi(\ve z)=-\sum_{k=1}^Kr_k\log\sir_k(\ve z)-\sum_{k=1}^K (c_k-r_k)\log z_k-\sum_{k=K+1}^N c_k\log z_k.
\end{equation}
By this form, $-\varphi$ can be recognized as the weighted throughput function of the described interference channel with additional cost functions. When $c_k\geq r_k\geq 0$, $k\in\Ks$, such cost functions penalize logarithmically the excessive use of transmit power by the users. 
By the proof of Lemma \ref{lm3}, the weighted throughput function (\ref{ifu}) is known to be convex as a function of the logarithmic power vector $\ve v=\log\ve z$, $\ve z\in\R_{++}^N$ (e.g. power allocation in dB).
Lemma \ref{lm3} and the above interpretation
lead to the conclusion that $(\ma X,\ma Y)$, with
$\ma Y\ve 1=\ve y$ and $\ma X\ve 1=\ve x$, is an $(\ve r,\ve c)$-scaling of an ($(\ve r,\ve c)$-scalable) $\ma A$ if $\ve y$ represents a power allocation which globally minimizes the penalized weighted throughput function (\ref{ifu}) in the described interference channel and $\ve x$ is determined by $\ve y$ via (\ref{xy}). 

Finally, we need the following scaling-related function.

\begin{Def}
\label{mum}
\textit{Given $\ve r\in\R_{++}^K$, let the map $\ma V\mapsto\mu(\ma V)$, $\ma V\in\R_+^{K\times N}$, be defined as\footnote{We omit here the indication of the dependence on $\ve r$, since it does not introduce any ambiguities in the remainder.}
\begin{equation*}
\mu(\ma V)=\max_{(\ve x,\ve y)\in\Xs(\ma V,\ve r,\ve 1)}\min_{(n,k)\in\Ns\times\Ks}{(x_ky_n)}^2.
\end{equation*}}
\end{Def}
Such function represents the minimum squared geometric mean of pairs of diagonal entries of an $(\bve r,\bve c)$-scaling of a given matrix, achievable among all $(\bve r,\bve c)$-scalings no larger than $(\ve r,\ve c)$. In the spirit of \cite{RoS80}, we can regard $\mu$ as a (kind of) metric, or measure, of the entire class of such scalings of a given matrix.

As the simple property of later interest, we observe that if the row and column sum vectors of $\ma V$ do not exceed $(\ve r,\ve 1)$, i.e. $\ma V\ve 1\leq \ve r$ and $\ve 1'\ma V\leq \ve 1'$, then $\mu(\ma V)\geq 1$. 

\subsection{Lower bounds}

Using Definition \ref{mum} we can formulate the following lower bound on the max-min fair performance in parallel channels under fixed sharing graph. 

\begin{Prop}
\label{ubprop1}
\textit{Given any $G=(\Ks,\Es)$ and $\ve r\in\R_{++}^K$, we have
\begin{equation}
\label{ub}
\max_{(\ma A,\ma P)\in\As(G,\ve r)\times\Ps(\ma A)}\min_{k\in\Ks}
\frac{\langle\ve a_k,f_k(\ve p_k)\rangle^2}{\gamma_k^2}
\geq
\max_{\substack{\ma B\in\Bs^3(G,\ve w),\text{ }\ma V\in\R_+^{K\times N}:\\\ma V\ma V'=\lmax^{-1}(\ma B)(\ve w\sr{\ve w\sr}'-\ma B)+\ma I}}
\frac{\mu(\ma R\ma V)\max_{\ve f\in\hat{\Fs}(G,\ve r)}
\langle\ve f,\ve f\rangle}{\lmax(\ma B)},
\end{equation}
where $\ve w$ is such that (\ref{wgr}), where
\begin{equation}
\label{nkp}
N\geq\frac{K(K+1)}{2},
\end{equation}
and where, with $\hat{\ma A}$ defined as in Corollary \ref{lbcorf}, 
\begin{equation*}
\begin{split}
\hat{\Fs}(G,\ve r)=\{\ve f\in\R_+^N: 
&\ve f=f_k(\ve p_k),\quad k\in\Ks,\quad\text{for some}\quad\ma P\in\Ps(\hat{\ma A}),\\
&\ve f=\arg\max_{\bve f\in\R_+^N}\max_{\ma A\in\As(G,\ve r)}\min_{k\in\Ks}
\frac{\la\ve a_k,\bve f\ra^2}{\gamma_k^2\la \bve f,\bve f\ra}\}.
\end{split}
\end{equation*}
Moreover, for a particular 
\begin{equation}
\label{blb}
\ma B=\arg\min_{\bma B\in\Bs^3(G,\ve w)}\lmax(\bma B)
\end{equation}
(\ref{ub}) implies further
\begin{equation}
\label{ub_}
\max_{(\ma A,\ma P)\in\As(G,\ve r)\times\Ps(\ma A)}\min_{k\in\Ks}
\frac{\langle\ve a_k,f_k(\ve p_k)\rangle^2}{\gamma_k^2}
\geq
\underset{\substack{\ma V\in\R_+^{K\times N}:\\\ma V\ma V'=\lmax^{-1}(\ma B)(\ve w\sr{\ve w\sr}'-\ma B)+\ma I}}{\max}
\frac{\mu(\ma R\ma V)\max_{\ve f\in\hat{\Fs}(G,\ve r)}\langle\ve f,\ve f\rangle}{\theta^3(G,\ve w)}.
\end{equation}}
\end{Prop}

\begin{proof}
Let any $G=(\Ks,\Es)$ and any $\ma B\in\Bs^3(G,\ve w)$ be given,
and let $\ma V=(\ve v_1,\ldots,\ve v_K)'$ satisfy
\begin{equation}
\label{vfac}
\ma V\ma V'=\lmax^{-1}(\ma B)(\ve w\sr{\ve w\sr}'-\ma B)+\ma I,\quad\ma V\in\R_+^{K\times N},
\end{equation}
where by the definition (\ref{B3}) it is known that such $\ma V$ exists whenever $N=N(\ma B)$ satisfies $N\geq\phi(\lmax^{-1}(\ma B)(\ve w\sr{\ve w\sr}'-\ma B)+\ma I)$.
Defining $\ma W\sr=\diag(\ma W\sr)$ as ${(\ma W\sr)}_{kk}={(\ve w\sr)}_k$, $k\in\Ks$, the right-hand side of (\ref{vfac}) can be rewritten 
due to $-\ma B+\lmax(\ma B)\ma I\succeq 0$ as
\begin{equation}
\label{lwc}
\lmax^{-1}(\ma B)(\ve w\sr{\ve w\sr}'-\ma B)+\ma I=
\lmax^{-1}(\ma B)\ma W\sr\ma C\ma C'\ma W\sr+\ma X\ma X',
\end{equation}
with any $\ma C=(\ve c,\ldots,\ve c)'\in\R^{K\times N}$ such that $\la\ve c,\ve c\ra=1$ and with any $\ma X=(\ve x_1,\ldots,\ve x_K)'\in\R^{K\times N}$ 
which satisfies $\ma X\ma X'=-\ma B+\lmax(\ma B)\ma I$ and ${(\ve x_k)}_n=0$, $n>K$, $k\in\Ks$. Letting now $N\geq\max\{K+1,\phi(\lmax^{-1}(\ma B)(\ve w\sr{\ve w\sr}'-\ma B)+\ma I)\}$, we can find for any such $\ma X$ some $\ma C=\ma C(\ma X)\in\R_+^{K\times N}$ satisfying
\begin{equation}
\label{cx0}
\la\ve c,\ve x_k\ra=0,\quad k\in\Ks,
\end{equation} 
so that (\ref{lwc}) can be rewritten as
\begin{equation}
\label{lwc2}
\lmax^{-1}(\ma B)(\ve w\sr{\ve w\sr}'-\ma B)+\ma I=(\lmax^{-\frac{1}{2}}(\ma B)\ma W\sr\ma C\pm\ma X){(\lmax^{-\frac{1}{2}}(\ma B)\ma W\sr\ma C\pm\ma X)}'.
\end{equation}
By (\ref{vfac}), it follows now that any vector tuple
$\ve x_k+\sqrt{\frac{w_k}{\lmax(\ma B)}}\ve c$, $k\in\Ks$, yielding (\ref{lwc2}) satisfies 
\begin{equation*}
\la\ve x_k+\sqrt{\frac{w_k}{\lmax(\ma B)}}\ve c,\ve x_l+\sqrt{\frac{w_l}{\lmax(\ma B)}}\ve c\ra=\la\ve v_k,\ve v_l\ra,\quad k,l\in\Ks,
\end{equation*}
i.e., vector tuple $\ve x_k+\sqrt{\frac{w_k}{\lmax(\ma B)}}\ve c$, $k\in\Ks$, has the same lengths and mutual angles as any vector tuple $\ve v_k$, $k\in\Ks$, yielding (\ref{vfac}). As a consequence, for any tuple $\ve v_k$, $k\in\Ks$, satisfying (\ref{vfac}) and for any $\ve c$ and $\ve x_k$, $k\in\Ks$, from (\ref{lwc2}), there exists a rotation matrix $\ma Q\in\R^{N\times N}$ (a real-valued orthogonal matrix with unit determinant) for which \cite{Mur62}
\begin{equation*}
\ve v_k=\ma Q(\ve x_k+\sqrt{\frac{w_k}{\lmax(\ma B)}}\ve c),\quad k\in\Ks.
\end{equation*}
By orthogonality of $\ma Q$ we have $\la\ma Q\ve c,\ma Q\ve c\ra=\la\ve c,\ve c\ra$ and $\la\ma Q\ve x_k,\ma Q\ve x_l\ra=\la\ve x_k,\ve x_l\ra$, $k,l\in\Ks$, and (\ref{cx0}) implies $\la\ma Q\ve c,\ma Q\ve x_k\ra=0$, $k\in\Ks$.
Thus, it follows now that any factor in (\ref{vfac})
can be written as
\begin{equation}
\label{vcx}
\ma V=\sqrt{\lmax^{-1}(\ma B)}\ma W\sr\ma C\pm\ma X
\end{equation}
for some $\ma C=(\ve c,\ldots,\ve c)'\in\R_+^{K\times N}$, $\la\ve c,\ve c\ra=1$, and for some $\ma X\in\R^{K\times N}$ satisfying (\ref{cx0}) (where $\ma X$ is such that $\ma X\ma X'=-\ma B+\lmax(\ma B)\ma I$).
This further yields that
\begin{equation}
\label{vcw}
\la\ve v_k,\ve c\ra=
\sqrt{\frac{w_k}{\lmax(\ma B)}},\quad k\in\Ks,
\end{equation}
and, by the Definition (\ref{B3}), also
\begin{equation}
\label{vcw2}
\la\ve v_k,\ve v_k\ra=1,\quad k\in\Ks,
\quad\quad\quad
\la\ve v_k,\ve v_l\ra=0, k\neq l,\quad\text{if}\quad(k,l)\notin\Es.
\end{equation} 
By (\ref{vcw}) we have 
\begin{equation}
\label{lf1}
\frac{\lmax(\ma B)}{\la\ve f,\ve f\ra}=\frac{w_k}{\la\ve v_k,\ve f\ra^2},\quad k\in\Ks,
\end{equation}
for any $\ve f\in\R_+^N$ chosen to satisfy $\frac{\ve f}{\sqrt{\la\ve f,\ve f\ra}}=\ve c$ for the particular vector $\ve c$ in (\ref{vcw}).
By Lemma \ref{lm4} it is further implied that there exist $(\ma Z,\ma Y)\in\R_+^{K\times K}\times\R_+^{N\times N}$ which represent an $(\bve r,\bve c)$-scaling of $\ma R\ma V$ such that $(\bve r,\bve c)\leq(\ve r,\ve 1)$: By setting $\ma Z\ve 1=\ve z$, $\ma Y\ve 1=\ve y$ this means that we can take any $(\ve z,\ve y)\in\Xs(\ma R\ma V,\ve r,\ve 1)$, so that by (\ref{vcw2}) and Definition \ref{or} it follows that $\ma U=\ma Z\ma R\ma V\ma Y$ satisfies $\ma U\in\As(G,\ve r)$. 
Furthermore, writing $\ma U=(\ve u_1,\ldots,\ve u_K)'$, we  have then by the definition of $\ve w$ that
\begin{equation}
\label{lf1_}
\frac{w_k}{\la\ve v_k,\ve f\ra^2}=
\frac{\gamma_k^2}{(\sum_{n\in\Ns}r_k{(\ve v_k)}_nf_n)^2}=
\frac{\gamma_k^2z_k^2}{(\sum_{n\in\Ns}\frac{1}{y_n}{(\ve u_k)}_nf_n)^2}\geq\frac{\gamma_k^2
\min_{(k,n)\in\Ks\times\Ns}(z_ky_n)^2}{\la\ve u_k,\ve f\ra^2},\quad k\in\Ks,
\end{equation}
which implies with (\ref{lf1}) and Definition \ref{mum} finally that
\begin{equation}
\label{luf}
\frac{\lmax(\ma B)}{\mu(\ma R\ma V)\la\ve f,\ve f\ra}\geq
\frac{\gamma_k^2}{\langle\ve u_k,\ve f\rangle^2},\quad k\in\Ks.
\end{equation}
Note now that (\ref{luf}) holds for any $\ve f\in\R_+^N$ with $\frac{\ve f}{\sqrt{\la\ve f,\ve f\ra}}=\ve c$ for the particular $\ve c$ in (\ref{vcw}) and, by the 
assumption (\ref{cse}) and the assumptions with respect to $f_k$, $k\in\Ks$, we can always find a particular $\ve f$ such that additionally $\ve f=f_k(\ve p_k)$, $k\in\Ks$, for an arbitrary $\ma A\in\As(G,\ve r)$ and for some
$\ma P\in\Ps(\ma A)$ \footnote{Equivalently, by these 
assumptions,
$\{\ve f\in\R_+^{N}:\ve f=f_k(\ve p_k),k\in\Ks,\quad\ma P\in\Ps(\ma A)\}$ has a nonempty intersection with the ray $\{\ve f\in\R_+^N:\frac{\ve f}{\sqrt{\la\ve f,\ve f\ra}}=\ve c\}$ for any $\ve c\in\R_+^{N}$, and $\ma A\in\As(G,\ve r)$.}.
Consequently, it is further implied that 
\begin{equation}
\label{luf_}
\frac{\lmax(\ma B)}{\mu(\ma R\ma V)}\geq\min_{\substack{\ma A\in\As(G,\ve r),\ve f\in\R_+^N:\\\ve f=f_k(\ve p_k),k\in\Ks,\text{ for some }\ma P\in\Ps(\hat{\ma A})}}\max_{k\in\Ks}\frac{\gamma_k^2\la\ve f,\ve f\ra}{\la\ve a_k,\ve f\ra^2}=
\max_{k\in\Ks}\frac{\gamma_k^2\la\hat{\ve f},\hat{\ve f}\ra}{\la{\bve a}_k,\hat{\ve f}\ra^2},\quad\hat{\ve f}\in\hat{\Fs}(G,\ve r),
\end{equation}
where $\hat{\ma A}$ is defined as in Corollary \ref{lbcorf} and ${\bma A}=\arg\min_{\ma A\in\As(G,\ve r)}\max_{k\in\Ks}\frac{\gamma_k^2\la\hat{\ve f},\hat{\ve f}\ra}{\la\ve a_k,\hat{\ve f}\ra^2}$.
Thus, finally it is immediate that
\begin{equation}
\label{lmf}
\frac{\lmax(\ma B)}{\mu(\ma R\ma V)\la\ve f,\ve f\ra}\geq
\min_{(\ma A,\ma P)\in\As(G,\ve r)\times\Ps(\ma A)}\max_{k\in\Ks}\frac{\gamma_k^2}{\la\ve a_k,f_k(\ve p_k)\ra^2},
\quad\ve f\in\hat{\Fs}(G,\ve r),
\end{equation}
for any $\ma B\in\Bs^3(G,\ve w)$, for any $\ma V$ satisfying (\ref{vfac}) and $N\geq\max\{K+1,\phi(\lmax^{-1}(\ma B)(\ve w\sr{\ve w\sr}'-\ma B)+\ma I)\}$. According to Lemmas \ref{lm1}, \ref{lm2}, the latter condition is satisfied regardless of $\ma B\in\Bs^3(G,\ve w)$ if $N\geq K(K+1)/2$.
As (\ref{lmf}) is satisfied in particular for $\bma B$ such that $\lmax(\bma B)=\max_{\ma B\in\Bs^3(G,\ve w)}\lmax(\ma B)=\theta^3(G,\ve w)$, the proof is completed.
\end{proof}

The proposition says that the squared max-min fair performance achieved in parallel channels under fixed sharing topology is guaranteed to be no worse than the maximum ratio of some two expressions. The denominator expression is the maximum eigenvalue of a matrix $\ma B$ from $\Bs^3(G,\ve w)$, which is determined by the given sharing graph $G$ and the vector of squared user performance requirements normalized by assigned channel fractions.
The numerator corresponds to the squared $2$-norm of a vector from the set $\hat{\Fs}(G,\ve r)$ multiplied by the value of the metric $\mu$ of the class of $(\bve r,\bve c)$-scalings no larger than $(\ve r,\ve 1)$, of a nonnegative
factor of 
\begin{equation}
\label{rlr}
\ma R(\lmax^{-1}(\ma B)(\ve w\sr{\ve w\sr}'-\ma B)+\ma I)\ma R.
\end{equation}
By Lemma \ref{lm2}, the matrix (\ref{rlr}) represents a particular feasible matrix of a subgraph of the sharing graph. Obviously, the looser bound (\ref{ub}) is obtained by replacing the maximization of ${\mu(\ma R\ma V)}/{\lmax(\ma B)}$, conducted over $\ma B\in\Bs^3(G,\ve w)$ and the factors of (\ref{rlr}), by the minimization of the eigenvalue only. 
By Corollary \ref{lbcorf} and Proposition \ref{ubprop1} we have now
\begin{equation}
\label{jbound1}
\begin{split}
\frac{\min_{\ve f\in\check{\Fs}(G,\ve r)}\la\ve f,\ve f\ra}{\theta^2(G,\ve w)}
&=
\max_{\ma B\in\Bs^2(G,\ve w)}
\frac{\min_{\ve f\in\check{\Fs}(G,\ve r)}\la\ve f,\ve f\ra}{\lmax(\ma B)}\\
&\geq
\max_{(\ma A,\ma P)\in\As(G,\ve r)\times\Ps(\ma A)}\min_{k\in\Ks}
\frac{\langle\ve a_k,f_k(\ve p_k)\rangle^2}{\gamma_k^2}
\geq\\
\max_{\substack{\ma B\in\Bs^3(G,\ve w),\text{ }\ma V\in\R_+^{K\times N}:\\\ma V\ma V'=\lmax^{-1}(\ma B)(\ve w\sr{\ve w\sr}'-\ma B)+\ma I}}
&\frac{\mu(\ma R\ma V)\max_{\ve f\in\hat{\Fs}(G,\ve r)}\la\ve f,\ve f\ra}{\lmax(\ma B)}
\geq
\frac{\mu(\ma R\ma V)\max_{\ve f\in\hat{\Fs}(G,\ve r)}\la\ve f,\ve f\ra}{\theta^3(G,\ve w)},
\end{split}
\end{equation}
where $\ma R\ma V$ in the outer lower bound denotes any nonnegative factor of (\ref{rlr}) 
for the particular matrix (\ref{blb}), achieving the value of the $\theta^3$ function (see (\ref{lfd3})).
While $\check{\Fs}(G,\ve r)$ was shown to be the set of dominating performance function values for some policy $(\hat{\ma A},\hat{\ma P})$, set $\hat{\Fs}(G,\ve r)$ includes precisely those QoS function values which
\begin{itemize}
\item are equal for any user accessing the parallel channels,
\item are attainable by some allowable power allocation (under some $\hat{\ma A}\in\As(G,\ve r)$) and,
\item optimize the worst user performance under fixed sharing graph $G$ and under QoS function values normalized to unit $2$-norm and equal for all users. 
\end{itemize}
It is immediate that $\hat{\Fs}(G,\ve r)$ is included in the feasible performance set (\ref{qosr}) of the parallel channels and has the property that $\ve f\in\hat{\Fs}(G,\ve r)$ implies $\alpha\ve f\in\hat{\Fs}(G,\ve r)$, $\alpha<1$.

The inequality (\ref{jbound1}) contains the tightest proposed bounds which utilize the extensions $\theta^2,\theta^3$ of the Lovasz function and Delsarte number.
Since the intricacy of these bounds lies evidently in the structure of the sets $\check{\Fs}(G,\ve r)$, $\hat{\Fs}(G,\ve r)$, we proceed by proving some loosened lower bounds which together with the looser lower bound (\ref{lbff}) lead to our central insights.

\begin{Cor}
\label{ubcor1}
\textit{Given any $G=(\Ks,\Es)$ and $\ve r\in\R_{++}^K$, 
we have
\begin{equation*}
\max_{(\ma A,\ma P)\in\As(G,\ve r)\times\Ps(\ma A)}\min_{k\in\Ks}
\frac{\langle\ve a_k,f_k(\ve p_k)\rangle^2}{\gamma_k^2}
\geq
\max_{\substack{\ma B\in\Bs^3(G,\ve w),\text{ }\ma V\in\R_+^{K\times N}:\\\ma V\ma V'=\lmax^{-1}(\ma B)(\ve w\sr{\ve w\sr}'-\ma B)+\ma I}}
\frac{\mu(\ma R\ma V)\max_{\ve f\in\bar{\Fs}(G,\ve r)}\langle\ve f,\ve f\rangle}{\lmax(\ma B)},
\end{equation*}
with $\ve w$ such that (\ref{wgr}), with $N\in\N$ satisfying (\ref{nkp}), and, given $\hat{\ma A}$ defined as in Corollary \ref{lbcorf},
\begin{equation*}
\bar{\Fs}(G,\ve r)=
\{\ve f\in\R_+^N: \la\bve f,\bve f\ra\leq\la\ve f,\ve f\ra
\Rightarrow\bve f= f_k(\ve p_k),k\in\Ks,\quad\textit{for some}\quad\ma P\in\Ps(\hat{\ma A})\}.
\end{equation*}
Moreover, given a particular (\ref{blb}), this further implies
\begin{equation*}
\max_{(\ma A,\ma P)\in\As(G,\ve r)\times\Ps(\ma A)}\min_{k\in\Ks}
\frac{\langle\ve a_k,f_k(\ve p_k)\rangle^2}{\gamma_k^2}
\geq
\underset{\substack{\ma V\in\R_+^{K\times N}:\\\ma V\ma V'=\lmax^{-1}(\ma B)(\ve w\sr{\ve w\sr}'-\ma B)+\ma I}}{\max}
\frac{\mu(\ma R\ma V)\max_{\ve f\in\bar{\Fs}(G,\ve r)}\langle\ve f,\ve f\rangle}{\theta^3(G,\ve w)}.
\end{equation*}}
\end{Cor}

\begin{proof}
The definition of $\bar{\Fs}(G,\ve r)$ implies that $\ve f\in\bar{\Fs}(G,\ve r)$ if and only if
\begin{equation}
\label{fmd1}
\la\ve f,\ve f\ra\leq\max_{\delta>0}\delta\quad\text{subject to}\quad
\Bs(\delta)
\subseteq\cap_{k\in\Ks}\{\ve f=f_k(\ve p_k):\ma P\in\Ps(\hat{\ma A})\},
\end{equation}
where we define $\Bs(\delta)=\{\ve f\in\R_+^{N}:\la\ve f,\ve f\ra\leq\delta\}$.  
On the other hand, we can write by the definition of $\hat{\Fs}(G,\ve r)$ (quite redundantly) 
\begin{equation}
\begin{split}
\label{fmd2}
\max_{\ve f\in\hat{\Fs}(G,\ve r)}\la\ve f,\ve f\ra=
\min_{\bar{\delta}>0}\bar{\delta}
\quad\text{subject to}\quad
\Bs(\bar{\delta})\supseteq
(&\cap_{k\in\Ks}\{\ve f=f_k(\ve p_k):\ma P\in\Ps(\hat{\ma A})\}\\
&\cap\{\ve f=\alpha\hat{\ve f}:\alpha\geq 0\}),
\end{split}
\end{equation}
where $\hat{\ve f}=\arg\min_{\ve f\in\R_+^N}\min_{\ma A\in\As(G,\ve r),\ve f\in\R_+^N}\max_{k\in\Ks}\frac{\gamma_k^2\la\ve f,\ve f\ra}{\la\ve a_k,\ve f\ra^2}$.
Now, as any $\hat{\ve f}$ is arbitrarily nonnegatively scalable (that is, the latter set in the constraints in (\ref{fmd2}) is a ray in $\R_+^N$), 
it is implied further by (\ref{fmd2}) that
\begin{equation}
\begin{split}
\label{fmd3}
\max_{\ve f\in\hat{\Fs}(G,\ve r)}\la\ve f,\ve f\ra\geq
\min_{\bar{\delta}>0}\bar{\delta}
\quad\text{subject to}\quad
\Bs(\bar{\delta})\supseteq\Bs(\delta),
\end{split}
\end{equation}
for any $\delta$ satisfying the constraints in (\ref{fmd1}).
Thus, by (\ref{fmd1}), (\ref{fmd3}) it follows finally
\begin{equation*}
\la\ve f,\ve f\ra\leq\max_{\bve f\in\hat{\Fs}(G,\ve r)}\la\bve f,\bve f\ra,\quad\ve f\in\bar{\Fs}(G,\ve r),
\end{equation*}
which is, according to Proposition \ref{ubprop1}, sufficient for the proof for any given $G=(\Ks,\Es)$ and $\ve r\in\R_{++}^K$.  
\end{proof}

By (\ref{lbff}) and Corollary \ref{ubcor1} we have now
\begin{equation}
\label{jbound2}
\begin{split}
\frac{\min_{\ve f\in\tilde{\Fs}(G,\ve r)}\la\ve f,\ve f\ra}{\theta^2(G,\ve w)}
&=
\max_{\ma B\in\Bs^2(G,\ve w)}
\frac{\min_{\ve f\in\tilde{\Fs}(G,\ve r)}\la\ve f,\ve f\ra}{\lmax(\ma B)}\\
&\geq
\max_{(\ma A,\ma P)\in\As(G,\ve r)\times\Ps(\ma A)}\min_{k\in\Ks}
\frac{\langle\ve a_k,f_k(\ve p_k)\rangle^2}{\gamma_k^2}
\geq\\
\max_{\substack{\ma B\in\Bs^3(G,\ve w),\text{ }\ma V\in\R_+^{K\times N}:\\\ma V\ma V'=\lmax^{-1}(\ma B)(\ve w\sr{\ve w\sr}'-\ma B)+\ma I}}
&\frac{\mu(\ma R\ma V)\max_{\ve f\in\bar{\Fs}(G,\ve r)}\la\ve f,\ve f\ra}{\lmax(\ma B)}
\geq
\frac{\mu(\ma R\ma V)\max_{\ve f\in\bar{\Fs}(G,\ve r)}\la\ve f,\ve f\ra}{\theta^3(G,\ve w)},
\end{split}
\end{equation}
with $\ma R\ma V$ in the last expression as any nonnegative factor of the particular matrix (\ref{rlr}), with (\ref{blb}).
From the proof of the corollary it is evident that $\bar{\Fs}(G,\ve r)$ can be interpreted as the largest, say, ball (in the $2$-norm) of performance function values, equal for all users, included in each user dimension of the feasible QoS set (\ref{qosr}) of the parallel channels. On the other hand, recall that the hull $\tilde{\Fs}(G,\ve r)$, determining the optimistic bound in (\ref{jbound2}),
contains any such user dimension of the feasible QoS set.
Thus, the ball $\bar{\Fs}(G,\ve r)$ and the hull $\tilde{\Fs}(G,\ve r)$ determine the interval (\ref{jbound2}) of candidate max-min fair performance values in terms of the structure of the feasible performance set of parallel channels; that is, in terms of the structure of the set of allowable power allocations $\Ps(\hat{\ma A})$ and the features of the QoS functions $f_k$, $k\in\Ks$.  
In precise terms, the only such feature which is decisive for the bounds (\ref{jbound2}) is the (squared $2$-norm of the) minimum gap between $\tilde{\Fs}(G,\ve r)$ and $\bar{\Fs}(G,\ve r)$, measured as
\begin{equation*}
\min_{\ve f\in\tilde{\Fs}(G,\ve r)}\la\ve f,\ve f\ra-\max_{\ve f\in\bar{\Fs}(G,\ve r)}\la\ve f,\ve f\ra.
\end{equation*}
Such gap is visualized, together with the hull $\tilde{\Fs}(G,\ve r)$ and the ball $\bar{\Fs}(G,\ve r)$ for exemplary instance of parallel channels in Figs. \ref{fig:figfour} and \ref{fig:figfive}\footnote{Note here that the main results of this work are \textit{not} proven to hold for the parallel channels instances from Figs. \ref{fig:figfour}, \ref{fig:figfive} as the condition $N\geq K$ is violated in these cases. Figs. \ref{fig:figfour}, \ref{fig:figfive} serve, however, only as an exemplary visualization of the notions.}.

\begin{figure}
\centering
\scalebox{0.63}{\hspace{3 cm}\input{pcatt4_snrvar005_cap.pstex_t}}
\caption{The user dimensions of the feasible QoS set of parallel channels under the per-user power constraints (4b) and capacity (8) as performance function (left hand side), as well as
the resulting hull $\tilde{\Fs}(\ve r)$, the ball $\bar{\Fs}(\ve r)$ and the gap $\delta=\min_{\ve f\in\tilde{\Fs}(\ve r)}\la\ve f,\ve f\ra-\max_{\ve f\in\bar{\Fs}(\ve r)}\la\ve f,\ve f\ra$ (right hand side).
We simulated the parallel channels with $K=4$ users $k=1,2,3,4$ accessing $N=2$ channels $n=1,2$. The channels $h_{kn}$ and the variances $\sigma_{kn}$ were picked randomly from uniform distributions resulting in an average Signal-to-Noise Ratio of 6 dB.}
\label{fig:figfour}
\end{figure}

\begin{figure}
\centering
\scalebox{0.63}{\hspace{3 cm}\input{pcatt8_snrvar001_mse.pstex_t}}
\caption{The user dimensions of the feasible QoS set of parallel channels under the per-user power constraints (4b) and mean square reliability (7) as performance function (left hand side), as well as
the resulting hull $\tilde{\Fs}(\ve r)$, the ball $\bar{\Fs}(\ve r)$ and the gap $\delta=\min_{\ve f\in\tilde{\Fs}(\ve r)}\la\ve f,\ve f\ra-\max_{\ve f\in\bar{\Fs}(\ve r)}\la\ve f,\ve f\ra$ (right hand side).
We simulated the parallel channels as in Fig. 4, but for an average Signal-to-Noise Ratio of 9 dB.}
\label{fig:figfive}
\end{figure}

Consider now specifically the case of limitations of transmit powers at any time (in a frame), e.g. constrained transmit power of any user or constrained joint power budget of all users resulting in (\ref{ppc}), respectively.
As in such case $\Ps(\ma A)=\Ps$, $\ma A\in\As(\ve r)$, it is readily seen that also $\bar{\Fs}(G,\ve r)=\bar{\Fs}(\ve r)$ for any sharing graph $G$ (i.e., $\bar{\Fs}(G,\ve r)$ is independent of the induced sharing graph). As a consequence, the influence of the features of the channel sharing policy on the interval (\ref{jbound2}) of candidate values of max-min fair performance is in such case completely separated from the impact of the allowable power allocations. The combinatorial properties of the sharing graph $G$ govern the inner bounds in (\ref{jbound2}) via the minimum achievable eigenvalues $\lmax(\ma B)$ among matrices $\ma B\in\Bs^2(G,\ve w)$ and the normalized eigenvalues $\lmax(\ma B)/\mu(\ma R\ma V)$ among matrices 
$\ma B\in\Bs^3(G,\ve w)$; the normalization is by the (values of) the metrics $\mu$ of the associated factors of (\ref{rlr}).
Analogously, the outer bound behavior is described by the function values $\theta^2(G,\ve w)$ and normalized function values $\theta^3(G,\ve w)/\mu(\ma R\ma V)$, for the sharing graph $G$ and the vector of squared user performance requirements per assigned channel fraction $\ve w$, where the normalization is now by the metric $\mu$ of the corresponding factor of (\ref{rlr}) such that $\lmax(\ma B)=\theta^3(G,\ve w)$. 
Thus, the tightest pessimistic bound is obtained for a matrix $\ma B\in\Bs^3(G,\ve w)$ and a nonnegative factor $\ma R\ma V$ of (\ref{rlr}) which provide the minimum normalized eigenvalue $\lmax(\ma B)/\mu(\ma R\ma V)$. Similarly, the outer bounds in (\ref{jbound2}) are tightest for a factor $\ma R\ma V$ of the particular (\ref{rlr}), with (\ref{blb}), which maximizes metric $\mu$.
 
The outer bounds in (\ref{jbound2}) can be made in some sense symmetric whenever there exists a nonnegative factor $\ma R\ma V$ of the matrix (\ref{rlr}) satisfying (\ref{blb}) which has row sums not exceeding $\ve r$ and each column sum no larger than $1$: In fact, as it is immediate from the Definition \ref{mum} that then $\mu(\ma R\ma V)\geq 1$, we can embrace the max-min fair performance according to
\begin{equation}
\label{symbds}
\frac{\min_{\ve f\in\tilde{\Fs}(G,\ve r)}\la\ve f,\ve f\ra}{\theta^2(G,\ve w)}
\geq
\max_{(\ma A,\ma P)\in\As(G,\ve r)\times\Ps(\ma A)}\min_{k\in\Ks}
\frac{\langle\ve a_k,f_k(\ve p_k)\rangle^2}{\gamma_k^2}
\geq
\frac{\max_{\ve f\in\bar{\Fs}(G,\ve r)}\la\ve f,\ve f\ra}{\theta^3(G,\ve w)},	
\end{equation}
where we have purely spectral dependence on the sharing graph $G$ in the form of functions (\ref{lfd2}), (\ref{lfd3}). 
By the theory of matrix scaling \cite{Lon71}, \cite{RoS89}, the existence of such particular factor depends on the pattern of its zero entries, which is shown by the following paraphrased result from \cite{Bru68}.

\begin{Prop}[\cite{Bru68}]
\label{bruprop}
\textit{Let denote by $\ma R\ma V(\Ls \vert \Ms)$, with $\Ls\subset\Ks$, $\Ms\subset\Ns$, the submatrix of $\ma R\ma V\in\R_+^{K\times N}$ which is obtained by deleting  all rows $k\in\Ls$ and all columns $n\in\Ms$ from $\ma R\ma V$. Then, we have $\mu(\ma R\ma V)\geq 1$ if
\begin{equation}
\label{bruconds}
\begin{split}
&\sum_{k\in\Ls}\bar{r}_k<
\sum_{n\in\Ms}\bar{c}_n\quad\text{if}\quad
\ma R \ma V(\Ks\setminus\Ls\vert\Ms)=\ma 0,\quad
\ma R \ma V(\Ls\vert\Ns\setminus\Ms)\neq\ma 0,\\
&\sum_{k\in\Ls}\bar{r}_k=
\sum_{n\in\Ms}\bar{c}_n\quad\text{if}\quad
\ma R \ma V(\Ks\setminus\Ls\vert\Ms)=\ma 0,\quad
\ma R \ma V(\Ls\vert\Ns\setminus\Ms)=\ma 0,
\end{split}
\end{equation}
holds for some $(\bve r,\bve c)\leq(\ve r,\ve 1)$. Thus, the max-min fair performance satisfies (\ref{symbds}) if there exists a nonnegative factor $\ma R\ma V$ of the matrix (\ref{rlr}), such that (\ref{blb}) and (\ref{bruconds}) for some $(\bve r,\bve c)\leq(\ve r,\ve 1)$.}
\end{Prop}  
Recall here from Lemma \ref{lm2} that any matrix (\ref{rlr}) is a feasible matrix of some subgraph of the sharing graph. Thus, by the above proposition, the existence of a factor ensuring $\mu(\ma R\ma V)\geq 1$, depends on the existence/nonexistence of certain edges in the sharing graph.

\subsection{Role of scalings}

The row and column sums of factors of the certain feasible matrix (\ref{rlr}) of some sharing subgraph influence the max-min fair performance in a specific way, which we show more explicitly here.
Proposition \ref{ubprop2} in Appendix \ref{ares} provides a technical alternative version of the bounds from Corollary \ref{ubcor1} and we simplify it in the following. By the proof, one can readily see that the bounds from Proposition \ref{ubprop2} are slightly tighter than those from Corollary \ref{ubcor1}, at the expense of higher complexity\footnote{Note also that the proof of Proposition \ref{ubprop2} allows for an even tighter bound formulation which generalizes Proposition \ref{ubprop1}; set $\bar{\Fs}(G,\ve r,\ve y)$ has to be merely replaced by $\hat{\Fs}(G,\ve r,\ve y)$ given in (\ref{fhaty}).} 
Under apriori setting $\ve y=\ve 1$ in Proposition \ref{ubprop2} and using the definition of $\ve w(\ve x)$ and (\ref{B3}), we obtain a more insightful, loosened version of the bounds: Together with (\ref{lbff}), we yield then precisely
\begin{equation}
\label{jbound3}
\begin{split}
\frac{\min_{\ve f\in\tilde{\Fs}(G,\ve r)}\la\ve f,\ve f\ra}{\theta^2(G,\ve w)}
&=
\max_{\ma B\in\Bs^2(G,\ve w)}
\frac{\min_{\ve f\in\tilde{\Fs}(G,\ve r)}\la\ve f,\ve f\ra}{\lmax(\ma B)}\\
&\geq
\max_{(\ma A,\ma P)\in\As(G,\ve r)\times\Ps(\ma A)}\min_{k\in\Ks}
\frac{\langle\ve a_k,f_k(\ve p_k)\rangle^2}{\gamma_k^2}
\geq\\
\max_{\substack{\ma B\in\Bs^3(G,\ve w),\\(\ve x,\ve 1)\in\Xs(\ma R\ma V,\ve r,\ve 1),\text{ }\ma V\in\R_+^{K\times N}:\\\ma V\ma V'=\lmax^{-1}(\ma B)(\ve w\sr{\ve w\sr}'-\ma B)+\ma I}}
&\frac{\min_{k\in\Ks}x_k\max_{\ve f\in\bar{\Fs}(G,\ve r)}\la\ve f,\ve f\ra}{\lmax(\ma B)}
\geq
\frac{\min_{k\in\Ks}x_k\max_{\ve f\in\bar{\Fs}(G,\ve r)}\la\ve f,\ve f\ra}{\theta^3(G,\ve w)},
\end{split}
\end{equation}
where in the last expression we can take any $\ve x$ satisfying $(\ve x,\ve 1)\in\Xs(\ma R\ma V,\ve r,\ve 1)$ for any nonnegative factor $\ma R\ma V$ of the matrix (\ref{rlr}) for (\ref{blb}) (note here that for $\tilde{\Fs}(G,\ve r,\ve y)$, $\ve y\in\R_{++}^N$, defined in Proposition \ref{ubprop2} we have $\bar{\Fs}(G,\ve r,\ve 1)=\bar{\Fs}(G,\ve r)$).
If the constraints of transmit power at any time in a frame are considered (e.g. either of (\ref{ppc})), then 
one can see the same separate impact of allowable power allocations and the channel sharing combinatorics on the bounds (\ref{jbound3}) as in the case of (\ref{jbound2}):
In terms of $\Ps(\ma A)=\Ps$, $\ma A\in\As(G,\ve r)$, the interval of max-min fair performance values is determined by
the distance between the largest included ball $\bar{\Fs}(G,\ve r)$ and the hull $\tilde{\Fs}(G,\ve r)$ of each user dimension of the feasible QoS set. Independently, the
minimum achievable eigenvalues $\lmax(\ma B)$ within sets (\ref{B2}) and (\ref{B3}), or the spectral characterizations $\theta^2$ and $\theta^3$, govern the bounds in terms of the channel sharing topology expressed by the sharing graph $G$.
By the definition of $\Xs(\ma R\ma V,\ve r,\ve 1)$, it is further evident that the outer lower bound in (\ref{jbound3}) is a linear function of the minimum row scaling factor which is required to scale a nonnegative factor $\ma R\ma V$ of (\ref{rlr}) down, until each column sum does not exceed unity and the vector of row sums is no larger than $\ve r$. This leads to a conclusion that the outer bounds (\ref{jbound3}) embrace the max-min fair performance value as tightly as possible if such a nonnegative factor $\ma R\ma V$ is taken which has componentwise smallest row sum vector relative to $\ve r$.
Note that, as (\ref{rlr}) represents some feasible matrix of some sharing subgraph $G'\subset G$, the row sums of a factor of (\ref{rlr}) are determined by the channel sharing combinatorics, that is, by the existence/nonexistence of certain edges in the sharing graph (and by the vector $\ve w$ of squared user performance requirements per assigned channel fraction).

By the bounds (\ref{jbound3}) it can be again seen that a symmetric embracing of max-min fair performance according to (\ref{symbds}) is implied whenever there exists a factor $\ma R\ma V$ of (\ref{rlr}), for the particular (\ref{blb}), which has all column sums no larger than unity and all row sums componentwise not exceeding $\ve r$ (see Proposition \ref{bruprop}): In fact, in such case we can find a particular $\ve x$ such that $\min_{k\in\Ks} x_k\geq 1$ among all $(\ve x,\ve 1)\in\Xs(\ma R\ma V,\ve r,\ve 1)$.

Consider now the complementary simplification of Proposition \ref{ubprop2}, where $\ve x=\ve 1$ is set apriori. Then, together with (\ref{lbff}) we yield immediately
\begin{equation}
\label{jbound4}
\begin{split}
\frac{\min_{\ve f\in\tilde{\Fs}(G,\ve r)}\la\ve f,\ve f\ra}{\theta^2(G,\ve w)}
&=
\max_{\ma B\in\Bs^2(G,\ve w)}
\frac{\min_{\ve f\in\tilde{\Fs}(G,\ve r)}\la\ve f,\ve f\ra}{\lmax(\ma B)}\\
&\geq
\max_{(\ma A,\ma P)\in\As(G,\ve r)\times\Ps(\ma A)}\min_{k\in\Ks}
\frac{\langle\ve a_k,f_k(\ve p_k)\rangle^2}{\gamma_k^2}
\geq\\
\max_{\substack{\ma B\in\Bs^3(G,\ve w),\text{ }(\ve 1,\ve y)\in\Xs(\ma R\ma V,\ve r,\ve 1),\text{ }\ma V\in\R_+^{K\times N}:\\\ma V\ma V'=\lmax^{-1}(\ma B)(\ve w\sr{\ve w\sr}'-\ma B)+\ma I}}
&\frac{\max_{\ve f\in\bar{\Fs}(G,\ve r,\ve y)}\la\ve y\circ\ve f,\ve y\circ\ve f\ra}{\lmax(\ma B)}
\!\geq\!
\frac{\max_{\ve f\in\bar{\Fs}(G,\ve r,\ve y)}\la\ve y\circ\ve f,\ve y\circ\ve f\ra}{\theta^3(G,\ve w)},
\end{split}
\end{equation}
where in the outer lower bound we can choose any $\ve y$
such that $(\ve y,\ve 1)\in\Xs(\ma R\ma V,\ve r,\ve 1)$, with $\ma R\ma V$ as any nonnegative factor of (\ref{rlr}), where (\ref{blb}).
It is evident from the definition that the set $\bar{\Fs}(G,\ve r,\ve y)$ consists of performance function values which  
\begin{itemize}
\item are equal to, say, $\ve f\in\R_+^N$ for all users accessing the parallel channels and,
\item when weighted by $\ve y$ in the sense $\ve y\circ\ve f$, are included in each user dimension of the feasible performance set of parallel channels.
\end{itemize}
By analogy to $\bar{\Fs}(G,\ve r)$, we can interpret the set $\bar{\Fs}(G,\ve r,\ve y)$ as a kind of largest ball which is included in each user dimension of set (\ref{qosr}), but which size is measured in a weighted (by $\ve y$) Euclidean norm. Thus, the interval of max-min fair performance values (\ref{jbound4}) is influenced by the structure of the set $\Ps(\hat{\ma A})$ and functions $f_k$, $k\in\Ks$,
through the included weighted-norm ball $\bar{\Fs}(G,\ve r,\ve y)$ and the hull $\tilde{\Fs}(G,\ve r)$ of any user dimension of set (\ref{qosr}); the impact is purely via the weighted norm gap
\begin{equation*}
\min_{\ve f\in\tilde{\Fs}(G,\ve r)}\la\ve f,\ve f\ra-\max_{(\ve 1,\ve y)\in\Xs(\ma R\ma V,\ve r,\ve 1),\bar{\Fs}(G,\ve r,\ve y)}\la\ve f,\ve f\ra,
\end{equation*}
where $\ma R\ma V$ is a factor of (\ref{rlr}) for $\ma B\in\Bs^3(G,\ve w)$ achieving the tighter lower bound, or
a factor of (\ref{rlr}) for (\ref{blb}) when the outer lower bound is considered.

Recall that under constraints on transmit power at any time in a frame, such as (\ref{ppc}), we have $\bar{\Fs}(G,\ve r)=\bar{\Fs}(\ve r)$ and $\tilde{\Fs}(G,\ve r)=\tilde{\Fs}(\ve r)$ regardless of $G$, and thus the power allocations and channel sharing graph influence separately the numerator and denominator of the bounds (\ref{jbound4}).
In addition to the impact of channel sharing combinatorics through the minimum of $\lmax(\ma B)$ within (\ref{B2}) and (\ref{B3}) (respectively, via the Lovasz function and Delsarte bound extensions $\theta^2$, $\theta^3$), we see that the looser lower bound in (\ref{jbound4}) is proportional to the weighted squared $2$-norm of $\ve y$ subject to $(\ve 1,\ve y)\in\Xs(\ma R\ma V,\ve r,\ve 1)$. Thus, the lower bound scales bilinearly with the vector of scaling factors which are needed in column-wise scaling of a nonnegative factor $\ma R\ma V$ of (\ref{rlr}) to attain row sums and column sums componentwise not exceeding $(\ve r,\ve 1)$. Again, as (\ref{rlr}) is a feasible matrix of a certain subgraph of the sharing graph, the column sums of $\ma R\ma V$ are determined by the channel sharing topology and by the vector $\ve w$.
It can be observed that a factor $\ma R\ma V$ which achieves smallest possible column sums is desired to provide as tight as possible outer interval of max-min fair performance values in (\ref{jbound4}).
The bounds (\ref{jbound4}) confirm the conclusion that
we have the symmetric bounds (\ref{symbds}) whenever matrix (\ref{rlr}) satisfying (\ref{blb}) has a nonnegative factor with row sum vector no larger than $\ve r$ and no column sum exceeding unity (in this case (\ref{symbds}) is implied by (\ref{jbound4}) by taking $\ve y=\ve 1$, and we also have $\bar{\Fs}(G,\ve r,\ve 1)=\bar{\Fs}(G,\ve r)$).

We can finally conclude that each of the inequalities proposed so far allows us to embrace by bounds also the nonrestricted max-min fair performance of parallel channels, i.e. the max-min fair performance when no 
sharing graph is given apriori. For instance, (\ref{jbound2}) implies
\begin{equation*}
\begin{split}
\frac{\min_{\ve f\in\tilde{\Fs}(\bar{G},\ve r)}\la\ve f,\ve f\ra}{\theta^2(\hat{G},\ve w)}
&=
\max_{\ma B\in\Bs^2(\hat{G},\ve w)}
\frac{\min_{\ve f\in\tilde{\Fs}(\bar{G},\ve r)}\la\ve f,\ve f\ra}{\lmax(\ma B)}\\
&\geq
\max_{(\ma A,\ma P)\in\As(\ve r)\times\Ps(\ma A)}\min_{k\in\Ks}
\frac{\langle\ve a_k,f_k(\ve p_k)\rangle^2}{\gamma_k^2}
\geq\\
\max_{\substack{\ma B\in\Bs^3(\hat{G},\ve w),\text{ }\ma V\in\R_+^{K\times N}:\\\ma V\ma V'=\lmax^{-1}(\ma B)(\ve w\sr{\ve w\sr}'-\ma B)+\ma I}}
&\frac{\mu(\ma R\ma V)\max_{\ve f\in\bar{\Fs}(\bar{G},\ve r)}\la\ve f,\ve f\ra}{\lmax(\ma B)}
\geq
\frac{\mu(\ma R\ma V)\max_{\ve f\in\bar{\Fs}(\bar{G},\ve r)}\la\ve f,\ve f\ra}{\theta^3(\hat{G},\ve w)},
\end{split}
\end{equation*}
where $\hat{G}$ is a max-min fair sharing graph, i.e. a graph induced by the max-min fair sharing policy $\hat{\ma A}$, such that (\ref{polglob}) (that is, we have $\hat{G}=G(\hat{\ma A})$ and $\hat{\ma A}\in\As(\hat{G},\ve r)$). Clearly, the other bounds (\ref{jbound1}), (\ref{symbds}), (\ref{jbound3}), (\ref{jbound4}) give rise to analogous enclosing of graph-nonrestricted max-min fair performance, when a max-min fair sharing graph is incorporated. 

\subsection{Relation to the interference channel}

By (\ref{jbound4}), one can recognize an interesting relation between max-min fair performance in parallel channels and the (weighted) throughput optimization in the interference channel considered in Section \ref{scals}. Lemma \ref{lm3} and the definition of $\Xs(\ma R\ma V,\ve r,\ve 1)$
make evident that vector $\ve y$ in the lower bounds in (\ref{jbound4}) corresponds to a certain power allocation in the associated interference channel.

\begin{Cor}
\textit{Let an interference channel with user population $\Ns$ have an interference matrix $((\ma R\ma V)'\text{ }\ma 0)'\in\R_+^{N\times N}$, describing the interference among users according to \cite{Lnc06}, which corresponds to any nonnegative factor of (\ref{rlr}) such that (\ref{blb}). Then, $\ve y$ in the outer lower bound (\ref{jbound4}) is a power allocation in such interference channel which maximizes the weighted throughput 
\begin{equation*}
\sum_{k\in\Ks}\bar{r}_k\log\sir_k(\ve z),\quad\ve z\in\R_{++}^N,
\end{equation*} 
with additive logarithmic power penalty terms $(\bar{c}_k-\bar{r}_k)\log z_k$, $k\in\Ks$, and $\bar{c}_k\log z_k$, $k\in\Ns\setminus\Ks$, for some $(\bve r,\bve c)\leq(\ve r,\ve 1)$.}
\end{Cor}
Precisely, by Lemma \ref{lm3}, the weight vectors $(\bve r,\bve c)$ in the throughput function collect row and column sums obtained under columnwise scaling of $\ma R\ma V$ by $\ve y$.
The interesting point is that the throughput-optimal power allocation $\ve y$ in the described interference channel influences the pessimistic bounds on max-min fair performance in the related parallel channels.
For instance, the outer lower bound (\ref{jbound4}) becomes tighter if the taken nonnegative factor $((\ma R\ma V)'\text{ }\ve 0)'$ of the matrix (\ref{rlr}) for the particular (\ref{blb}) represents such an interference matrix of the associated interference channel, which enforces higher user powers for optimizing the  weighted throughput from the corollary. 
Recall here that (\ref{rlr}) is a feasible matrix of some sharing subgraph $G'\subset G$, so that the candidate interference matrices of the associated interference channel depend on the channel sharing topology in the original parallel channels. 

\subsection{Role of sharing graph cycles}
\label{cyc}

What is apparent in all proposed inequalities enclosing the max-min fair performance so far, is the difference in the dependence on the channel sharing combinatorics between the upper and lower bounds. Upper bounds depend on the given sharing graph $G$ (and weight vector $\ve w$) through
the minimum eigenvalue $\lmax(\ma B)$ among matrices $\ma B\in\Bs^2(G,\ve w)$, respectively through the associated value of the function $\theta^2$.
The lower bounds depend on the channel sharing policy via the minimum of $\lmax(\ma B)$ among matrices $\ma B$ from the smaller set $\Bs^3(G,\ve w)$, respectively via the 
value which the function $\theta^3$ assumes for $G$ and $\ve w$.
By the recent results on completely positive graphs, we can, however, unify the dependence on the sharing graph for a large class of sharing graphs/topologies.

\begin{Prop}
\label{prop23}
\textit{Let $G=(\Ks,\Es)$ be any sharing graph with $K\leq 4$ or including no odd cycles longer than $4$. Then, the bounds from Propositions \ref{ubprop1}, \ref{ubprop2} and Corollary \ref{ubcor1} and the bounds
(\ref{jbound1}), (\ref{jbound2}), (\ref{symbds}), (\ref{jbound3}), (\ref{jbound4}) are satisfied with
\begin{equation*}
\Bs^2(G,\ve w)=\Bs^3(G,\ve w),\quad\textit{and thus,}\quad\theta^3(G,\ve w)=\theta^2(G,\ve w).
\end{equation*}}
\end{Prop}

The proposition is an immediate consequence of Lemma \ref{lm2} and the definitions (\ref{lfd2}), (\ref{lfd3}). The key to the above identity of $\theta^2$ and $\theta^3$ is that, for any sharing graph $G$ with no more than $K=4$ vertices or no odd cycles longer than $4$, 
any feasible matrix (\ref{rlr}) of a sharing subgraph, for any $\ma B\in\Bs^2(G,\ve w)$, is completely positive and not only doubly nonnegative (see proof of Lemma \ref{lm2}).
Proposition \ref{prop23} implies that whenever the parallel channels are accessed by no more than $K=4$ users, the value of the function $\theta^2$ assumed for the sharing graph $G$ (and vector $\ve w$) is a sufficient 
characterization of the sharing policy
for enclosing the max-min fair performance from above and from below, according to (\ref{jbound1}), (\ref{jbound2}), (\ref{symbds}), (\ref{jbound3}) or (\ref{jbound4}). 
Similarly, the value $\theta^2(G,\ve w)$, for the given sharing graph $G$, 
is a sufficient description of the channel sharing for the proposed bounds (\ref{jbound1}), (\ref{jbound2}), (\ref{symbds}), (\ref{jbound3}), (\ref{jbound4})  on max-min fair performance, when there is no odd chain of more than $K=4$ users such that any two subsequent users share some channel and the last user shares a channel with the first one (this makes up a cycle in the sharing graph).
In particular, we have such property when the users accessing the parallel channels can be partitioned into no more than $M=4$ groups such that no pair of users within one group is allowed (or able) to share a channel; for instance, due to certain 
constraints on traffic class processing or hardware.
The channel sharing is represented in such case by an $M$-partite sharing graph, $M=2,3,4$, with particular examples depicted in Fig. \ref{fig:figthree}.
Two parallel channel instances of this type, and thus such that the bounds (\ref{jbound1}), (\ref{jbound2}), (\ref{symbds}), (\ref{jbound3}), (\ref{jbound4}) are determined solely be the function $\theta^2$, were presented in Examples \ref{ex7} and \ref{ex8}: In the multi-user multi-carrier channel from Example \ref{ex7} certain user constellations are not allowed to    
share channels due to regulations on traffic processing, while in Example \ref{ex8} the sharing of channels within some user classes is prevented/undesired because of excessive difference of delay times.

To summarize, we note that whenever the parallel channels are shared according to any sharing graph $G$ from Proposition \ref{prop23}, the proposed bounds enclosing the max-min fair performance
are determined by
the \textit{spectral} properties of the 
channel sharing combinatorics via some value of $\lmax(\ma B)$ among matrices $\ma B\in\Bs^2(G,\ve w)$, respectively via the value of $\theta^2$ assumed by graph $G$ and vector $\ve w$ collecting squared user performance requirements per assigned channel fraction.
The \textit{structural} features of the sharing topology have impact on the bounds  through the metric $\mu$, or row-sums, or column sums of a nonnegative factor of a feasible matrix (\ref{rlr}) of some sharing subgraph.
As far as transmit power constraints at any time (in a frame) are considered, e.g. (\ref{ppc}), the
impact of the (set of) allowable power allocations and the curvature of QoS functions is decoupled from the influence of the sharing graph; it is mirrored by the gap separating the hull $\tilde{\Fs}(\ve r)$ from the largest included ball $\bar{\Fs}(\ve r)$ of each user dimension of the feasible QoS set.

We close the discussion on the max-min fair performance by discussing the issue of the channel ensemble. It is evident from Corollary \ref{lbcorf} and Propositions \ref{ubprop1}, \ref{ubprop2} that the max-min fair performance of parallel channels can be enclosed by bounds (\ref{jbound1}), (\ref{jbound2}), (\ref{symbds}), (\ref{jbound3}), (\ref{jbound4}) whenever the number of accessed parallel channels satisfies (\ref{nkp}).
This means that the proposed bounds apply to non-overloaded parallel channels for which the (cardinality of) channel population exceeds the (cardinality of) user population $K$ at least by the factor $(K+1)/2$.
From the proofs of Propositions \ref{ubprop1}, \ref{ubprop2} it is evident that 
such condition
results from the use of the general \textit{nontight} bound on cp-rank of matrix (\ref{rlr}) implied by Lemmas \ref{lm1}, \ref{lm2}. As a consequence, the class of parallel channels instances satisfying (\ref{nkp})
can be generalized, depending on the 
particular matrices $\ma B\in\Bs^3(G,\ve w)$
achieving the lower bounds in Propositions \ref{ubprop1}, \ref{ubprop2}. Precisely, the bounds (\ref{jbound1}), (\ref{jbound2}), (\ref{symbds}), (\ref{jbound3}), (\ref{jbound4}) apply, more generally, when
\begin{equation*}
N\geq\max\{K+1,\phi(\lmax^{-1}(\ma B)(\ve w\sr{\ve w\sr}'-\ma B)+\ma I)\}
\end{equation*}
is satisfied for the corresponding matrices $\ma B\in\Bs^3(G,\ve w)$ in the lower bounds. In other words, the proposed bounds apply, more generally, when the channel population exceeds the user population $K$ by a factor no smaller than 
$\max\{K+1,\phi\}/K$, with $\phi$ as the cp-rank of matrix (\ref{rlr}), for $\ma B$ achieving the lower bound of interest.

\section{Characterization of some fair policies}
\label{policies}

The proofs of the lower bounds from Propositions \ref{ubprop1}, \ref{ubprop2} 
are constructive, that is, they contain implicit specifications of certain parallel channels policies.
This allows us in this section to derive some algorithms 
for the computation of \textit{fair policies} in the case of predefined topology, or equivalently graph, of parallel channels sharing.
A fair policy is understood here as a policy which ensures user performance of any user be no worse than some specified pessimistic bound. According to Definition \ref{or}, a predefined sharing graph means predetermined binary relations consisting in sharing/no sharing of channels by the single user pairs. We already explained in Section \ref{sec_4} that the predetermination of channel sharing topology can be motivated by regulations on processing of different traffic classes, e.g. in the manner as in the multi-user multi-carrier channel from Example \ref{ex7}.
The fixing of a channel sharing graph can be also necessary under certain constraints on hardware and/or signal processing, similarly to the Example \ref{ex8} of parallel channels.

In order to simplify the presentation, 
we assume that the predefined sharing topology results in a sharing graph with no odd cycles longer than $4$, so that we have the equivalence from Proposition \ref{prop23} throughout this section. 
Also, we restrict our attention to constraints on transmit power, e.g. by assuming individually constrained user power or constrained joint power budget of users at any time (in a frame) according to (\ref{ppc}): As a consequence, in what follows we have $\tilde{\Fs}(G,\ve r)=\tilde{\Fs}(\ve r)$ and $\bar{\Fs}(G,\ve r)=\bar{\Fs}(\ve r)$ regardless of the sharing graph $G$. It is, however, easily verified that all the algorithmic concepts proposed in the following are straightforwardly extendable to the case of energy constraints (per frame).

\subsection{Fair policy as orthonormal-like representation}
\label{fol}

Using the conventional optimization formulation, the problem of ensuring max-min fairness under given channel sharing topology can be written as
\begin{equation}
\label{clprb_}
\min_{(\ma A,\ma P)}\max_{k\in\Ks}-
\frac{\langle\ve a_k,f_k(\ve p_k)\rangle}{\gamma_k},
\quad\text{subject to}\quad
\begin{cases}
&(\ma A,\ma P)\in\As(\ve r)\times\Ps\\
&\la\ve a_k,\ve a_l\ra\leq 0,\quad (k,l)\notin\Es,
\end{cases}
\end{equation}
where the set $\Es$ is such that $\Ks^2\setminus\Es$ collects all user pairs which are not allowed to share a channel or, equivalently (Definition \ref{or}), $G=(\Ks,\Es)$ is a given sharing graph\footnote{Note, that the inequality in the second constraint in (\ref{clprb_}) is equivalent to equality as nonnegativity is implicit from $\ma A\in\As(\ve r)$.}. 
Conventional optimization methods (e.g. interior point methods \cite{Ber99}) allow for a global solution basically in the case of convexity of the problem. Such feature is prevented in (\ref{clprb_}) since a bilinear form, used in the constraints, is not a convex function.
Additionally, we consider very general performance functions $f_k$, $k\in\Ks$, and arbitrary constraints on transmit power, so that a standard method solution of (\ref{clprb_}) is expected, in general, to be only local. 
In this light, resorting to efficient computation methods of (suboptimal) fair policies seems to be an attractive alternative. 

One possible method is implied in the proof of Proposition \ref{ubprop2} by the inequality (a reformulation of the first inequality in (\ref{luf_2}))
\begin{equation}
\label{luf__}
\begin{split}
&\frac{\la\ve y\circ\hat{\ve f},\ve y\circ\hat{\ve f}\ra}{\lmax(\ma B)}
\leq\min_{k\in\Ks}\frac{\la\hat{\ve a}_k,\hat{\ve f}\ra^2}{\gamma_k^2},\\
&(\hat{\ma A},\frac{\hat{\ve f}}{\sqrt{\la\ve y\circ\hat{\ve f},\ve y\circ\hat{\ve f}\ra}})=\arg\min_{(\ma A,\ve c)}\max_{k\in\Ks}-\frac{\la\ve a_k,\ve c\ra}{\gamma_k}\quad\text{subject to}\quad
\begin{cases}
&(\ma A,\ve c)\in\As(\ve r)\times\R_+^N\\
&\la\ve a_k,\ve a_l\ra\leq 0,\quad (k,l)\notin\Es\\
&\la\ve y\circ\ve c,\ve y\circ\ve c\ra\leq 1,
\end{cases}
\end{split}
\end{equation}
given any $\ma B\in\Bs^2(G,\ve w(\ve x))$, with map
$\ve z\mapsto\ve w(\ve z)$, $\ve z\in\R_{++}^K$, defined in Proposition\footnote{From the objective of the problem it is readily seen that the last inequality constraint can be replaced by equality.} \ref{ubprop2}. Hereby, any vectors $\ve x,\ve y$ satisfying $(\ve x,\ve y)\in\Xs(\ma R\ma V,\ve r,\ve 1)$ for (\ref{vfac}) can be chosen. 
It is evident by (\ref{lfd}) that the problem in (\ref{luf__}) is closely related to the computation of an orthonormal representation and a unit vector which achieve the value of the Lovasz function (\ref{lfd}) (recall the definition of orthonormal representation from Section \ref{sec_4}): In (\ref{luf__}), the unit vector $\ve c$ is, however, considered in weighted norm and is additionally restricted to be nonnegative, while the constraints on $\ma A$ are expressed in $1$-norm. 
The complexity of the problem in (\ref{luf__}) is significantly reduced in relation to the original problem (\ref{clprb_}). 
We can restate this problem as an instance of so-called \textit{bilinear program} by replacing the objective by some variable $s$ and by adding the inequalities $\la\ve a_k,\ve c\ra/\gamma_k-s\leq 0$, $k\in\Ks$, to the constraints. 
Although a bilinear program does not represent a convex problem, there exists a variety of efficient methods for its global and local solution; without giving further details we refer for a selection of such methods to \cite{Whi92}, \cite{ShS80} and references therein.

Clearly, in the orthonormal-like representation $(\hat{\ma A},\frac{\hat{\ve f}}{\sqrt{\la\ve y\circ\hat{\ve f},\ve y\circ\hat{\ve f}\ra}})$ obtained from the bilinear program in (\ref{luf__}) vector $\hat{\ve f}$ is arbitrarily scalable
by $\alpha>0$. 
Due to our assumption (\ref{cse}),
a power allocation $\hat{\ma P}\in\Ps$ satisfying 
\begin{equation}
\label{fha}
f_k(\hat{\ve p}_k)=\alpha\hat{\ve f},
\quad k\in\Ks,
\end{equation}
always exists and is trivially constructed whenever  $\alpha>0$ is chosen sufficiently small:
Under an appropriate $\alpha$, any user $k\in\Ks$ accessing the parallel channels simply assigns on any channel $n\in\Ns$
a transmit power $\hat{p}_{kn}$ which achieves performance $\alpha\hat{f}_n$ and the resulting power allocation $\hat{\ma P}$ remains allowable.
By iterative increasing of $\alpha$ in suitably small steps, we achieve, with some accuracy, the particular largest $\alpha$ for which (\ref{fha}) is yet fulfilled for some $\hat{\ma P}\in\Ps$. Under a simple structure of the set of allowable power allocations, e.g. (\ref{ppc}), such value of $\alpha$ is often computable directly/non-iteratively once the performance functions $f_k$, $k\in\Ks$, are known.
For such particular $\alpha$ 
we achieve the tightest lower bound in (\ref{luf__}) among all 
$\alpha\hat{\ve f}$ inside the set $\hat{\Fs}(\ve r,\ve y)$
which is further smaller than the corresponding bound for any $\alpha\hat{\ve f}\in\bar{\Fs}(\ve r,\ve y)$ (recall (\ref{fhaty}), (\ref{lfo})).

By (the proof of) Lemma \ref{lm3}, a candidate vector $\ve y$ in (\ref{luf__}) is computable as a solution of an unconstrained convex problem.
As a first approach we prefer, however, to apply the simplification $\ve y=\ve 1$, which implicitly enforces $\ve x$ to satisfy $(\ve x,\ve 1)\in\Xs(\ma R\ma V,\ve r,\ve 1)$. This results in the following simple procedure, for which $\ve r$, $\gamma_k$, $k\in\Ks$, and
the set $\Es$ of user pairs not allowed to share a channel (equivalently, sharing graph $G=(\Ks,\Es)$) are given as input parameters along with some suitably small $\alpha,\delta>0$. 
\textit{\begin{Alg}
\label{alg1}
\text{ }\vspace{-14.0 pt}\\
\begin{algorithmic}[1]
\STATE Find a sharing matrix $\hat{\ma A}$ and vector $\hat{\ve f}$ from (\ref{luf__}), $\ve y=\ve 1$, by any bilinear programming method \cite{Whi92}, \cite{ShS80}.
\STATE Compute a power allocation $\hat{\ma P}$ from (\ref{fha}).
\STATE If $\hat{\ma P}\in\Ps$ then set $\alpha\mapsto\alpha+\delta$ and go to step 2, otherwise stop.
\end{algorithmic}
\end{Alg}}
With Proposition \ref{prop23}, the user performance of the obtained fair policy $(\hat{\ma A},\hat{\ma P})$ is immediately evident from the proof of Proposition \ref{ubprop2} (see bounds (\ref{jbound3})).

\begin{figure}
\centering
\scalebox{0.59}{\hspace{3 cm}\input{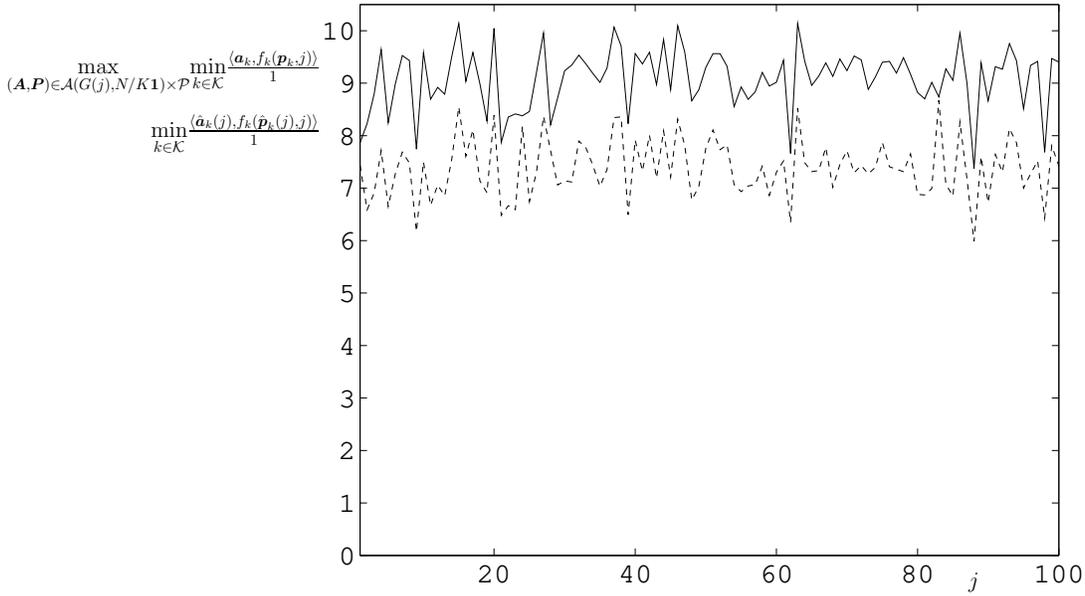}}
\caption{The comparison of user performance under the policy $(\hat{\ma A},\hat{\ma P})$ from Algorithm 1 (dashed line) with the max-min fair performance (solid line), with sum-power constraint (4a) and with the capacity (8) as performance function. 
We simulated parallel channels with $K=4$ users accessing $N=6$ channels, equal user performance requirements $\gamma_k=1$, $k\in\Ks$, and $\ve r=N/K\ve 1$. The sharing graphs $G(j)$, $1\leq j\leq 100$, were picked randomly from all graphs with vertex set $\Ks$ and edges occurring independently with probability $0.5$. The channels $h_{kn}(j)$ and the variances $\sigma_{kn}(j)$, $1\leq j\leq 100$ were picked randomly from uniform distributions resulting in an average Signal-to-Noise Ratio of 20 dB.}
\label{fig:figsix}
\end{figure}

\begin{Cor}
\label{conve1}
\textit{Given $\Es$ such that $G=(\Ks,\Es)$ has no odd cycles longer than $4$, the policy $(\hat{\ma A},\hat{\ma P})$ from Algorithm \ref{alg1} satisfies the bounds from Proposition \ref{ubprop2} for $\ve y=\ve 1$, which implies
\begin{equation*}
\frac{\min_{\ve f\in\tilde{\Fs}(\ve r)}\la\ve f,\ve f\ra}{\theta^2(G,\ve w)}
\geq 
\min_{k\in\Ks}\frac{\la\hat{\ve a}_k,f_k(\hat{\ve p}_k)\ra^2}{\gamma_k^2}
\geq
\frac{\min_{k\in\Ks}x_k\max_{\ve f\in\bar{\Fs}(\ve r)}\la\ve f,\ve f\ra}{\theta^2(G,\ve w)},
\end{equation*}
where $(\ve x,\ve 1)\in\Xs(\ma R\ma V,\ve r,\ve 1)$
subject to (\ref{vfac}) and
\begin{equation}
\label{bprb}
\ma B=\arg\min_{\bma B\in\Bs^2(G,\ve w)}\lmax(\bma B).	
\end{equation}
Thus, $\min_{k\in\Ks}\frac{\la\hat{\ve a}_k,f_k(\hat{\ve p}_k)\ra^2}{\gamma_k^2}$ is at most
\begin{equation}
\label{conveq1}
\frac{\min_{\ve f\in\tilde{\Fs}(\ve r)}\la\ve f,\ve f\ra -\min_{k\in\Ks}x_k \max_{\ve f\in\bar{\Fs}(\ve r)}\la\ve f,\ve f\ra}{\theta^2(G,\ve w)},
\end{equation}
away from the max-min fair performance under given $\Es$.}
\end{Cor}
Fig. \ref{fig:figsix} provides an exemplary comparison between the user performance achieved by the policy from Algorithm \ref{alg1} and the max-min fair performance.
For the evaluated ensemble of parallel channels (with their sharing graphs) we observe a loss of about 20 \% to the max-min fair performance. One can also show by simulation that such loss decreases if the differences between the user channel vectors $\ve h_k$, $k\in\Ks$, and the differences between the variance ensembles $\sigma_{kn}$, $n\in\Ns$, of users $k\in\Ks$ diminish. In fact, this behavior can be recognized already from the feature (\ref{fha}) of the policy from Algorithm \ref{alg1}, which means that the resulting performance function value is the same for all users.
Using Corollary \ref{conve1}, the same limit behavior can be also deduced from Figs. \ref{fig:figfour} and \ref{fig:figfive} since in the case of similar channel vectors and variance ensembles the (forms of) user dimensions of the feasible QoS set become similar as well and make the $2$-norm gap between the hull $\tilde{\Fs}(\ve r)$ and the ball $\bar{\Fs}(\ve r)$ vanish. 
As can be expected conversely, under variations between the channel vectors and variance ensembles of users becoming more severe, the loss of the policy from Algorithm \ref{alg1} increases. 

\begin{figure}
\centering
\scalebox{0.59}{\hspace{3 cm}\input{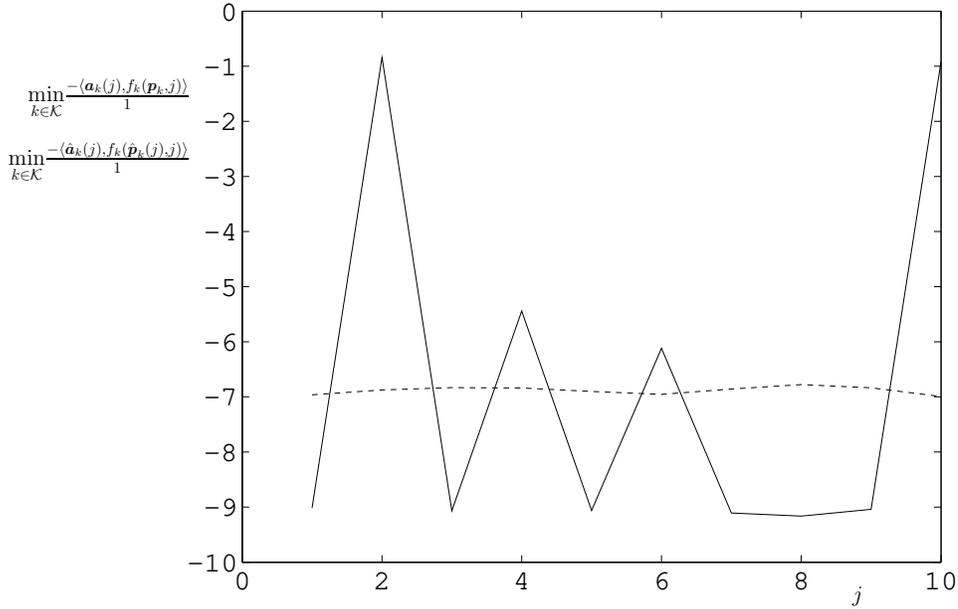}}
\caption{The comparison of user performance under the policy $(\hat{\ma A},\hat{\ma P})$ from Algorithm 1  using the BFGS method in step 1 (dashed line), and under policy $(\ma A,\ma P)$ as a local BFGS solution to problem (67) (solid line), with sum-power constraint (4a) and with the capacity (8) as performance function. We simulated the parallel channels as in Fig. 6, but for $K=3$, $N=4$ and an average Signal-to-Noise Ratio of 10 dB.}
\label{fig:figseven}
\end{figure}

The advantageous complexity-performance trade off of  Algorithm \ref{alg1} becomes evident when the bilinear program in (\ref{luf__}) and the original problem solution (\ref{clprb_}) are both computed by the same \textit{local} optimization method. As shown in Fig. \ref{fig:figseven} for some selected parallel channel instances (and sharing graphs), the efficient and widely used Broyden-Fletcher-Goldfarb-Shanno (BFGS) method may be attracted by highly suboptimal local optima of the original nonlinear problem. On the other hand, the values of the local optima of the bilinear program in (\ref{luf__}) are apparently much less scattered, so that the same BFGS method is able to find a good (local) solution (\ref{luf__}) quite reliably.
As a result, the worst user performance under policy from Algorithm \ref{alg1} happens to be superior to the locally computed max-min fair policy (under given sharing graph).

As a second approach to the scaling of $\ma R\ma V$, instead of the simplification $\ve y=\ve 1$ we can find a scalar scaling so that $(\ve 1,y\ve 1)\in\Xs(\ma R\ma V,\ve r,\ve 1)$ for some $y>0$. 
In this case, $\ve w(\ve x)$ reduces to $\ve w$ defined as (\ref{wgr}) and thus matrix $\ma B$, which gives rise to the nonnegative factorization (\ref{vfac}), needs to satisfy $\ma B\in\Bs^2(G,\ve w)$. In the best case, a particular matrix (\ref{bprb}) is desired.
Since the constraints determining the set (\ref{B2}) are linear, the problem in (\ref{bprb}) corresponds to eigenvalue minimization over a polyhedron, which is a canonical problem in optimization theory and a variety of efficient solution methods exists \cite{Ber99}.
For the nonnegative factorization of any given $\ma B\in\Bs^2(G,\ve w)$, or the particular (\ref{bprb}), we use one of the two celebrated methods which are proposed in \cite{LeS00} and are further extended and analyzed e.g. in \cite{Hoy04}, \cite{Lin07}.
Precisely, for any $m\in\N$, we apply the particular form 
\begin{equation}
\label{lusfac}
{(\ma V(m+1))}_{kn}\!=\!\frac{{(\ma V(m))}_{kn}}{{(\ve 1'\ma V(m))}_n}
\sum_{l\in\Ks}\frac{{(\ma V(m))}_{ln}{(\lmax^{-1}(\ma B)(\ve w\sr{\ve w\sr}'-\ma B)+\ma I)}_{kl}}{{(\ma V(m)\ma V'(m))}_{kl}},
\quad\!\!(k,n)\!\in\!\Ks\times\Ns,
\end{equation}
of the factorization iteration from Theorem 2 in \cite{LeS00}. The sequence $\ma R\ma V(m)$, $m\in\N$, obtained by (\ref{lusfac}) converges monotonically to a matrix $\ma R\ma V$ which achieves a stationary point of the generalized Kullback-Leibler (KL) divergence between $\ma R\ma V\ma V'\ma R$ and (\ref{rlr}) (for the definition of this divergence and further discussion we refer to \cite{LeS00}). The minimization of the generalized KL divergence between a matrix and its factorization
is an intricate problem with multiple local minima, so that $\ma R\ma V\ma V'\ma R$ obtained from (\ref{lusfac}) can happen to remain at a nonzero, but relatively small, generalized KL divergence to (\ref{rlr}). For this reason, we can resort also to alternative factorization iterations, such as the gradient descent method, which seem, however, to be inferior to the methods from \cite{LeS00} in terms of complexity-convergence trade off \cite{LeS00}, \cite{Lin07}.

The above discussion leads to the following second procedure which uses $\ve r$, $\gamma_k$, $k\in\Ks$, and $\Es$ as input data and some sufficiently small parameters $\alpha,\delta>0$.
\textit{\begin{Alg}
\label{alg2}
\text{ }\vspace{-14.0 pt}\\
\begin{algorithmic}[1]
\STATE Compute a matrix (\ref{bprb}) by any convex eigenvalue minimization method \cite{Ber99}.
\STATE Compute $\ma V$ by the iteration (\ref{lusfac}).
\STATE Compute the largest solution $y>0$ of the inequalities $y\ve 1'\ma R\ma V\leq\ve 1'$, $y\ma R\ma V\ve 1\leq\ve r$.
\STATE Find a sharing matrix $\hat{\ma A}$ and vector $\hat{\ve f}$ from (\ref{luf__}), $\ve y=y\ve 1$, by any bilinear programming method \cite{Whi92}, \cite{ShS80}.
\STATE Compute a power allocation $\hat{\ma P}$ from (\ref{fha}).
\STATE If $\hat{\ma P}\in\Ps$ then set $\alpha\mapsto\alpha+\delta$ and go to step 5, otherwise stop.
\end{algorithmic}
\end{Alg}}
By Proposition \ref{prop23}, Theorem 2 in \cite{LeS00} and the proof of Proposition \ref{ubprop2} we have the following result on the user performance under the fair policy $(\hat{\ma A},\hat{\ma P})$ from Algorithm \ref{alg2}.

\begin{Cor}
\label{conve2}
\textit{Assume the generalized KL divergence between (\ref{rlr}) and $\ma R\ma V$, with $\ma V$ computed in step $3$ of Algorithm \ref{alg2}, be zero and let $\Es$ such that $G=(\Ks,\Es)$ has no odd cycles longer than $4$ be given.
Then, the policy $(\hat{\ma A},\hat{\ma P})$ from Algorithm \ref{alg2} satisfies the bounds from Proposition \ref{ubprop2} for $\ve y=y\ve 1$, with $y$ computed in step $3$, which implies
\begin{equation}
\label{conveq2_}
\frac{\min_{\ve f\in\tilde{\Fs}(\ve r)}\la\ve f,\ve f\ra}{\theta^2(G,\ve w)}
\geq 
\min_{k\in\Ks}\frac{\la\hat{\ve a}_k,f_k(\hat{\ve p}_k)\ra^2}{\gamma_k^2}
\geq
\frac{\max_{\ve f\in\bar{\Fs}(\ve r,y\ve 1)}\la y\ve f,y\ve f\ra}{\theta^2(G,\ve w)}.
\end{equation}
Thus, $\min_{k\in\Ks}\frac{\la\hat{\ve a}_k,f_k(\hat{\ve p}_k)\ra^2}{\gamma_k^2}$ is at most
\begin{equation}
\label{conveq2}
\frac{\min_{\ve f\in\tilde{\Fs}(\ve r)}\la\ve f,\ve f\ra -\max_{\ve f\in\bar{\Fs}(\ve r,y\ve 1)}\la y\ve f,y\ve f\ra}{\theta^2(G,\ve w)}
\end{equation}
away from the max-min fair performance under given $\Es$.}
\end{Cor}
Obviously, by adding more technicality, Corollary \ref{conve2} can be extended to the case when the factorization iteration in step $3$ happens to converge only locally, i.e., when the KL divergence between $\ma R\ma V\ma V'\ma R$ and (\ref{rlr}) does not vanish.

The performance and complexity-performance trade off of the policy from Algorithm \ref{alg2} behaves, essentially, quite identically to the policy from Algorithm \ref{alg2} (see Figs. \ref{fig:figsix}, \ref{fig:figseven}). The potential nonzero KL divergence remaining after iteration by (\ref{lusfac}) is hereby hardly visible.

\subsection{Fair policy from factorization}

To summarize so far, by the Algorithms \ref{alg1}, \ref{alg2},  the solution of the original intricate problem (\ref{clprb_}) is replaced by some algebraic operations and the solution of canonical, more efficiently solvable optimization problems: The sharing matrix is obtained directly from the solution of a bilinear program, while the power allocation results from simple scaling (Algorithm \ref{alg1}), respectively, from the solution of eigenvalue minimization, nonnegative factorization and scaling (Algorithm \ref{alg2}). 
As the price payed for this simplification, the resulting fair parallel channels policies are suboptimal, but achieve the worst user performance within the distances (\ref{conveq1}) and (\ref{conveq2}), respectively, from the optimum under given sharing graph $G$.

The proof of Proposition \ref{ubprop2} implies, however,  that a bilinear program can be further exchanged here by nonnegative factorization and a solution of a simple equation system. The key step of the proof which gives rise to such alternative algorithm is the equality (see (\ref{luf2}))
\begin{equation}
\label{fha_}
\frac{\la\ve y\circ\hat{\ve f},\ve y\circ\hat{\ve f}\ra}{\lmax(\ma B)}=
\frac{\la\hat{\ve a}_k,\hat{\ve f}\ra^2}{\gamma_k^2},\quad k\in\Ks,\quad\ma B\in\Bs^2(G,\ve w(\ve z)),
\end{equation}
for the given sharing graph $G$, for any $\ma B\in\Bs^2(G,\ve w(\ve z))$, for map $\ve x\mapsto\ve w(\ve x)$, $\ve x\in\R_{++}^K$, defined in Proposition \ref{ubprop2} and for some $\ve z\in\R_{++}^K$, $\hat{\ma A}\in\As(G,\ve r)$, $\hat{\ve f}\in\R_+^N$ related as follows.
\begin{itemize}
\item Any sharing matrix $\hat{\ma A}$ results from scaling of a factor $\ma R\ma V$, such that (\ref{vfac}), by a scaling $(\ma Z,\ma Y)$ with $(\ve z,\ve y)\in\Xs(\ma R\ma V,\ve r,\ve 1)$, where $\ma Z\ve 1=\ve z$, $\ma Y\ve 1=\ve y$.
\item Factor $\ma R\ma V$ can be split as (\ref{vcx}), $\ve w=\ve w(\ve z)$, with $\ma X=(\ve x_1,\ldots,\ve x_K)'$ as a factor of $\ma I-\lmax^{-1}(\ma B)\ma B$, where $\ma C=(\ve c,\ldots,\ve c)'\in\R_+^{K\times N}$ is orthogonal according to (\ref{cx0}) and determines $\hat{\ve f}$ as $\frac{\ve y\circ\hat{\ve f}}{\sqrt{\la\ve y\circ\hat{\ve f},\ve y\circ\hat{\ve f}\ra}}=\ve c$.
\end{itemize}

Again, in the best case, a nonnegative factor $\ma R\ma V$ of a particular matrix 
(\ref{bprb}), $\ve w=\ve w(\ve z)$, obtained from canonical eigenvalue minimization \cite{Ber99}, is desired. 
As above, such factor is computable by the version (\ref{lusfac}) of a factorization method from \cite{LeS00}.
Once a factor $\ma R\ma V$ is computed, 
vector $\hat{\ve f}$ follows as a solution of a simple vector equation. Precisely, combining (\ref{vcx}) with the definition $\frac{\ve y\circ\hat{\ve f}}{\sqrt{\la\ve y\circ\hat{\ve f},\ve y\circ\hat{\ve f}\ra}}=\ve c$ and the orthogonality condition (\ref{cx0}) shows that $\hat{\ve f}$ is a solution to the equation
\begin{equation}
\label{csplit}
\ma V\frac{\ve y\circ\ve f}{\sqrt{\la\ve y\circ\ve f,\ve y\circ\ve f\ra}}-\lmax^{-1}(\ma B)\ma W\sr(\ve z)=\ve 0,\quad\ve f\in\R_+^N,
\end{equation}
where the definition $\ma W\sr(\ve z)=\diag(\ma W\sr(\ve z))$, ${(\ma W\sr(\ve z))}_{kk}={\ve w\sr(\ve z)}_k$, $k\in\Ks$, is obvious. 

We are free to solve the equation (\ref{csplit}) by any available numerical method; we refer here to \cite{OrR70} for a wide selection of such methods.
Any solution to (\ref{csplit}) is arbitrarily scalable by a positive $\alpha$ and from the discussion in Section \ref{fol} it is clear how a power allocation $\hat{\ma P}\in\Ps$ satisfying (\ref{fha}) is constructed for a sufficiently small $\alpha$. 
Again, by gradual increasing the particular largest $\alpha$ is found, for which (\ref{fha}) yet holds for some allowable power allocation $\hat{\ma P}\in\Ps$. For such an $\alpha$, $\alpha\hat{\ve f}$ achieves the value of the left hand side of (\ref{fha_}), which is further no smaller than the corresponding maximum value among all $\alpha\hat{\ve f}\in\bar{\Fs}(\ve r,\ve y)$.

As a simplified approach to the scaling of $\ma R\ma V$, we find a scalar scaling which yields $(\ve 1,y\ve 1)\in\Xs(\ma R\ma V,\ve r,\ve 1)$, for some $y>0$. Since then $\ve w(\ve z)$ reduces to $\ve w$ given by (\ref{wgr}) and matrices $\ma B\in\Bs^2(G,\ve w)$ have to be considered, the above discussion results in the following procedure (as above, $\Es$, $\ve r$, and $\gamma_k$, $k\in\Ks$, together with suitably small $\alpha,\delta>0$ are given as input parameters).
\textit{\begin{Alg}
\label{alg3}
\text{ }\vspace{-14.0 pt}\\
\begin{algorithmic}[1]
\STATE Compute a matrix (\ref{bprb}) by any convex eigenvalue minimization method \cite{Ber99}.
\STATE Compute $\ma V$ by the iteration (\ref{lusfac}).
\STATE Compute the sharing matrix $\hat{\ma A}=y\ma R\ma V$, for the largest solution $y>0$ of the inequalities $y\ve 1'\ma R\ma V\leq \ve 1$, $y\ma R\ma V\ve 1\leq\ve r$.
\STATE Compute a solution $\hat{\ve f}$ to equation (\ref{csplit}) by any numerical method \cite{OrR70}.
\STATE Compute a power allocation $\hat{\ma P}$ from (\ref{fha}).
\STATE If $\hat{\ma P}\in\Ps$ then set $\alpha\mapsto\alpha+\delta$ and go to step 5, otherwise stop.
\end{algorithmic}
\end{Alg}}

According to Proposition \ref{prop23}, Theorem 2 in \cite{LeS00} and Proposition \ref{ubprop2}, the fair policy 
$(\hat{\ma A},\hat{\ma P})$ computed by Algorithm \ref{alg3} achieves the following user performance.

\begin{figure}
\centering
\scalebox{0.59}{\hspace{3 cm}\input{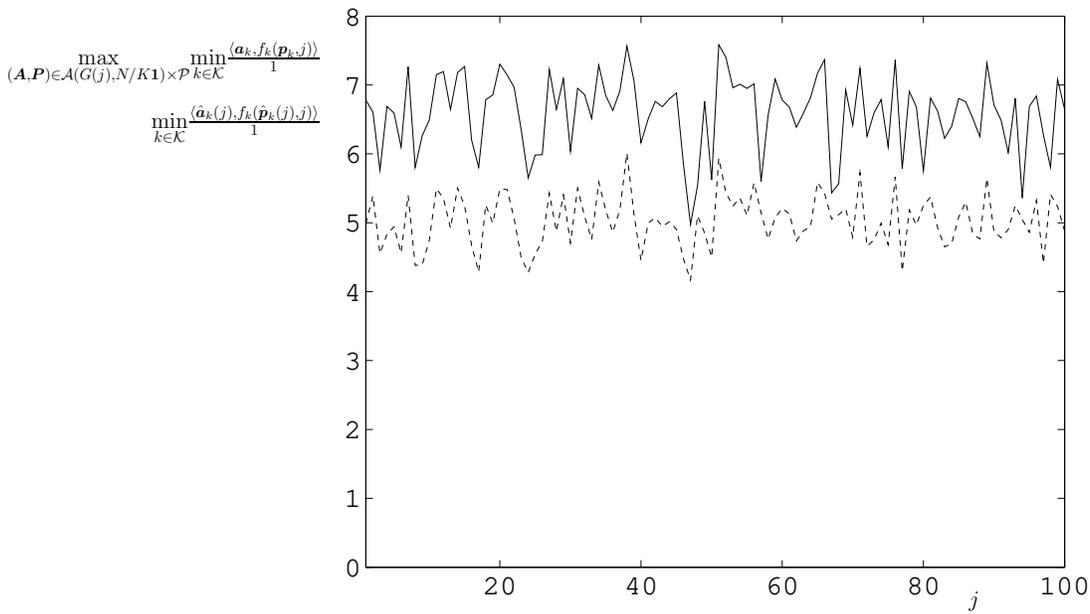}}
\caption{The comparison of user performance under the policy $(\hat{\ma A},\hat{\ma P})$ from Algorithm 3 (dashed line) with the max-min fair performance (solid line), with sum-power constraint (4a) and with the capacity (8) as performance function. We simulated the parallel channels as in Fig. 6, but for $K=6$ and $N=7$.}
\label{fig:figeight}
\end{figure}

\begin{Cor}
\label{conve3}
\textit{Assume the generalized KL divergence between (\ref{rlr}) and $\ma R\ma V$, with $\ma V$ computed in step $2$ of Algorithm \ref{alg3}, be zero and let $\Es$ such that $G=(\Ks,\Es)$ has no odd cycles longer than $4$ be given.
Then, the policy $(\hat{\ma A},\hat{\ma P})$ from Algorithm \ref{alg3} satisfies the bounds from Proposition \ref{ubprop2} for $\ve y=y\ve 1$, with $y$ computed in step $3$, which implies (\ref{conveq2_}).
Thus, 
$\min_{k\in\Ks}\frac{\la\hat{\ve a}_k,f_k(\hat{\ve p}_k)\ra^2}{\gamma_k^2}$ is at most (\ref{conveq2})
away from the max-min fair performance under given $\Es$.}
\end{Cor}
Fig. \ref{fig:figeight} shows an exemplary comparison of user performance achieved under the policy from Algorithm \ref{alg3} and the max-min fair performance. It is evident that the average loss to the max-min fair performance is about 23 \% for the simulated instances of parallel channels and their sharing graphs (thus, the potential nonzero KL divergence remaining after iteration (\ref{lusfac}) does hardly manifest itself in a gap to the performance of Algorithm \ref{alg1}).
By the feature (\ref{fha}), or by Corollary \ref{conve3} and the Figs. \ref{fig:figfour}, \ref{fig:figfive}, we recognize again that such loss evolves analogously as in the case of 
Algorithms \ref{alg1} and \ref{alg2}; it decreases with the user channel vectors and user noise variance ensembles converging to common values, and increases with the corresponding variations becoming stronger. 

\section{Conclusions}
\label{concl}

This work allows for several novel conclusions on the behavior of the max-min fair performance in parallel channels, understood as the maximum attainable worst
user performance. We assumed a very general performance function which is subject to the max-min fairness criterion; it includes the most celebrated functions in communications and information theory (capacity, spectral efficiency, decoder reliability) as very special cases.
We succeeded in embracing the max-min fair performance by  optimistic and pessimistic bounds which show, under constraints on transmit power, the 
same behavior as functions of the channel sharing topology.
This lead to the first central conclusion that the max-min fair performance in parallel channels behaves as a special
extension of the Lovasz function, or Delsarte bound, of a certain graph $G$ (the sharing graph) characterizing the combinatorial topology of channel sharing among the users. 
An essential role is played hereby by the minimum spectral characterization $\lmax(\ma B)$ achievable within certain $G$-dependent sets $\Bs^2(G,\ve w)$, $\Bs^3(G,\ve w)$ with vector $\ve w$ as a parameterizing vector determined by the user performance requirements. When such spectral characterization of the channel sharing topology is obtained, the characterization of the real-valued subproblem of power allocation to users and shared channels by a simple $2$-norm distance is sufficient 
for embracing the max-min fair performance by the proposed bounds: The influence of all properties of the allowable power allocations and all analytic features of the used QoS functions on the max-min fair performance is accumulated in a simple $2$-norm gap between a certain hull and a certain included ball of the feasible QoS set of the parallel channels.

Our results showed also that a key role is played by the existence/nonexistence of cycles in the sharing graph, interpretable as closed chains of users such that any two subsequent users in such a chain share a channel: We showed that under nonexistence of long odd chains of such type, the max-min fair performance is characterized by the minimum achievable $\lmax(\ma B)$ in the specific set $\Bs^2(G,\ve w)$ as a function of the channel sharing topology (and the gap between the proposed bounds is equal precisely to the $2$-norm gap between some hull and some included ball of the feasible QoS set).
As a byproduct of our calculations, we also
illustrated a relation of the max-min fair performance in parallel channel to the graph capacity and independence number of the graph describing the channel sharing topology.

The constructive proofs of our bounds allowed further for the formulation of three novel power and time allocation algorithms for parallel channels with predefined channel sharing topologies (which is the case, e.g., under certain regulations/constraints on QoS class processing).
The algorithms offer a nice performance-complexity trade off and incorporate some surprising techniques, such as nonnegative factorization.

\appendix

\subsection{Doubly nonnegative and completely positive matrices}
\label{app}

\begin{Def}[\cite{GrW80}]
\label{dnmd}
\textit{A matrix $\ma X\in\R^{K\times K}$ is said to be doubly nonnegative, and we write $\ma X\in\D^K$, if $\ma D\in\R_+^{K\times K}$ and $\ma D\succeq 0$.}
\end{Def}
In simple words, a matrix $\ma X\in\D^K$ is nonnegative in the conventional order $\geq$ on $\R^{K\times K}$ and in the partial order $\succeq$ on the set of symmetric matrices in $\R^{K\times K}$.

\begin{Def}[\cite{GrW80}]
\label{cpmd}
\textit{A matrix $\ma X\in\R^{K\times K}$ is said to be completely positive, and we write $\ma X\in\P^K$, if there exists some $N\in\N$ such that 
\begin{equation}
\label{cpmdef}
\ma X=\ma Y\ma Y'\quad\text{for some}\quad\ma Y\in\R_+^{K\times N}.
\end{equation}
The smallest number $N$ for which we have (\ref{cpmdef}) is referred to as the cp-rank of $\ma X$ and is denoted as $N=\phi(\ma X)$.}
\end{Def}
Condition (\ref{cpmdef}) is frequently used in its equivalent form as
\begin{equation*}
\ma X=\sum_{i=1}^N\ve y_i\ve y_i'\quad\text{for some}\quad\ve y_i\in\R_+^K,\quad 1\leq i\leq N,
\end{equation*}
where $\ma Y=(\ve y_1,\ldots,\ve y_N)$ is assumed.

By Definition \ref{cpmd}, it is readily seen that $\ma X\in\P^K$ implies $\ma X\in\D^K$ ($\P^K\subset\D^K$). By the celebrated result from \cite{GrW80} it is further known that $\P^K=\D^K$ whenever $K\leq 4$, while otherwise examples of matrices $\ma X\in\D^K$ such that $\ma X\notin\P^K$ can be constructed.

\subsection{Association schemes}
\label{app_}

From the view of graph theory, the most accessible definition of an association scheme is based on the notion of \textit{edge coloring} of a graph, as a partition of its edge set into vertex-disjoint edge classes. Precisely, an edge $M$-coloring of a graph $G=(\Ks,\Es)$ corresponds to the tuple $(\Ks,\{\Es_i\}_{i=1}^M)$, where $(k,l),(m,n)\in\Es_i$ implies that $k\neq m$ and $l\neq n$ \cite{Bol98}.

\begin{Def}[\cite{Bai04}]
\label{assch}
\textit{An association scheme with $M$ associate classes on a set $\Ks$ is an edge $M$-coloring of a (complete) graph $G=(\Ks,\Ks^2)$ such that\\
i.) for any $1\leq k,l,m\leq M$ there exists so called intersection number $p_{kl}^m\in\N$ such that $p_{kl}^m=\vert\{n\in\Ks: (i,n)\in\Es_k, (n,j)\in\Es_l\}\vert$ whenever $(i,j)\in\Es_m$,\\
ii.) for any $1\leq k\leq M$ there exists $q_k\in\N$ such that $q_k=\vert\{(i,j)\in\Es_k:i=n\}\vert$ for any $n\in\Ks$.\\
iii.) $\Es_k\neq\o$, $1\leq k\leq M$.}
\end{Def}

\begin{Def}
\label{cli}
\textit{Given an association scheme $(\Ks,\{\Es_i\}_{i=1}^M)$ and any $\Ms\subset\{1,\ldots,M\}$, we refer to $\Ls\subset\Ks$ as an $\Ms$-clique of the association scheme if $i,j\in\Ls$, $i\neq j$, implies $(i,j)\in\cup_{m\in\Ms}\Es_m$.}
\end{Def}

For any association scheme $(\Ks,\{\Es_i\}_{i=1}^M)$ and any its $\Ms$-clique $\Ls\subset\Ks$, the unweighted Delsarte number can be formulated as the map
\begin{equation}
\label{dbd}
((\Ks,\{\Es_i\}_{i=1}^M),\Ls,\Ms)\mapsto
\max_{\substack{\ve a\in\As^1((\Ks,\{\Es_i\}_{i=1}^M),\Ls,\Ms):\\1+\langle\ve a,\bs\sigma\rangle\geq 0}}1+\langle\ve a,\ve 1\rangle,
\end{equation}
where $\As^1((\Ks,\{\Es_i\}_{i=1}^M),\Ls,\Ms)$ denotes the set of so-called inner distributions of the $\Ms$-clique $\Ls$ and $\bs\sigma=(\sigma_1,\ldots,\sigma_M)\in\R^M$ collects especially normalized eigenvalues of the adjacency matrices of graphs $(\Ks,\Es_m)$, $1\leq m\leq M$, having a common eigenvector \cite{Del73}.

\subsection{Additional bound formulations}
\label{ares}

\begin{Cor}
\label{lbcor}
\textit{Given $N\geq K$, any $G=(\Ks,\Es)$ and any 
$(\ma A,\ma P)\in\As(G,\ve r)\times\Ps(\ma A)$, $\ve r\in\R_{++}^K$, we have
\begin{equation}
\label{lb_}
\min_{k\in\Ks}\frac{\langle\ve a_k,f_k(\ve p_k)\rangle^2}{\gamma_k^2}\leq
\frac{\min_{\ve f\in\R_+^N}\langle\ve f,\ve f\rangle }{\theta^i(G(\ma A),\ve w(\ve f))},\quad i=0,1,2,
\end{equation}
where $\ve f\mapsto\ve w(\ve f)$, $\ve f\in\R_+^N$, is such that, for any $k\in\Ks$,
\begin{equation*}
w_k(\ve f)\leq\frac{\gamma_k^2\langle\ve a_k,\ve f\rangle^2}{r_k^2\langle\ve a_k,f_k(\ve p_k)\rangle^2}
\quad\text{if}\quad i=1,\quad\quad\quad
w_k(\ve f)=\frac{\gamma_k^2\langle\ve a_k,\ve f\rangle^2}{r_k^2\langle\ve a_k,f_k(\ve p_k)\rangle^2}
\quad\text{if}\quad i=0,2.
\end{equation*}}
\end{Cor}

\begin{proof}
The proof is a slight modification of the proof of Proposition \ref{lbprop}. For any $\ma A\in\As(G,\ve r)$, $\ve f\in\R_+^N$ and $\ma P\in\Ps(\ma A)$ such that $\la\ve a_k,f_k(\ve p_k)\ra\neq 0$, $k\in\Ks$ (which by our assumptions on $\Ps(\ma A)$ and $f_k$, $k\in\Ks$, exists), let $\ma Z=(\ve z_1,\ldots,\ve z_K)'\in\R^{K\times N}$, be given as
\begin{equation*}
\ve z_k=\sqrt{\frac{w_k(\ve f)}{\langle\ve f,\ve f\rangle}}\ve f-
\frac{\gamma_k\sqrt{\langle\ve f,\ve f\rangle}}{r_k\langle\ve a_k,f_k(\ve p_k)\rangle}\ve a_k,\quad k\in\Ks.
\end{equation*}
Then, 
\begin{equation*}
\label{ }
\langle\ve z_k,\ve z_l\rangle\!\!=\!\!\sqrt{w_k(\ve f)w_l(\ve f)}
-\frac{\gamma_k\sqrt{w_l(\ve f)}\langle\ve a_k,\ve f\rangle}{r_k\langle\ve a_k, f_k(\ve p_k)\rangle}
-\frac{\gamma_l\sqrt{w_k(\ve f)}\langle\ve a_l,\ve f\rangle}{r_l\langle\ve a_l, f_l(\ve p_l)\rangle}
+\frac{\gamma_k\gamma_l\langle\ve f,\ve f\rangle\langle\ve a_k,\ve a_l\rangle}{r_kr_l\langle\ve a_k,f_k(\ve p_k)\rangle\langle\ve a_l, f_l(\ve p_l)\rangle},
\end{equation*}
for any $k,l\in\Ks$, 
so that by the definition of $\ve w$ in the case $i=0,2$ we yield again (\ref{zklin}), with $\ve w=\ve w(\ve f)$ and with $\la\ve z_k,\ve z_k\ra=-w_k(\ve f)+\frac{\gamma_k^2\la\ve a_k,\ve a_k\ra\la\ve f,\ve f\ra}{r_k^2\la\ve a_k,f_k(\ve p_k)\ra^2}$ in particular, while by Definition \ref{or} again (\ref{zkleq}) for $\ve w=\ve w(\ve f)$ is satisfied. In the case of $i=1$, the definition of $\ve w$ implies $\la\ve z_k,\ve z_k\ra\leq-w_k(\ve f)+\frac{\gamma_k^2\la\ve a_k,\ve a_k\ra\la\ve f,\ve f\ra}{r_k^2\la\ve a_k,f_k(\ve p_k)\ra^2}$ and together with Definition \ref{or} also 
\begin{equation*}
\la\ve z_k,\ve z_l\ra\leq-\sqrt{w_k(\ve f)w_l(\ve f)},\quad (k,l)\notin\Es,\quad k\neq l.
\end{equation*}
Thus, given $i=0,2$, we can write (\ref{bz}) with $\bma A=\ma A$ and $\ve w=\ve w(\ve f)$ by the definition (\ref{B2}), while $-\ma B\succeq\ma Z\ma Z'-\la\ve f,\ve f\ra\ma G(\ma A)$ is satisfied for some $\ma B\in\Bs^1(G,\ve w(\ve f))$ in the case $i=1$.
In either case (\ref{giz}) is implied and up from (\ref{giz}) the proof goes as the proof of Proposition \ref{lbprop}.
\end{proof}

\begin{Prop}
\label{ubprop2}
\textit{Given any $G=(\Ks,\Es)$ and $\ve r\in\R_{++}^K$, we have
\begin{equation*}
\max_{(\ma A,\ma P)\in\As(G,\ve r)\times\Ps(\ma A)}\min_{k\in\Ks}
\frac{\langle\ve a_k,f_k(\ve p_k)\rangle^2}{\gamma_k^2}
\geq
\max_{\substack{\ma B\in\Bs^3(G,\ve w(\ve x))\\(\ve x,\ve y)\in\Xs(\ma R\ma V,\ve r,\ve 1),\text{ }\ma V\in\R_+^{K\times N}:\\\ma V\ma V'=\lmax^{-1}(\ma B)(\ve w\sr{\ve w\sr}'-\ma B)+\ma I}}
\!\!\frac{\max_{\ve f\in\bar{\Fs}(G,\ve r,\ve y)}\la\ve y\circ\ve f,\ve y\circ\ve f\ra}{\lmax(\ma B)},
\end{equation*}
where $\ve x\mapsto\ve w(\ve x)$, $\ve x\in\R_{++}^{K}$, is such that
\begin{equation*}
w_k(x_k)=\frac{\gamma_k^2}{(x_kr_k)^2},\quad k\in\Ks,
\end{equation*}
where $N\in\N$
satisfies (\ref{nkp}), and where, given $\hat{\ma A}$ defined as in Corollary \ref{lbcorf}, we defined
\begin{equation*}
\begin{split}
\bar{\Fs}(G,\ve r,\ve y)=\{\ve f\in\R_+^N:&\la\ve y\circ\bve f,\ve y\circ\bve f\ra\leq\la\ve y\circ\ve f,\ve y\circ\ve f\ra\Rightarrow\\
&\bve f=f_k(\ve p_k),k\in\Ks,\quad\text{for some}\quad\ma P\in\Ps(\hat{\ma A})\},
\end{split}
\end{equation*}
$\ve y\in\R_{++}^N$.
Moreover, given a particular 
\begin{equation*}
\ma B=\arg\min_{\bma B\in\Bs^3(G,\ve w(\ve x))}\lmax(\bma B),
\end{equation*}
this further implies
\begin{equation*}
\max_{(\ma A,\ma P)\in\As(G,\ve r)\times\Ps(\ma A)}\min_{k\in\Ks}
\frac{\langle\ve a_k,f_k(\ve p_k)\rangle^2}{\gamma_k^2}
\geq
\max_{\substack{(\ve x,\ve y)\in\Xs(\ma R\ma V,\ve r,\ve 1),\text{ }\ma V\in\R_+^{K\times N}:\\\ma V\ma V'=\lmax^{-1}(\ma B)(\ve w\sr{\ve w\sr}'-\ma B)+\ma I}}
\!\!\frac{\max_{\ve f\in\bar{\Fs}(G,\ve r,\ve y)}\la\ve y\circ\ve f,\ve y\circ\ve f\ra}{\theta^3(\ma B,\ve w(\ve x))}.
\end{equation*}}
\end{Prop}

\begin{proof}
Under the substitution $\ve w=\ve w(\ve x)$, with an arbitrary $\ve x\in\R_{++}^K$, the proof goes exactly as the proof of Proposition \ref{ubprop1} up to the implication (\ref{lf1}) for any $\ve f\in\R_+^N$ such that $\frac{\ve f}{\sqrt{\la\ve f,\ve f\ra}}=\ve c$ is satisfied for the particular $\ve c$ from (\ref{vcw}). Again, by Lemma \ref{lm4}, it follows that we can always find some $(\ve z,\ve y)\in\Xs(\ma R\ma V,\ve r,\ve 1)$ such that (\ref{vfac}), $\ve w=\ve w(\ve x)$, and thus we have $\ma U\in\As(G,\ve r)$ for $\ma U=\ma Z\ma R\ma V\ma Y$ with $\ma Z\ve 1=\ve z$, $\ma Y\ve 1=\ve y$. Writing now $\ve f=\ve y\circ\bve f$, for some $\bve f\in\R_+^N$, and $\ma U=(\ve u_1,\ldots,\ve u_K)'$ and setting $\ve x=\ve z$, we have by the definition of the map $\ve x\mapsto\ve w(\ve x)$, $\ve x\in\R_{++}^K$, that
\begin{equation*}
\frac{w_k(z_k)}{\la\ve v_k,\ve y\circ\bve f\ra^2}=\frac{\gamma_k^2}{(\sum_{n\in\Ns}z_kr_k{(\ve v_k)}_ny_n\bar{f}_n)^2}
=
\frac{\gamma_k^2}{(\sum_{n\in\Ns}{(\ve u_k)}_n\bar{f}_n)^2}
=
\frac{\gamma_k^2}{\la\ve u_k,\bve f\ra^2},\quad k\in\Ks.
\end{equation*}
With (\ref{lf1}), $\ve w=\ve w(\ve z)$, this yields
\begin{equation}
\label{luf2}
\frac{\lmax(\ma B)}{\la\ve y\circ\bve f,\ve y\circ\bve f\ra}=\frac{\gamma_k^2}{\la\ve u_k,\bve f\ra^2},\quad k\in\Ks,
\end{equation}
for any $\bve f\in\R_+^N$ such that $\frac{\ve y\circ\bve f}{\sqrt{\la\ve y\circ\bve f,\ve y\circ\bve f\ra}}=\ve c$
is satisfied for $\ve c$ from (\ref{vcw}).
By our assumption (\ref{cse}) and the assumptions on $f_k$, $k\in\Ks$, we can always find a particular $\bve f$ such that $\bve f=f_k(\ve p_k)$, $k\in\Ks$, for an arbitrary $\ma A\in\As(G,\ve r)$ and some $\ma P\in\Ps(\ma A)$, so that it is implied then with (\ref{luf2}) that
\begin{equation}
\label{luf_2}
\begin{split}
\lmax(\ma B)\geq 
&\min_{(\ma A,\ve f)\in\As(G,\ve r)\times\R_+^N}
\max_{k\in\Ks}\frac{\gamma_k^2\la\ve y\circ\ve f,\ve y\circ\ve f\ra}{\la\ve a_k,\ve f\ra^2}\\
=&\min_{\substack{\ma A\in\As(G,\ve r),\ve f\in\R_+^N:\\\ve f=f_k(\ve p_k),k\in\Ks,\text{ for some }\ma P\in\Ps(\hat{\ma A})}}\max_{k\in\Ks}\frac{\gamma_k^2\la\ve y\circ\ve f,\ve y\circ\ve f\ra}{\la\ve a_k,\ve f\ra^2}
=
\min_{\ma A\in\As(G,\ve r)}\max_{k\in\Ks}\frac{\gamma_k^2\la\ve y\circ\hat{\ve f},\ve y\circ\hat{\ve f}\ra}{\la\ve a_k,\hat{\ve f}\ra^2}
\end{split}
\end{equation}
for any $\hat{\ve f}\in\hat{\Fs}(G,\ve r,\ve y)$.
Hereby, we defined (as a straight generalization of $\hat{\Fs}(G,\ve r)$)
\begin{equation}
\label{fhaty}
\begin{split}
\hat{\Fs}(G,\ve r,\ve y)=\{\ve f\in\R_+^N: 
&\ve f=f_k(\ve p_k),k\in\Ks,\quad\text{for some}\quad\ma P\in\Ps(\hat{\ma A}),\\
&\ve f=\arg\max_{\bve f\in\R_+^N}\max_{\bma A\in\As(G,\ve r)}
\min_{k\in\Ks}
\frac{\la\bve a_k,\bve f\ra^2}{\gamma_k^2\la\ve y\circ\bve f,\ve y\circ\bve f\ra}\}.
\end{split}
\end{equation}
Thus, it follows finally that
\begin{equation*}
\frac{\lmax(\ma B)}{\la\ve y\circ\ve f,\ve y\circ\ve f\ra}
\geq
\min_{(\ma A,\ma P)\in\As(G,\ve r)\times\Ps(\hat{\ma A})}\max_{k\in\Ks}
\frac{\gamma_k^2}{\la\ve a_k,f_k(\ve p_k)\ra^2},
\end{equation*}
for any $\ma B\in\Bs^3(G,\ve w(\ve z))$, any $\ma V$ such that (\ref{vfac}), $\ve w=\ve w(\ve z)$, any $\ve y$ satisfying $(\ve z,\ve y)\in\Xs(\ma R\ma V,\ve r,\ve 1)$ and $N\geq\max\{K+1,\phi(\lmax^{-1}(\ma B)(\ve w\sr(\ve z){\ve w\sr(\ve z)}'-\ma B)+\ma I)\}$.
Hereby, by Lemmas \ref{lm1}, \ref{lm2}, the last condition is implied by $N\geq K(K+1)/2$ and additionally, along exactly the same lines as in the proof of Corollary \ref{ubcor1}, it is readily shown that
\begin{equation}
\label{lfo}
\la\ve y\circ\ve f,\ve y\circ\ve f\ra\leq\max_{\bve f\in\hat{\Fs}(G,\ve r,\ve y)}\la\ve y\circ\bve f,\ve y\circ\bve f\ra,\quad\ve f\in\bar{\Fs}(G,\ve r,\ve y).
\end{equation}
This completes the proof of the first inequality of the proposition, while the second inequality is obtained by taking a particular $\bma B$ with $\lmax(\bma B)=\max_{\ma B\in\Bs^3(G,\ve w(\ve z))}\lmax(\ma B)=\theta^3(G,\ve w(\ve z))$.  
\end{proof}


\end{document}